\documentclass[10pt,a4paper]{article}
\usepackage[utf8]{inputenc}
\usepackage[a4paper, total={6in, 8in}, top=1in, bottom=1in, left=1in, right=1in]{geometry}
\usepackage{amsmath}
\usepackage{amsfonts}
\usepackage{amssymb}
\usepackage{physics}
\usepackage{dsfont}
\usepackage{subcaption}
\usepackage{xcolor}
\usepackage{verbatim}
\usepackage{slashed}
\usepackage{comment}
\usepackage{multirow}
\usepackage{mathtools}
\usepackage{longtable}
\usepackage{soul}
\usepackage{listings}

\usepackage{tikz}
\usepackage{tikz-feynman}
\tikzfeynmanset{warn luatex=false}

\usepackage{color}
\definecolor{DeclarationColor}{rgb}{0.0,0.5,0.1}
\definecolor{ExcutableColor}{rgb}{0.1,0.2,0.8}
\definecolor{ModuleColor}{rgb}{0.8,0.1,0.1}
\definecolor{DirectiveColor}{rgb}{0.8,0.1,0.8}
\definecolor{FunctionColor}{rgb}{0.0,0.5,0.9}
\definecolor{StringColor}{rgb}{0.8,0.4,0.0}
\definecolor{CommentColor}{rgb}{0.5,0.5,0.5}
\newcommand{\pfl}[3]{\left\langle #1 \middle| #2 \middle|  #3\right\rangle}

\newcommand{\bx}{\mathbf{x}}
\newcommand{\bxbar}{\mathbf{\bar{x}}}
\newcommand{\by}{\mathbf{y}}
\newcommand{\bybar}{\mathbf{\bar{y}}}

\chardef\MyArticleWithColor=\pdfcolorstackinit page direct{0 g}

\usepackage{fontawesome}

\usepackage{jheppub} %

\gdef\@fpheader{Prepared for submission to JHEP}
\gdef\@journal{jhep}

\title{Manifestly Causal Loop-Tree Duality}

\author[]{Zeno Capatti,}
\author[]{Valentin Hirschi,}
\author[]{Dario Kermanschah,}
\author[]{Andrea Pelloni}
\author[]{and Ben Ruijl}

\affiliation[]{ETH Z\"urich,\\
R\"amistrasse 101, %
8092 Z\"urich, Switzerland}

\emailAdd{zeno.ca@gmail.com}
\emailAdd{valentin.hirschi@gmail.com}
\emailAdd{d.kermanschah@gmail.com}
\emailAdd{apelloni90@gmail.com}
\emailAdd{benruyl@gmail.com}

\abstract{
Loop-Tree Duality (LTD) is a framework in which the energy components of all loop momenta of a Feynman integral are integrated out using residue theorem, resulting in a sum over tree-like structures. Originally, the LTD expression exhibits cancellations of non-causal thresholds between summands, also known as dual cancellations. As a result, the expression exhibits numerical instabilities in the vicinity of non-causal thresholds and for large loop momenta.
In this work we derive a novel, generically applicable, Manifestly Causal LTD (\cLTD{}) representation whose only thresholds are causal thresholds, i.e. it manifestly realizes dual cancellations. Consequently, this result also serves as a general proof for dual cancellations.
We show that LTD, \cLTD{}, and the expression stemming from Time Ordered Perturbation Theory (TOPT) are locally equivalent. TOPT and \cLTD{} both feature only causal threshold singularities, however \cLTD{} features better scaling with the number of propagators.
On top of the new theoretical perspectives offered by our representation, it has the useful property that the ultraviolet (UV) behaviour of the original 4D integrand is maintained for every summand.
We show that the resulting \cLTD{} integrand expression is completely stable in the UV region which is key for practical applications of LTD to the computation of amplitudes and cross sections.
We present explicit examples of the \cLTD{} expression for a variety of up to four-loop integrals and show that its increased computational complexity can be efficiently mitigated by optimising its numerical implementation. Finally, we provide a computer code that automatically generates the \cLTD{} expression for an arbitrary topology.
}

\begin{document}

\newcommand{\cLTD}{cLTD}

\maketitle

\section{Introduction}
The computation of multi-scale multi-loop amplitudes in Quantum Field Theories remains to this day a challenging feat. Traditional analytic methods are based on the reduction of amplitudes to a set of master integrals
~\cite{Tkachov:1981wb,Chetyrkin:1981qh,Gehrmann:1999as,Baikov:1996iu,Smirnov:1999wz,Anastasiou:2000mf,Laporta:2001dd,Anastasiou:2004vj,Smirnov:2005ky,Lee:2008tj,Kant:2013vta,vonManteuffel:2014ixa,Smirnov:2014hma,Ruijl:2017cxj,Maierhoefer:2017hyi,Smirnov:2019qkx,Peraro:2019svx,Frellesvig:2019uqt,Boehm:2017wjc,Kosower:2018obg}
, whose value can be obtained by solving systems of differential equations~\cite{Kotikov:1990kg,Papadopoulos:2014hla,Remiddi:1999ew,Goncharov:1998kja,Brown:2004ugm,Duhr:2012fh,Peraro:2016wsq,Mastrolia:2016dhn,Broedel:2017siw,Passarino:2016zcd,Ablinger:2017bjx,Broedel:2019hyg,Remiddi:2017har,Bonciani:2019jyb,Francesco:2019yqt}
. The property that integrands of amplitudes are rational functions in the integration variables is used to determine the class of functions whose linear combinations reproduce a Feynman diagram.

This same property is also taken advantage of by numerical methods, although in different capacities. At their core, numerical methods perform a Monte-Carlo integration relying on an efficient evaluation and sampling of the integrand, whether in Feynman-parameters space
~\cite{Hepp:1966eg,Roth:1996pd,Binoth:2000ps,Anastasiou:2007qb,Lazopoulos:2007ix,Anastasiou:2005pn,Anastasiou:2008rm,Carter:2010hi,Borowka:2017idc, Smirnov:2008py}
, four dimensional Minkowski space~\cite{Gong:2008ww,Becker:2011vg, Becker:2012bi}, or three-dimensional Euclidean space using Loop-Tree Duality (LTD)~\cite{Soper:1999xk,Catani:2008xa,Bierenbaum:2010cy,Capatti:2019ypt}.
In the LTD approach, the 4D integrand is rewritten as a sum of functions (referred to as \emph{dual integrands}) in the \emph{spatial} components of the loop momenta, and the singularities are defined by the roots of their denominator; polynomials in the on-shell energies of the virtual particles involved in the process together with the energies of the external kinematics.
Such polynomials can be factorized into linear combinations of its variables and the singular surfaces defined by each of these factors are represented by a (convex) manifold embedded in the integration space. 
These manifolds can be divided into two groups, namely H- and E-surfaces.
The H-surfaces are characterized by mixed signs in the linear combination of energies, whereas the E-surfaces have a common sign.

Previous work on the analysis of the singular structure of LTD expressions~\cite{Capatti:2019ypt,Catani:2008xa,Aguilera-Verdugo:2019kbz}
revealed that each of these dual integrands can contain poles defined by both H- and E-surface, whereas the sum of all the contributions only retains the singularities corresponding to E-surfaces.
The elimination of all the H-surfaces from the final expression is referred to as \textit{dual-cancellation}.
Each E-surface that satisfies the existence condition and is not pinched, requires an ad-hoc contour deformation~\cite{Capatti:2019edf}, or other means of regulation as e.g. subtraction of thresholds as in refs.~\cite{kilian2009numerical} and \cite{Kermanschah:2020toAppear} (to appear) and if it is pinched, requires counterterms~\cite{Nagy:2003qn,Assadsolimani:2009cz,Seth:2016hmv,Sborlini:2016gbr,Anastasiou:2018rib,Ma:2019hjq,Anastasiou:2020sdt} or some other local cancellation mechanism that will be the topic of a future publication~\cite{Capatti:2020toAppear}.
\par

Given the polynomial nature of these H-surface singularities, one can resolve the cancellations happening among different dual integrands through an algebraic manipulation: one can obtain a sum of terms which involve only E-surface kind of singularities that is locally equivalent to the original LTD expression. 
Such representation was studied previously for selected cases~\cite{Verdugo:2020kzh,Aguilera-Verdugo:2020kzc,Ramirez-Uribe:2020hes}, including three- and four-loop scalar topologies. Here we propose a systematic and algorithmic procedure that yields such a representation for \emph{any} multi-scale, multi-loop Feynman diagram with an arbitrary numerator. We refer to this new representation as \emph{Manifestly Causal LTD} (\cLTD{}).
\par
We observe that the expression stemming from Time Ordered Perturbation Theory (TOPT) locally evaluates to the same quantity as the \cLTD{} (and LTD) expression, and that its denominator is also constructed from E-surfaces only. However, in general the TOPT expression has many more terms than the \cLTD{} expression.
\par

The general \cLTD{} representation introduced in this paper contains a sum of terms which grows exponentially with the number of propagators. However these many terms are functions of a limited number of combinations of on-shell energies thus offering great potential in optimising their numerical implementation by identifying common sub-expressions.
We demonstrate with benchmark loop integrals that this apparent growth in computational complexity can indeed be efficiently mitigated in this way.
\par
The \cLTD{} expression improves numerical stability w.r.t. that of the LTD representation whenever evaluated in the vicinity of dual-canceling H-surfaces, which is always the case in the ultraviolet (UV) region.
The UV scaling of each dual integrand of the original LTD expression is often many orders of magnitude larger than that of the original four-dimensional representation, and it is only due to dual cancellations that one eventually retrieves the original scaling of the 4D integrand. Especially for physical theories, dual cancellations occur across several orders of magnitude due to the presence of a numerator. 
For example, in the case of the NLO $q \bar q \rightarrow \gamma \gamma \gamma$ amplitude (see sect. 4.2.3 of ref.~\cite{Capatti:2019edf}), the cancellation occurs between diverging terms that scale as $k^2$, down to a scaling of $1/k^2$. 
In \cLTD{}, each summand has the same scaling as the four-dimensional counterpart and is therefore more numerically stable in the UV region.

The outline of this paper is as follows. In section~\ref{sec:generic_partial_fractioning} we will start by presenting a general framework for removing spurious divergences which arise from applying residue theorem to an integral with poles in the integration variable. In section~\ref{sec:causal_ltd} we will explain how this general formula can be applied to physical cases, realising \cLTD{}. In section~\ref{sec:examples} we present several examples illustrating the \cLTD{} procedure. In section~\ref{sec:results} we provide quantitative assessments of the \cLTD{} performance. We present our conclusions in section~\ref{sec:conclusion}.
Finally, we provide a \textsc{python} code that generates the \cLTD{} expression for general topologies as an ancillary file (see \texttt{cLTD.py}) and describe its usage in appendix~\ref{app:pyfractioning}.

\section{Algebraic cancellation of spurious singularities}
\label{sec:generic_partial_fractioning}
 
We consider the case of a function $F$ defined as the integral over the real line of an integrand consisting of a general \emph{regular} function $\mathcal{N}$ in the numerator and a polynomial with $n+\bar n$ distinct roots in the denominator.
We will now present a general procedure to rewrite $F$ as a function which is manifestly regular.
Within our procedure, we first perform the integral by using residue theorem. We then re-express the result by iteratively factoring each of its spurious poles. This yields terms involving divided differences of the numerator $\mathcal{N}$ which evaluate to a finite quantity on the spurious poles, thus establishing that each term of our resulting expression is regular at these locations.
In sect.~\ref{sec:polynomial_numerators} we specialise to the case of polynomial numerators for which the divided differences can be explicitly evaluated algebraically.
Finally, we show in sect.~\ref{sec:degenerate_case} how our procedure can accommodate higher-order poles in $F$ by regarding them as the limiting case of coinciding roots of $\mathcal{N}$.

\subsection{General case}
\label{sec:systematics}
We start by introducing a general function defined as an integral of a regular numerator and a polynomial in $z$ with roots not lying on the real line:
\begin{equation}
\label{eq:F_general}
    F\left(\mqty{\bx\\\bxbar}\,;\mathcal{N}\right)
    :=
    \frac{-1}{2\pi i}\int_{\mathbb{R}}\dd z \;
    \frac{\mathcal{N}(z)}{
        \prod_{i=1}^{n}(z-x_i)\;
        \prod_{i=1}^{\bar n}(z+\bar{x}_i)} \,,
\end{equation}
where $\mathbf{x}\in\left(\mathbb{H}^{*}\right)^{n}$ and $\mathbf{\bar{x}}\in\left(\mathbb{H}^{*}\right)^{\bar n}$, whose components lie in the lower complex half-plane $\mathbb{H}^{*}$.
As we intend to perform the integral using residue theorem, we assume the numerator $\mathcal{N}$ to be regular such that the integrand is holomorphic on $\mathbb{C}\setminus \{\bx,\bxbar\}$ with vanishing residue at infinity.
We can therefore close the contour arbitrarily on an arc in either the upper or the lower complex half-plane so as to correctly evaluate eq.~\eqref{eq:F_general}.

\par
If all components of $\bx$ as well as all components of $\bxbar$ are pairwise distinct, the integrand only has single poles in the complex plane. 
Otherwise, the integrand exhibits higher-order poles.
Although the treatment of higher-order poles may seem more involved at first, the expressions in this section are valid in this case as well, and have to be understood as the limit in which single poles merge (see sect.~\ref{sec:degenerate_case}).
\par
In order to keep our expressions compact we introduce the shorthand notation for a slice of a vector $\by$ and a concatenation with components $x_i$ as
\begin{align}
    \by_{(i,j)} &:= (y_{i},\dots,y_{j}),
    \quad
    \by_{(i,)} := \by_{(i,\dim(\by))},
    \quad
    [\mathbf{y},x_1,\ldots,x_n] := [y_1,\ldots,y_{\text{dim}(\by)}, x_1,\ldots,x_n ] ,
\end{align}
and the following two recurring expressions:
\begin{align}
    \label{eq: E_definition}
	E(x_i|\bybar)
	&:=\frac{1}{\prod_{j=1}^{\dim(\bybar)}(x_i+\bar y_j)},\\[1em]
    \label{eq: N_definition}
    \text{N}_\mathcal{F}([\by,x_i,x_j])
    &:=\frac{ \text{N}_\mathcal{F}([\by,x_j]) - \text{N}_\mathcal{F}([\by,x_i])}{x_j-x_i},
    \quad
    \text{N}_\mathcal{F}([x_i]):=\mathcal{F}(x_i),
\end{align}
for any regular function $\mathcal{F}$ and arbitrary vectors $\by \in \left(\mathbb{H}^*\right)^{\dim(\by)}$ and $\bybar \in \left(\mathbb{H}^*\right)^{\dim(\bybar)}$.
We point out that the recursion for $\text{N}_\mathcal{F}$ is commonly referred to as the algorithm of \emph{divided differences}.
Divided differences are symmetric in their arguments, meaning that $\text{N}_\mathcal{F}([\by])$ is independent of the order in $\by$.
Furthermore, we emphasise that the recursive step in eq.~\eqref{eq: N_definition} does not introduce a pole if $x_j=x_i$ but instead translates into a derivative at $x_i$, as we will see in more detail in sect.~\ref{sec:degenerate_case}.
Note as well that each factor in $	E(x_i|\bybar)$ with $x_i\in \mathbb{H}^*$ has a definite imaginary part, i.e. $\Im(x_i+\bar{y}_j)<0$ and therefore cannot be singular.\footnote{It can be singular in the limit where the imaginary part vanishes, as in the physical case (see sect.~\ref{sec:causal_ltd}, where $E$ will correspond to a product of E-surfaces), which however does not affect the discussion here.}
With these two observations, we can conclude that whereas $\text{N}_\mathcal{F}$ has spurious poles in its arguments, $E$ is manifestly free of spurious poles.

\par
We carry on with the calculation of the integral eq.~\eqref{eq:F_general} by applying residue theorem, such that we arrive at
\begin{equation}\label{eq: F_integrated}
    F\left(\mqty{\bx\\\bxbar}\,; \mathcal{N} \right)
    =
    \sum_{i=1}^{n}
    \frac{
        \text{N}_{\mathcal{N}}([x_i]) E(x_i|\mathbf{\bar x})
    }{
        \prod_{\substack{j\neq i}} (x_i-x_j)}\,,
\end{equation}
which sums the residues located in the lower-half complex plane.
If $n=1$ we define the empty product to be one.
Of course, one is free to close the contour in the upper complex half-plane, which yields a representation of eq.~\eqref{eq: F_integrated} as the sum of residues that is equivalent, yet formally different.%
\par

The poles at $x_i=x_j$, shown explicitly in the denominator of eq.~\eqref{eq: F_integrated}, are spurious and correspond to a higher-order pole in the integrand of $F$ in eq.~\eqref{eq:F_general} which still integrates to a finite quantity with residue theorem.
Such spurious poles can be removed from the integrated expression by means of partial fractioning and, as derived explicitly in appendix~\ref{sec:recursion-derivation}, this procedure can be summarized in a generic recursive step:
\begin{equation}
\label{eq:pf_general_step}
\begin{gathered}
    F\left(\mqty{\by\\\bybar}\,;\mathcal{F}(z)\right)
=   
-\text{N}_\mathcal{F}([y_1])
    \sum_{j=1}^{\dim(\bybar)} E\left(y_1|\bybar_{(j,)}\right) F\left(\mqty{\by_{(2,)}\\\bybar_{(1,j)}};1\right)
+F\left(\mqty{\by_{(2,)}\\\bybar};\text{N}_{\mathcal{F}}([y_1,z])\right),%
\end{gathered}
\end{equation}
which systematically removes all spurious poles in $y_1$, as each summand in the expression is manifestly free of denominators $(y_1-y_j)$ for $1<j\leq \dim(\mathbf{y})$.
Of course, such denominators are still hidden in the divided difference $\text{N}_\mathcal{F}([y_1,z])$. 
However, as established before, $\text{N}_\mathcal{F}$ is regular if $\mathcal{F}$ is and therefore does not introduce singularities to $F$.

This recursion holds for $\dim(\mathbf{y})>1$.
Then, one can iterate the recursion for the remaining instances of $F$ on the right-hand side of eq.~\eqref{eq:pf_general_step} so as to explicitly and successively remove each spurious poles in $y_j$.
Eventually one arrives at the case $\dim(\by)=1$, the boundary condition, where the resulting expression for eq.~\eqref{eq: F_integrated} is already manifestly free of spurious poles. Notice as well that the recursion simplifies whenever the numerator function is constant, since in that case the divided difference appearing in the last summand in eq.~\eqref{eq:pf_general_step} vanishes.
\par

We now apply the generic algorithm in eq.~\eqref{eq:pf_general_step} to the expression in eq. \eqref{eq: F_integrated} and unfold it into
\begin{align}
\label{eq:pf_1loop_compact}
    F\left(\mqty{\bx\\\bxbar}\;;\mathcal{N}\right)
    =&
    \sum_{i=1}^n
    (-1)^{n+i}
    \text{N}_{\mathcal{N}}([\bx_{(1,i)}])
    \sum_{\vec{j}\in T^{\bar n}_{n-i+1}}
    \prod_{k=1}^{n-i+1}
    E(x_{k}|\bxbar_{(j_k,j_{k-1})})
\end{align}
where we defined,
\begin{equation}
    \text{T}_{k}^{m} := \left\{\vec{j}\in \mathbb{N}^k 
    \middle| 
        j_{i+1}\leq j_i \leq m
        \;\textrm{and}\; j_k=1
        \right\}
,
\end{equation}
with the additional boundary term defined as $j_0=\bar{n}$.
We also use the property that when the function $\mathcal{F}(z)$ in \eqref{eq:pf_general_step} is itself a divided difference of the initial univariate numerator function $\mathcal{N}$ as in $\mathcal{F}(z)=\text{N}_\mathcal{N}([\by,z])$ it then fulfils,
\begin{align}
    \text{N}_{\mathcal{F}}([x_1,\dots,x_n])
    =\text{N}_{\mathcal{N}}([\by,x_1,\dots,x_n]) \,.
\end{align}
We emphasise that the expression in eq.~\eqref{eq:pf_1loop_compact} is the main result in this section.
It is an alternative representation of eq.~\eqref{eq: F_integrated} as well as eq.~\eqref{eq:F_general}.
Although eq.~\eqref{eq: F_integrated} provides a compact expression as the sum of residues for the integral representation in eq.~\eqref{eq:F_general}, it features spurious singularities that render each residue potentially singular.
Only in the sum of residues are these singularities subject to exact cancellations.
Our alternative representation in eq.~\eqref{eq:pf_1loop_compact} is manifestly free of those spurious singularities, as each individual summand cannot become singular.
This means that eq.~\eqref{eq:pf_1loop_compact} also does not rely on large cancellations around singularities unlike eq.~\eqref{eq: F_integrated}.
Although formally more complicated, the representation in eq.~\eqref{eq:pf_1loop_compact} can be evaluated in a straight-forward fashion, especially because the divided differences $\text{N}_\mathcal{F}$ are a well-known mathematical concept.
\par 

Furthermore, it is important to note that eq.~\eqref{eq:pf_1loop_compact} is not the only representation of eq.~\eqref{eq: F_integrated} that is free of spurious singularities.
Indeed, a different choice of order in $\bx$ and $\bxbar$ or contour closure will in general yield a different formal representation of eq.~\eqref{eq:pf_1loop_compact}. 
Such different functional representations locally evaluate to the same quantity and all possess the property of being explicitly free of spurious singularities.
\par

Despite ambiguities in the representation of eq.~\eqref{eq:pf_1loop_compact}, the algorithm is symmetric under the exchange of $\bx \leftrightarrow \bxbar$ when selecting the opposite contour closure.
This implies that if $\dim(\bx)=\dim(\bxbar)$ the final number of summands in eq.~\eqref{eq:pf_1loop_compact} is always the same given a particular numerator $\mathcal{N}$.
In general however, the total number of terms varies and depends on the particular choice of numerator.
When considering the case of a constant numerator, all but the first summand in equation \eqref{eq:pf_1loop_compact} drop out.
Also, for a polynomial $\mathcal{F}$ of degree $k$, the divided differences $\text{N}_\mathcal{F}([\by])$ with $\dim(\by)>k+1$ vanish.
The final number of terms generated with a given numerator function can be computed by noticing that
\begin{equation}
    \label{eq:no_elements_in_T_mn}
    \abs{T_{n}^{m}}= 	\mqty(n+m-2\\m -1).
\end{equation}
For example, if the numerator $\mathcal{N}$ is a polynomial of degree $k< n $ the expression in eq.~\eqref{eq:pf_1loop_compact} has a total number of 
\begin{align}
    \label{eq:total_number_of_terms}
   N_{\text{terms}}(n,{\bar n},k)=\sum_{i=1}^{k+1} |T_{n-i+1}^{\bar n}| = \frac{1}{{\bar n}!} \left( \frac{(n+{\bar n}-1)!}{(n-1)!}-\frac{(n-k+{\bar n}-2)!}{(n-k-2)!}\right)
\end{align}
summands, where setting $k=n-1$ establishes an upper bound for an arbitrary numerator $\mathcal{N}$ and yields
\begin{align}
    \label{eq:total_number_of_terms_upper_bound}
   \textrm{max}_k\left[N_{\text{terms}}\right](n,{\bar n})=\frac{(n+{\bar n}-1)!}{\bar{n}!(n-1)!} \,.
\end{align}
This bound is especially important because it highlights a potential computational challenge with the representation in eq.~\eqref{eq:pf_1loop_compact}.
As we see in eq.~\eqref{eq:total_number_of_terms}, the total number of terms grows exponentially in $n$ regardless of the value of $k$.
Regarding computational efficiency, this is an evident drawback of eq.~\eqref{eq:pf_1loop_compact} when comparing to the representation in eq.~\eqref{eq: F_integrated}, which only has $n$ summands.
On the other hand, we will demonstrate for polynomial numerators that eq.~\eqref{eq:pf_1loop_compact} is numerically much better behaved than eq.~\eqref{eq: F_integrated} and that the numerical implementation of the many summands can be drastically optimised by identifying common sub-expressions so that it can be made competitive for physical applications (see sect.~\ref{sec:results}).

\subsection{Degenerate case}\label{sec:degenerate_case}

As we have already touched upon in the previous section, the integrand as defined in eq.~\eqref{eq:F_general} exhibits higher-order poles if the components of $\bx$ or $\bxbar$ are not pairwise distinct but two or more components are equal.
This raises the question of whether the computation of the integral in eq.~\eqref{eq: F_integrated} can still be correct in this case, since it is given by a sum of single pole residues.
As we will see in this section, the case of higher-order poles, the \emph{degenerate case}, does not require special treatment but can be understood as the limit in which multiple single poles merge.
This means that equations in the previous sect.~\ref{sec:generic_partial_fractioning} are still valid when considered in this limit.
Although the limit is not straight-forward to apply to the sum of residues in eq.~\eqref{eq: F_integrated}, it can easily be computed in the final expression in eq.~\eqref{eq:pf_1loop_compact}.
We will now explicitly evaluate the limit of the degenerate case and show that the result is the same as if residues of higher-order poles were used directly to compute the integral in eq.~\eqref{eq:F_general}.
\par

We start by computing the divided difference in eq.~\eqref{eq: N_definition} in the case where $r$ of its arguments are equal.
Since divided differences are symmetric in their arguments, we can arrange them such that the equal components are last.
It follows from the mean value theorem that the degenerate divided difference evaluates to
\begin{align}\label{eq:divided_difference_derivative}
    \text{N}_\mathcal{N}([\bx,\underbrace{y,\dots,y}_{r-\text{times}}])
    =
    \frac{1}{(r-1)!}
    \dv[r-1]{}{y}
    \text{N}_{\mathcal{N}}([\bx,y]),
\end{align}
the $(r-1)$\textsuperscript{th} derivative of the divided difference without degenerate arguments.
In the special case where all of the $n$ single poles merge into one pole located at $x$ of order $n$, eq.~\eqref{eq:pf_1loop_compact} becomes
\begin{align}
\begin{split}\label{eq:pf_fully_degenerate}
    F\left(\mqty{x\cdot\vec{1}\\\bxbar}\;;\mathcal{N}\right)
=&
 \frac{1}{(n-1)!}\sum_{r=0}^{n-1}\mqty(n-1\\r)\left[\dv[r]{\mathcal{N}(x)}{x}\right]\left[\dv[n-1-r]{}{x} E\left(x\middle|\bxbar\right) \right]
\\
=&
 \frac{1}{(n-1)!}\dv[n-1]{}{x}\left[\phantom{\frac{1}{1}}\hspace{-.75em}
    \mathcal{N}(x)\;E\left(x\middle|\bxbar\right) 
    \right],
\end{split}
\end{align}
where we used Leibniz's rule and
\begin{equation}\label{eq:e-surf-derivatives}
     \sum_{\vec{j}\,\in\,\text{T}_{m}^{\bar n}}\;\prod_{k=1}^{m} E\left(x\middle|\bxbar_{(j_{k},j_{k-1})}\right) 
     = \frac{(-1)^{m-1}}{(m-1)!}\dv[m-1]{}{x} E\left(x\middle|\bxbar\right) \,,
\end{equation}
which follows from eq.~\eqref{eq:pf_1loop_compact} when one considers all $x_j$ to be equal, which allows to identify the overlaps of denominators in the product of functions $E$ as raised powers induced by the action of the derivative on $x$. 

Notice that eq.~\eqref{eq:pf_fully_degenerate} shows that the integral commutes with the limit so that we can treat the degenerate case with higher-order poles in the same fashion as the non-degenerate one with single poles only.
This however raises a question regarding numerical evaluation in the degenerate limit.
In general, one then still expects to rely on the computation of derivatives,
unless the numerator is a polynomial as we will now demonstrate.

\subsection{Manifest cancellations for polynomial numerators}
\label{sec:polynomial_numerators}
For most physical applications of our procedure, the numerator $\mathcal{N}$ is a polynomial. For example, loop integrals that arise from gauge theories such as the Standard Model are analogous to $F$ and have a non-trivial polynomial numerator $\mathcal{N}$, as in eq.~\eqref{eq:F_general}. More specifically, as we will see in sect.~\ref{sec:causal_ltd}, the numerator of a loop integral will translate into a polynomial numerator in the loop energy variables, while the variables $\bx$ will correspond to on-shell energies.

As discussed in the previous sect.~\ref{sec:degenerate_case}, the divided differences defined in eq.~\eqref{eq: N_definition} are not singular for degenerate arguments as they correspond to a derivative in this limit.
This however means that a naive evaluation at $x_j=x_i$ of the divided difference in eq.~\eqref{eq: N_definition} is not possible without computing the derivative.
Furthermore, even an evaluation close to the degenerate case, i.e.
close to $x_j=x_i$ is numerically unstable, such that the expression eq.~\eqref{eq: N_definition} is not particularly suitable for numerical evaluations.
We will now demonstrate that this does not pose an issue whenever we deal with polynomial numerators, as we can evaluate divided differences algebraically.
More explicitly, if the numerator $\mathcal{N}$ is a polynomial of finite degree with coefficients $\alpha_i$, we can always explicitly cancel the potentially singular denominator $(x_j-x_i)$ from  eq.~\eqref{eq: N_definition} and rewrite the divided differences as polynomials of smaller degree, depending on the same coefficients $\alpha_i$.
This implies that in practice, using the following manipulations for polynomial numerators, the treatment of the degenerate as well as the non-degenerate case is identical.
For the application of loop integrals this means that raised propagators as well as intersections of singular surfaces do not require special attention regarding the derivation of the \cLTD{} expression.
\par

We start by writing a generic polynomial numerator of degree $r$ as
\begin{equation*}
	\mathcal{N}(z)=\sum_{i=0}^r \alpha_i z^i \,.
\end{equation*}
Making use of the geometric sequence identity
\begin{align}
    (x_1^n-x_2^n)=(x_1-x_2)\sum_{k=0}^{n-1} x_1^{n-k-1} x_2^{k}
\end{align}
one can arrange the difference $\mathcal{N}(x_1)-\mathcal{N}(x_2)$, as it appears in the divided difference $\text{N}_\mathcal{N}([x_1,x_2])$, such that the denominator $x_1-x_2$ can be explicitly cancelled.
Performing all cancellations analogously in the recursive definition in eq.~\eqref{eq: N_definition}, we can express the divided difference $\text{N}_\mathcal{N}([\bx])$ directly as
\begin{align}
    \label{eq:pf_poly_numerator}
    \text{N}_\mathcal{N}([x_1,\dots,x_n]) &= \sum_{s=0}^{r-(n-1)}\alpha^{(1,..,n-1)}_s x_{n}^s \,,
\end{align}
where
\begin{align}
\label{eq:pf_poly_numerator2}
    \alpha^{(1,..,m)}_s	&=\sum_{k=s+1}^{r-(m-1)}\alpha^{(1,..,m-1)}_k x_{m}^{k-s-1},\qquad 
    \alpha^{(1)}_s	=\sum_{k=s+1}^{r}\alpha_k x_{1}^{k-s-1}.
\end{align}
The iteration presented above can be completely unfolded to yield
\begin{align*}
\text{N}_\mathcal{N}([x_1,...,x_n])=
	\sum_{i_0=0}^{r}
	\sum_{i_1=1}^{i_0}
	\sum_{i_2=1}^{i_0-i_1}
	\dots{}
	\sum_{i_{n}=1}^{i_0-\sum_{m=1}^{n-1}i_m}\alpha_{i_0} x_n^{i_0-\sum_{m=1}^{n}i_m}\prod_{m=1}^{n-1}  x_m^{i_m-1} 
\end{align*}
On a side note, another interesting property of a divided difference of polynomials is its correspondence to the quotient $\mathcal{Q}$ in Euclid's division theorem
$\mathcal{N}(x_1)=\mathcal{Q}(x_1) (x_1-x_2) + \mathcal{R},$
which allows to identify $\mathcal{R}=\mathcal{N}(x_2)$ and therefore $\mathcal{Q}(x_2)= \text{N}_\mathcal{N}([x_1,x_2])$.
\par

In general the numerator may not be a polynomial, for example if one applies the partial fractioning procedure to manifestly remove infrared singularities in cross-sections, in which case the numerator can include an observable function, as in ref.~\cite{Capatti:2020toAppear}.
If the numerator function is not a polynomial, the explicit cancellation of factors cannot be carried out explicitly and the divided differences will be subject to the same numerical instabilities as the original expression in eq.~\eqref{eq: F_integrated}.
One can nevertheless cure those cases locally by performing a Taylor expansion around such points to analytically remove such instabilities with arbitrary precision. Alternatively, the degenerate case of raised propagators can be treated globally by computing derivatives, as we showed in sect.~\ref{sec:degenerate_case}.
\par

\section{Manifestly Causal LTD}\label{sec:causal_ltd}

In this section we present an alternative to the multi-loop LTD expression presented in refs.~\cite{Capatti:2019ypt, Bierenbaum:2010cy}, which is manifestly free of spurious singularities (denoted as H-surfaces in ref.~\cite{Capatti:2019ypt} or non-causal threshold singularities in refs.~\cite{Aguilera-Verdugo:2020kzc,Ramirez-Uribe:2020hes}).
The procedure outlined in this section will explicitly perform cancellations of spurious singularities (known as dual cancellations~\cite{Catani:2008xa}), such that the expression does not contain any H-surfaces but only contains causal threshold singularities, known as E-surfaces~\cite{Capatti:2019ypt}.

The extension to the multi-loop case uses the procedure laid out in sect.~\ref{sec:generic_partial_fractioning} for the integration over one loop energy.
Then, both this integration step and the extraction of spurious poles can be repeated for the integrals over each of the remaining loop energies.
We formally write this iterative procedure by showing how the individual terms generated at each step satisfy the necessary requirements for the identity of eq.~\eqref{eq:pf_1loop_compact} to apply again.
Although the formal derivation of the iterative procedure may seem involved, the algorithm can be expressed in a short \textsc{python} code (see the ancillary file \texttt{cLTD.py}).

\par

\subsection{Multi-loop derivation}
\label{sec:multi_loop_derivation}
We consider an $L$-loop integrand with $P$ propagators and perform $j$ loop energy integrations, as in
\begin{equation}
\label{eq:energy_integral_real}
	\mathcal{I}_j = \left(\prod_{i=1}^j \frac{-1}{2\pi \mathrm{i}}\int \dd k^0_i\right) \frac{\mathcal{N}(k_1^0,\dots,k_L^0)}{\prod_{i=1}^P D_i},
\end{equation}
which, for $j=L$ corresponds to same analytically performed integration when deriving the multi-loop LTD expression as in ref.~\cite{Capatti:2019ypt}.
We define the numerator $\mathcal{N}$ as a regular function in the loop energies, such that the integrand has a vanishing residue at infinity, and rewrite the Feynman propagator as
\begin{align}
\label{eq:Feynman_propagator}
\begin{split}
	\frac{1}{D_i}
	    &= \frac{1}{(\ell_i+p_i)^2-m_i^2+\mathrm{i}\delta}
		= \frac{1}{2E_i}\left(\frac{1}{\ell^0_i+p_i^0-E_i}-\frac{1}{\ell^0_i+p_i^0+E_i}\right)
		\\
		&=  \frac{1}{2E_i}\left(\pfl{-\hat{e}_i}{\hat{e}_i}{\vec\lambda_i} -\pfl{\hat{e}_i}{\hat{e}_i}{\vec\lambda_i} \right)
		= \frac{1}{2E_i}\sum_{\sigma_i\in\{-1,1\}}\sigma_i\pfl{-\sigma_i\hat{e}_i}{\hat{e}_i}{\vec\lambda_i},
\end{split}
\end{align}
where we introduced the loop line momentum $\ell_i = \sum_{j=1}^L \lambda_{ij} k_j$, with $\lambda_{ij}\equiv (\vec{\lambda}_{i})_j$ as a linear combination of the loop integration momenta, the shift of the propagator $p_i$ as a linear combination of the external momenta, the mass $m_i$ and the canonical basis vector $\hat e_i$, i.e. $(\hat e_i)_j=\delta_{ij}$.
The on-shell energy 
$E_i=\sqrt{(\vec{l}_i+\vec{p}_i)^2+m_i^2-\mathrm{i}\delta}$ satisfies $\Im E_i<0$ since the causal prescription satisfies $\delta>0$.
Furthermore, we introduced the generic linear propagator $i$ with the notation
\begin{equation*}
		\pfl{\vec{v}^{E}_i}{\vec{v}^{S}_i}{\vec{\lambda}_i} 
		:= \frac{1}{\left(\vec{\lambda}_i\cdot\vec{k}_0 + \vec{v}^E_i\cdot\vec{E} + \vec{v}^S_i\cdot\vec{p}_0\right)},
    \qquad
	\vec{k}_0= \mqty(k_1^0\\ \vdots \\ k_L^0),\qquad
	\vec{p}_0=  \mqty(p_1^0\\ \vdots \\ p_P^0),\qquad
	\vec{E}=  \mqty(E_1\\ \vdots \\ E_P),
\end{equation*}
where $\vec{v}^E_i\in \mathbb{R}^P$ and $\vec{v}^S_i\in \mathbb{R}^P$ define the linear combination of on-shell energies and shifts of propagators, respectively.
The vector $\vec{\lambda}_i\in \mathbb{R}^L$ encodes the flow of the loop energies, introduced in ref.~\cite{Capatti:2019ypt} as the \emph{signature} of the propagator.
Note that the linear combination $\vec{v}^S_i\cdot\vec{p}_0$ has a unique representation if expressed in terms of external momenta, as they define a basis of propagator shifts.
In this derivation we however choose the particular representation above.
Furthermore, we point out that a linear propagator $i$ whose non-zero components of $\vec{v}_i^E$ are all equal corresponds to an E-surface. Otherwise it is an H-surface.

The product of Feynman propagators in eq.~\eqref{eq:energy_integral_real} can then be rewritten in terms of linear propagators as
\begin{equation}
\label{eq:prop_pf_starting_expression}
	 \frac{1}{\prod_{i=1}^P D_i}
		= \left(\prod_{i=1}^P\frac{1}{2E_i}\right)\sum_{\vec{\sigma}\in\{-1,1\}^P}\prod_{i=1}^P\sigma_i\pfl{-\sigma_i\hat{e}_i}{\hat{e}_i}{\vec\lambda_i},
\end{equation}
such that inserting this relation in eq.~\eqref{eq:energy_integral_real} yields an integral over a sum of terms, each containing a product of $P$ linear propagators.
In the next section we will see that each of those summands can be identified with a function $F$ as in eq.~\eqref{sec:generic_partial_fractioning} and show that this will eventually allow us to iteratively perform one integration after the other following the partial fractioning procedure in sect.~\ref{sec:generic_partial_fractioning}, effectively solving it loop-by-loop.
The final expression after carrying out this procedure is free of spurious H-surface singularities and only contains E-surfaces.

\par

\subsection{Loop-by-loop iteration}
\label{sec:loop-by-loop_iteration}

In order to apply the partial fractioning procedure of sect.~\ref{sec:generic_partial_fractioning} in an iterative fashion, we have to identify the linear propagators appearing at each step of the integration with those of eq.~\eqref{eq:F_general}.
We will characterise each iteration by a single integral over a loop energy and one integration after the other.
Therefore, it is necessary to pick an (arbitrary) order $k_1^0,\dots,k_L^0$, with their respective integration contours with winding numbers $\Gamma_j$, $j\in\{1,\dots,L\}$.
Evidently, the integral will be independent of the choice of integration order, as well as the contour closure.
Note however that the representation of the result, including the number of summands, will depend on those choices.
Furthermore, the representation of the result will also be affected by the order in which the recursion of eq.~\eqref{eq:pf_general_step} is applied to cancel the spurious singularities explicitly.
On top of this, there is a freedom of changing the loop momentum routing through the loop diagram.
This is of particular interest since, as we show in sect.~\ref{sec:res:computational_complexity}, it effects the number of terms generated in this recursion and can therefore be used to reduce the size of the final expression.

\par

\paragraph{Induction hypothesis}
In order to make the identification clear with the procedure given in sect.~\ref{sec:generic_partial_fractioning}, we will provide a systematic iteration step.
This iteration will eventually allow us to write the \cLTD{} expression.
We set out our induction hypothesis to be the following:
After explicitly performing $(j-1)$ loop energy integrations, $\mathcal{I}_j$ can be expressed as a sum of integrals over $k_j^0$ of the form
\begin{equation}
    \label{eq:generic_ltd_linear_prop_term}
    \mathcal{I}_j=
    \left(\prod_{i=1}^P\frac{1}{2E_i}\right)
    \sum_{\vec{h}\in\Omega_j} I_{j\vec{h}}, \qquad I_{j\vec{h}}= \frac{-1}{2\pi i}\int_{\mathbb{R}} \dd k_j^0 \mathcal{F}_{\vec{h}}(k_j^0) \prod_{i\in S_{j\vec{h}}} \pfl{\vec{v}^E_{\vec{h}i}}{\vec{v}^S_{\vec{h}i}}{\vec{\lambda}_{\vec{h}i}}.
\end{equation}
where $\Omega_j$ is a set of indices and $S_{j\vec{h}}\subseteq\{1,\ldots,P\}$. $I_{j\vec{h}}$ is the one-dimensional integral of a product between a regular function $\mathcal{F}$ in $k_j^0$ and $|S_{j\vec{h}}|$ linear propagators, whose argument $\vec{v}^E_{\vec{h}i}$ is such that all of its non-zero components have the same sign.
In other words, all linear propagators are E-surfaces.
Since H-surfaces are defined as linear propagators whose energy vectors do not have a consistent sign across their components, this induction will effectively allow to construct an iterative procedure yielding a representation of multi-loop amplitudes that is manifestly free of spurious H-surface singularities. In the following, we will perform the integration for $I_{j\vec{h}}$ and suppress the index $\vec{h}$ for simplicity.

\paragraph{}
If this assumption holds true, we can identify $I_{j}$ in eq.~\eqref{eq:generic_ltd_linear_prop_term} with the starting expression $F$ in eq.~\eqref{eq:F_general} in the partial fractioning procedure in sect.~\ref{sec:generic_partial_fractioning}.
To now make this identification explicit, we first have to split the product of the $|S_{j}|$ linear propagators into three factors,
\begin{equation*}
    \prod_{i\in S_j}
    \pfl{\vec{v}^E_i}{\vec{v}^S_i}{\vec{\lambda}_i}
    =
    \left(
        \prod_{i\in S^0_j}
        \pfl{\vec{v}^E_i}{\vec{v}^S_i}{\vec{\lambda}_i}
    \right)
    \left(
        \prod_{i\in S_j^+}
        \pfl{\vec{v}^E_i}{\vec{v}^S_i}{\vec{\lambda}_i}
    \right)
    \left(
        \prod_{i\in S_j^{-}}
        \pfl{\vec{v}^E_i}{\vec{v}^S_i}{\vec{\lambda}_i}
    \right),
\end{equation*}
characterised by the disjoint sets defined as
\begin{align}
    S^{0}_j&:=
    \left\{
        \vphantom{\frac{\Gamma_j}{\lambda_{ij}}}
        i\in S_j \;\middle|\; \lambda_{ij}=0
        \right\}, \\
    S^{\pm}_j&:=
    \left\{
        i\in S_j \;\middle|\; \lambda_{ij}\neq0
        \text{ and }
        \pm \frac{\Gamma_j}{\lambda_{ij}}\vec{v}^E_i \geq \vec{0}
    \right\},
\end{align}
which is sufficient as a characterisation of all possible factors due to our induction hypothesis requiring that the non-zero components of $\vec{v}^E_i$ all have the same sign. Indeed, because of the induction hypothesis it follows that $S_j=S_j^+\cup S_j^- \cup S_j^0$.
Notice that the linear propagators in $S^{0}_j$ are constant in $k_j^0$ and therefore they are to be treated as simple prefactors of the whole partial fractioning procedure.
Each linear propagator $i\in S^{\pm}_j$ however, has a single pole at $k_j^0=k_{ji}^0$, where
\begin{align}
\label{eq:pole_linear_propagator}
    k_{ji}^0
    \equiv k_j^0
    -
    \frac{1}{\lambda_{ij}}
    \left(
        \vec{\lambda}_i \cdot \vec{k}_0
        +\vec{v}^E_i \cdot \vec{E}
        +\vec{v}^S_i \cdot \vec{p}_0
    \right)
    =
    -\frac{1}{\lambda_{ij}}
    \left(
         \sum_{\substack{r=1\\r\neq j}}^L\lambda_{ir} k_r
         +\vec{v}^E_i\cdot \vec{E}
         + \vec{v}^S_i\cdot \vec{p}_{0}
    \right)
\end{align}
lies either in the lower or the upper complex half-plane. 
Note that $k_{ji}^0$ is independent of $k_j^0$ since the first term cancels against the $k_j^0$ dependency of the first scalar product inside the parenthesis.
Since the causal prescription $\delta>0$ implies that $\Im E_i<0$ for all on-shell energies $i\in\{1,\dots,P\}$ and since we assume that $\vec{v}_i^E$ have consistent signs, it follows that $\Im k_{ji}^0$ has a definite sign that fixes its location on either the lower or the upper complex-half plane, for all possible values of spatial loop momenta. This effectively makes for the absence of residues that are sometimes in and sometimes out of the integration contour, which appeared in the derivation of our original work~\cite{Capatti:2019ypt}, although they dropped out in the final expression.

We therefore conclude that $S_j^+$ and $S_j^-$ enumerate the linear propagators with poles in $k_j^0$ lying inside and outside the contour, respectively.
Each linear propagator $i\in S_j^\pm$ can now be written as
\begin{align}
\label{eq:linear_propagator_identification}
    \pfl{\vec{v}^E_i}{\vec{v}^S_i}{\vec{\lambda}_i}
    =
    \frac{1}{\lambda_{ij}}
    \frac{1}{k_j^0-k_{ji}^0}.
\end{align}
\par
We now explicitly express our induction hypothesis in eq.~\eqref{eq:generic_ltd_linear_prop_term} as the expression in eq.~\eqref{eq:F_general}. According to eq.~\eqref{eq:linear_propagator_identification}, we identify
\begin{equation}
    z \equiv k_j^0, 
    \quad
    \bx \equiv \mathbf{k}_j^0 := \left(k^0_{ji}\right)_{i \in S^+_j},
    \quad 
    \bxbar \equiv \mathbf{\bar{k}}_j^0 := \left(-k^0_{ji}\right)_{i \in S^{-}_j},
\end{equation}    
\begin{equation}
\label{eq:arguments_identifications}
    I_j=\Bigg(\prod_{i\in S_j^+ }\frac{1}{\lambda_{ij}}\Bigg)\Bigg(\prod_{i\in S_j^-}\frac{1}{\lambda_{ij}}\Bigg) 
    \Bigg(\prod_{i\in S^0_j} \pfl{\vec{v}^E_i}{\vec{v}^S_i}{\vec{\lambda}_i}
    \Bigg)
    F \left(\mqty{\mathbf{x}\\ \mathbf{\bar{x}}}\,;\mathcal{F}\right).
\end{equation}
where the notation $(x_i)_{i\in S}$ denotes a vector of components whose index runs over the (integer) elements of $S$ sorted in ascending order. Now that we have seen that the induction hypothesis takes the exact same form as the starting expression in eq.~\eqref{eq:F_general}, we are finally ready to carry out the partial fractioning procedure laid out in sect.~\ref{sec:generic_partial_fractioning}.

To prove the assumption we will first show that it is satisfied for the base case $j=1$.
Then we show that performing both the integration over $k_j^0$ and as the partial fractioning procedure for $I_{j}$ implies that $\mathcal{I}_{j+1}$ satisfies the induction hypothesis in eq.~\eqref{eq:generic_ltd_linear_prop_term}.
\par

\paragraph{Base case}
First we show that the assumption holds for the base case, where $j=1$.
We consider the expression in eq.~\eqref{eq:energy_integral_real}.
Notice that the very first integral over $k_1^0$ in eq.~\eqref{eq:energy_integral_real} together with the decomposition into linear propagators in eq.~\eqref{eq:prop_pf_starting_expression} is a sum of terms, each a product of a regular numerator, 
\begin{equation}
    \mathcal{F}=\Bigg(\prod_{i=1}^P \sigma_i \Bigg) \mathcal{N}
\end{equation}
as well as $P$ linear propagators. Furthermore, $\Omega_1=\{\pm 1\}^P$. To prove that the hypothesis in eq.~\eqref{eq:generic_ltd_linear_prop_term} is indeed satisfied for each of those summands, we have to show that all the non-zero components of $\vec{v}_i^E$ have the same sign for each linear propagator $i$.
For $j=1$ the linear propagators are given as in eq.~\eqref{eq:Feynman_propagator} and all satisfy $\vec{v}_i^E =-\sigma_i \hat e_i$.
Thus, all components of $\vec{v}_i^E$ are zero except for its $i$\textsuperscript{th} component, which proves our hypothesis for $j=1$.

\paragraph{Inductive step}
To prove the hypothesis for $j>1$, we assume that it holds true for $j$ and show that it holds true for $j+1$.
Thanks to the identifications in eq.~\eqref{eq:arguments_identifications} we can directly apply the partial fractioning procedure outlined in sect.~\ref{sec:generic_partial_fractioning}. We can now express eq.~\eqref{eq:pf_1loop_compact} again in linear propagators.
To achieve this, we have to introduce some more notation.
\par
We define $\vec{w}^{\pm}\in\mathbb{N}^{|S_j^{\pm}|}$ and $\vec{w}^{0}\in\mathbb{N}^{|S_j^{0}|}$ to be the vectors collecting the elements of $S_j^{\pm}$ and $S_j^{0}$ respectively as entries in an ascending order. That is, $\vec{w}^{\pm}=(i)_{i\in S_j^\pm}$ and $\vec{w}^{0}=(i)_{i\in S_j^0}$.
For simplicity, we write $s=|S_j^{+}|$ and $\bar{s}=|S_j^{-}|$. Then
\begin{equation}
I_{j}=\sum_{i=1}^s \sum_{\vec{m}\in T^{\bar{s}}_{s-i+1}} J_{j i \vec{m}}
\end{equation}
with
\begin{equation}\label{eq:iterative-step-multi-loop}
    J_{j i \vec{m}}=\frac{(-1)^{i-1}N_{\mathcal{F}}([\mathbf{k}_j^{(1,i)}])}{\prod_{i\in S_j^+} \lambda_{ij}\prod_{i\in S_j^-} \lambda_{ij}}
    \left[\prod_{k=1}^{s-i+1} 
    \prod_{l=m_k}^{m_{k-1}} \lambda_{w^-_{l}j}
	    \pfl{\vec{v}^E_{w^{-}_l}}{\vec{v}^S_{w^{-}_l}}{\vec{\lambda}_{w^{-}_l}}
    \Big|_{k_j^0=k_{j w^{+}_k}^0}\right]
    \prod_{i\in S_j^0}\pfl{\vec{v}^E_i}{\vec{v}^S_i}{\vec{\lambda}_i }
\end{equation}
and
$\mathbf{k}_j^{(1,i)}=((\mathbf{k}_j^0)_1,\ldots,(\mathbf{k}_j^0)_i)$ the analogue of $\bx_{(1,i)}$. For an arbitrary linear propagator with index $r$, %
the evaluation at $k_j^0=k_{ji}^0$
induces the transformation
\begin{equation}
\left.
	    \pfl{\vec{v}^{E\prime}_r}{\vec{v}^{S\prime}_r}{\vec{\lambda}^\prime_r}
    \right. \equiv
	\left.
	    \pfl{\vec{v}^E_r}{\vec{v}^S_r}{\vec{\lambda}_r}
    \right|_{k_j^0=k_{ji}^0}
	=
	\pfl{\vec{v}^E_r-\frac{\lambda_{rj}}{\lambda_{ij}}\vec{v}^E_i}
				    {\vec{v}^S_r-\frac{\lambda_{rj}}{\lambda_{ij}}\vec{v}^S_i\,}
				    {\vec{\lambda}_r-\frac{\lambda_{rj}}{\lambda_{ij}}\vec{\lambda}_i},
\end{equation}
which is trivial if $\lambda_{rj} = 0$.
For each linear propagator on the right-hand side of eq.~\eqref{eq:iterative-step-multi-loop}, since $w_l^-\in S_j^-$ and $w_k^+\in S_j^+$, it therefore follows that its energy vector $\vec{v}^{E\prime}_i$ has non-zero entries that have a consistent sign between them. 
Since the linear propagator transforms into another linear propagator with different coefficients, we conclude that it is a sum of terms just like the starting expression in eq.~\eqref{eq:generic_ltd_linear_prop_term} with $k_j^0\rightarrow k_{j+1}^0$ for $j<N$ replaced.
This implies that the final expression $\mathcal{I}_L$ involves denominators featuring at most $L+1$ on-shell energies.
In summary, re-instating the $\vec{h}$ index, we can thus see that $\mathcal{I}_{j+1}$ can be written as a sum of integrals
\begin{equation}
    \mathcal{I}_{j+1}=
    \left(\prod_{i=1}^P\frac{1}{2E_i}\right)
    \sum_{\vec{h}\in\Omega_{j+1}} I_{j+1\vec{h}},
\end{equation}
where $\Omega_{j+1}$ can be written in terms of $\Omega_{j}$ as
\begin{equation}
    \Omega_{j+1}=\bigg\{(\vec{h}, \vec{m}, i)
    \bigg|\; \vec{h}\in \Omega_{j}, \ \vec{m}\in T^{\bar{s}}_{s-i+1}, \ i=1,\ldots, s, \ s=|S_{j\vec{h}}^+|, \ \bar{s}=|S_{j\vec{h}}^-|\bigg\}.
\end{equation}
The inductive step together with the base case prove the induction hypothesis.

\paragraph{}
Finally, by construction, $\mathcal{I}_{j+1}$ depend on one less loop energy variable with respect to $\mathcal{I}_{j}$.
Ultimately $\mathcal{I}_L$ no longer depends on any loop integration energies, which terminates the recursion. After the final iteration, $\mathcal{I}_L$ will still be expressed as a sum of terms in the form of eq.~\eqref{eq:iterative-step-multi-loop}, where each linear propagator has a definite imaginary part, determined by the causal prescription $\delta>0$ alone.
These linear propagators define precisely the causal threshold singularities, i.e. the E-surfaces of the LTD integrand.
As each term in the final expression is free of non-causal singularities, i.e. H-surfaces,  we have hereby proven that those spurious singularities are indeed absent in the multi-loop LTD integrand.
\par

\subsection{Comparison of the \cLTD{} and LTD representations}
\label{sec:comparison_with_LTD}
Despite the different functional forms of LTD and \cLTD{}, all numerical tests up to four-loop Feynman integrals have shown that this \cLTD{} expression corresponds to the one previously conjectured in ref.~\cite{Capatti:2019ypt}.
This equivalence holds locally at any value of spatial integration momenta. The singular thresholds one needs to consider when constructing a complex contour deformation are therefore the same by construction in both expressions.
The \cLTD{} therefore confirms the hypothesis of dual cancellations although it does not offer an alternative algorithm to derive the cut structures appearing in the LTD expression.

Because the original LTD expression in ref.~\cite{Capatti:2019ypt} exhibits dual cancellations, each of its \emph{dual integrands}~\cite{Catani:2008xa} cannot be integrated individually. In contrast, each summand building the \cLTD\ expression can be integrated separately with a dedicated complex contour deformation and importance sampling.

The main practical advantage of the \cLTD{} expression regards its numerical stability in the vicinity of H-surfaces. In the UV region, one is always close to an H-surface, as reflected by a worsened UV scaling of the LTD expression compared to original the 4D integrand.
More specifically, the superficial degree of UV divergence of each dual integrand is $\lambda^{3-P}$ for a one-loop scalar integral with a constant numerator and $P$ loop propagators, whereas that of the original four-dimensional integrand is $\lambda^{4-2P}$.
Especially for physical theories, dual cancellations in the LTD expression occur across several orders of magnitude. For example, in the case of the NLO $q \bar q \rightarrow \gamma \gamma \gamma$ amplitude (see sect. 4.2.3 of ref.~\cite{Capatti:2019edf}), the cancellation is between summands that scale as $k^2$, across four orders of magnitude down to a scaling of $1/k^2$ in the sum.
As a consequence, an approximate stabilisation procedure based on a Taylor expansion in the UV region was introduced in ref.~\cite{Capatti:2019edf} to cope with the such numerical instabilities.
In contrast, each summand of the \cLTD{} expression directly reproduces the UV behaviour of the original Feynman integral and thus renders such approximate stabilisation unnecessary.
\par

A discussion about the numerical stability and the performance of \cLTD{} is given in sect.~\ref{sec:results}.

\subsection{Connection to Time Ordered Perturbation Theory}
\label{sec:topt}
The main motivation for algebraically manipulating the LTD expression is to eliminate the spurious H-surface singularities from the sum of dual integrands.
The result of this algebraic manipulation is the \cLTD{} expression, a sum of terms that only feature linear propagators involving E-surfaces.
The denominator of each linear propagator has a defining property: it is a sum of \emph{positive} on-shell energies of loop particles and some shift (of arbitrary sign), determined by energies of external particles.
This property is shared by the terms in Time Ordered Perturbation Theory (TOPT), also called Old-Fashioned Perturbation Theory (OFPT) (cf. \cite{sterman_1993,Schwartz:2013pla,Mantovani_2016,bourjaily2020sequential}).
In this section we lay out the first steps towards establishing a closer connection between TOPT and \cLTD{}.
\par
Despite the equivalence of TOPT and covariant perturbation theory, there is no one-to-one correspondence between a single TOPT diagram, an integral over spatial degrees of freedom, and a Feynman diagram, an integral over four-dimensional Minkowski space.
However, after integrating out the loop energy degrees of freedom, arriving at the LTD expression, we might expect a direct connection between a TOPT and an LTD expression.
Nevertheless, the naive comparison of a dual integrand in LTD, i.e. a cut diagram corresponding to a spanning tree of the original graph, with a diagram in TOPT does not reveal any direct connection.
Not only does the denominator structure look different, but also their transformation properties are different.
Whereas each dual integrand is invariant under orthochronous Lorentz transformations, a single TOPT diagram is not.
However, when one compares the \cLTD{} expression to TOPT, more similarities appear.
Indeed, the denominator structure of each term of the \cLTD{} expression mirrors that of the TOPT diagrams as both involve E-surfaces only.
We verified up to four loops that the TOPT and LTD expressions are locally identical.
We now elaborate further on this observation.
\par

In a TOPT diagram, each internal particle is physical, i.e. its four-momentum is on-shell with positive energy.
Whereas this diagram then violates conservation of energy, conservation of the spatial three-momentum is still satisfied at each vertex.
Furthermore, the vertices are ordered, so as to fix a time hierarchy between them.
Therefore the number of terms is given by $V!$, where $V$ is the number of vertices.
We then translate a one-loop TOPT diagram into an expression by cutting it in all ways that separate the initial from the final states and write for each cut line a linear propagator of Lippmann-Schwinger form:
\begin{align}
\label{eq:TOPT_linear_propagator}
    G_{\text{LS}}=\frac{1}{q^0 - E^\text{c}},
    \quad
    E^\text{c}=\sum_{j,\text{ cut}} E_j,
    \quad
    q^0=\sum_{j} \left(q^{\text{in}}_{j}\right)^0-\sum_{j,\text{ cut}} \left(q^{\text{ext}}_{j}\right)^0,
\end{align}
where $E^\text{c}$ is the sum over the (positive) on-shell energies of the internal cut lines and $q^0$ is the difference between the sum of energies of all incoming legs and the sum of energies of the external cut (both incoming and outgoing) legs.
The energies of internal lines $E_j$ are defined as in sect.~\ref{sec:multi_loop_derivation}, satisfying $\Im E_j < 0$ due to the causal prescription $\delta>0$.
Therefore, in TOPT as well as in \cLTD{}, the remaining denominators of linear propagators are sums of \emph{positive} energies, defining causal threshold singularities.
Furthermore, each internal line $j$ contributes a factor $1/(2E_j)$, in both TOPT and \cLTD{}.
\par
We will now provide two simple examples of scalar one-loop diagrams highlighting in more detail the connection between \cLTD{} and TOPT.
In fact, for the scalar triangle diagram there is a one-to-one correspondence between each summand in the \cLTD{} expression and a TOPT diagram.
For the scalar box diagram this is not the case anymore.
However, we show that in that case, TOPT diagrams can be grouped in such a way that they correspond to a single \cLTD{} summand again.
Nevertheless, in general the TOPT representation is different from a \cLTD{} expression, although locally equivalent.
A more general study of the connection between TOPT and (c)LTD, especially at higher loop orders and with numerators is beyond the scope of this paper and left to future work.

\paragraph{Scalar Triangle}
We consider the scalar triangle integral, where the integration over the spatial loop momentum $\vec{k}$ is left out
\begin{align}
    \begin{gathered}
        \vcenter{\hbox{\includegraphics[page=1,scale=0.4]{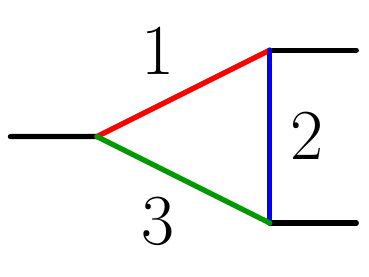}}}
    \end{gathered}
    =
    \frac{-1}{2\pi\mathrm{i}}\int_\mathbb{R} \mathrm{d}k^0
    \frac{1}{(k^0+p_1^0)^2-E_1^2}
    \frac{1}{(k^0+p_2^0)^2-E_2^2}
    \frac{1}{(k^0+p_3^0)^2-E_3^2},
    \quad
    \begin{array}{l}
        p_1 = q_1\\
        p_2 = q_1-q_2 \\
        p_3 = 0%
    \end{array} \,,
\end{align}
where $E_j = \sqrt{(\vec{k}+\vec{p}_j)^2+m_j^2-\mathrm{i}\delta}$, with arbitrary real masses $m_j$ and external momenta $q_j$ that satisfy energy momentum conservation.
The external particle with momentum $q_1$ is incoming and the other two with $q_2$ and $q_3$ are outgoing.
Performing the energy integration will yield the LTD expression when written as a sum of the three dual integrands, or the \cLTD{} representation after removing the H-surfaces explicitly.
Alternatively, we compute the integral using TOPT
\begin{align}
\label{eq:TOPT_triangle_sum}
    \begin{gathered}
        \vcenter{\hbox{\includegraphics[page=11,scale=0.4]{img/TOPT.pdf}}}
    \end{gathered}
    =
    \begin{gathered}
        \vcenter{\hbox{\includegraphics[page=2,scale=0.4]{img/TOPT.pdf}}}
    \end{gathered}
    +
    \begin{gathered}
        \vcenter{\hbox{\includegraphics[page=3,scale=0.4]{img/TOPT.pdf}}}
    \end{gathered}
    +
    \begin{gathered}
        \vcenter{\hbox{\includegraphics[page=4,scale=0.4]{img/TOPT.pdf}}}
    \end{gathered}
    +
    \begin{gathered}
        \vcenter{\hbox{\includegraphics[page=5,scale=0.4]{img/TOPT.pdf}}}
    \end{gathered}
    +
    \begin{gathered}
        \vcenter{\hbox{\includegraphics[page=6,scale=0.4]{img/TOPT.pdf}}}
    \end{gathered}
    +
    \begin{gathered}
        \vcenter{\hbox{\includegraphics[page=7,scale=0.4]{img/TOPT.pdf}}}
    \end{gathered} \,,
\end{align}
where the individual TOPT diagrams are given by
\begin{align}
\begin{split}
    \left( 2E_1 2E_2 2E_3 \right)
    \begin{gathered}
        \vcenter{\hbox{\includegraphics[page=2,scale=0.4]{img/TOPT.pdf}}}
    \end{gathered}
    &=
    \frac{1}{+q_1^0-(E_1+E_3)}
    \frac{1}{(+q_1^0-q_3^0)-(E_1+E_2)}, \\[-2pt]
    \left( 2E_1 2E_2 2E_3 \right)
    \begin{gathered}
        \vcenter{\hbox{\includegraphics[page=3,scale=0.4]{img/TOPT.pdf}}}
    \end{gathered}
    &=
    \frac{1}{+q_1^0-(E_1+E_3)}
    \frac{1}{(+q_1^0-q_2^0)-(E_2+E_3)}, \\[-2pt]
    \left( 2E_1 2E_2 2E_3 \right)
    \begin{gathered}
        \vcenter{\hbox{\includegraphics[page=4,scale=0.4]{img/TOPT.pdf}}}
    \end{gathered}
    &=
    \frac{1}{-q_2^0-(E_1+E_2)}
    \frac{1}{(+q_1^0-q_2^0)-(E_2+E_3)}, \\[-2pt]
    \left( 2E_1 2E_2 2E_3 \right)
    \begin{gathered}
        \vcenter{\hbox{\includegraphics[page=5,scale=0.4]{img/TOPT.pdf}}}
    \end{gathered}
    &=
    \frac{1}{-q_3^0-(E_2+E_3)}
    \frac{1}{(+q_1^0-q_3^0)-(E_1+E_2)}, \\[-2pt]
    \left( 2E_1 2E_2 2E_3 \right)
    \begin{gathered}
        \vcenter{\hbox{\includegraphics[page=6,scale=0.4]{img/TOPT.pdf}}}
    \end{gathered}
    &=
    \frac{1}{-q_2^0-(E_1+E_2)}
    \frac{1}{(-q_2^0-q_3^0)-(E_1+E_3)}, \\[-2pt]
    \left( 2E_1 2E_2 2E_3 \right)
    \begin{gathered}
        \vcenter{\hbox{\includegraphics[page=7,scale=0.4]{img/TOPT.pdf}}}
    \end{gathered}
    &=
    \frac{1}{-q_3^0-(E_2+E_3)}
    \frac{1}{(-q_2^0-q_3^0)-(E_1+E_3)}.
\end{split}
\end{align}
The \cLTD{} expression for the scalar triangle, obtained from the algorithm described in sect.~\ref{sec:causal_ltd}, takes the exact same form as eq.~\eqref{eq:TOPT_triangle_sum}.
Therefore, we can identify each \cLTD{} summand with a TOPT diagram, i.e. with a specific time-ordering of vertices.

\paragraph{Scalar Box}
When considering a scalar box diagram, some of its \cLTD{} summands correspond directly to one specific TOPT diagram.
However, such a one-to-one identification between \cLTD{} summands and TOPT diagrams is not always possible, as we will show with two examples.
Keep in mind that when comparing the following expressions with eq.~\eqref{eq:box_causal_LTD} for the scalar box diagram, the \cLTD{} and TOPT normalisations differ by an overall factor of $-1$.
Furthermore, all external momenta $q_j$ are understood to be incomming.
\par
This first TOPT expression below has a single equivalent summand counterpart in the \cLTD{} expression
\begin{align}
\label{eq:box_TOPT_is_cLTD}
    \bigg(\prod_{j=1}^4 2 E_j\bigg)
    \begin{gathered}
        \vcenter{\hbox{\includegraphics[page=8,scale=0.4]{img/TOPT.pdf}}}
    \end{gathered}
    =\frac{1}{q_1^0-(E_1+E_4)}
    \frac{1}{(q_1^0+q_2^0)-(E_2+E_4)}
    \frac{1}{(q_1^0+q_2^0+q_3^0)-(E_3+E_4)}.
\end{align}
More precisely, the term in eq.~\eqref{eq:box_TOPT_is_cLTD} corresponds to the last summand of the second row  in eq.~\eqref{eq:box_causal_LTD}, i.e. $1/(d_{41}d_{42}d_{43})$.
\par
We then consider the following two TOPT diagrams
\begin{align}
\label{eq:box_TOPT_1}
    \bigg(\prod_{j=1}^4 2 E_j\bigg)
    \begin{gathered}
        \vcenter{\hbox{\includegraphics[page=9,scale=0.4]{img/TOPT.pdf}}}
    \end{gathered}
    &=\frac{1}{q_1^0-(E_1+E_4)}
    \frac{1}{(q_1^0+q_3^0)-\sum_{j=1}^4 E_j}
    \frac{1}{(q_1^0+q_2^0+q_3^0)-(E_3+E_4)}, \\
\label{eq:box_TOPT_2}
    \bigg(\prod_{j=1}^4 2 E_j\bigg)
    \begin{gathered}
        \vcenter{\hbox{\includegraphics[page=10,scale=0.4]{img/TOPT.pdf}}}
    \end{gathered}
    &=\frac{1}{q_1^0-(E_1+E_4)}
    \frac{1}{(q_1^0+q_3^0)-\sum_{j=1}^4 E_j}
    \frac{1}{-q_2^0-(E_1+E_2)}.
\end{align}
It is already clear that none of the above two TOPT diagrams can have a corresponding \cLTD{} summand.
This is because the denominators in \cLTD{} summands can have at most $L+1$ on-shell internal energies at $L$-loops.
In TOPT however, eq.~\eqref{eq:box_TOPT_1} and eq.~\eqref{eq:box_TOPT_2} both come with a linear propagator featuring four on-shell energies.
Nevertheless, when rewriting the first diagram using partial fraction decomposition on its last two fractions, we can cancel the second diagram and obtain
\begin{align}
\label{eq:box_TOPT_sum}
    \bigg(\prod_{j=1}^4 2 E_j\bigg)
    \bigg(
    \begin{gathered}
        \vcenter{\hbox{\includegraphics[page=9,scale=0.4]{img/TOPT.pdf}}}
    \end{gathered}
    +
    \begin{gathered}
        \vcenter{\hbox{\includegraphics[page=10,scale=0.4]{img/TOPT.pdf}}}
    \end{gathered}
    \bigg)
    =
    \frac{1}{-q_2^0-(E_1+E_2)}
    \frac{1}{q_1^0-(E_1+E_4)}
    \frac{1}{(q_1^0+q_2^0+q_3^0)-(E_3+E_4)}.
\end{align}
The expression in eq.~\eqref{eq:box_TOPT_sum} corresponds exactly to a single summand in the \cLTD{} representation, more precisely, to the last summand of the first row of eq.~\eqref{eq:box_causal_LTD}, i.e. $1/(d_{21}d_{41}d_{43})$.
\par
The full \cLTD{} expression for the scalar box in eq.~\eqref{eq:box_TOPT_sum} can be obtained from TOPT diagrams in the following way:
In total, there are $4!=24$ TOPT diagrams and only $20$ \cLTD{} summands.
It turns out that $16$ TOPT diagrams are of the same form as eq.~\eqref{eq:box_TOPT_is_cLTD}.
Only $12$ of those are in one-to-one correspondence with a \cLTD{} summand.
The other $4$ have to be rewritten using the identity in eq.~\eqref{eq:box_identity} such that they can be identified with $4$ \cLTD{} summands.
The remaining $8$ diagrams are combined in $4$ pairs like in eq.~\eqref{eq:box_TOPT_sum} yielding a single \cLTD{} summand for each pair.

\newpage
\section{Examples}
\label{sec:examples}
In the following section we will showcase a variety of example Feynman diagrams whose \cLTD{} representation is computed explicitly. We will vary the number of loops, the number of external momenta and the rank of the numerator. In the following, we will make use of the notation presented in Appendix~\ref{sec:multi_loop_notation}.

In order to explicitly write out the \cLTD{} representations, we unfold the one-loop procedure discussed in sect.~\ref{sec:generic_partial_fractioning} and the multi-loop procedure of sect.~\ref{sec:loop-by-loop_iteration}, while also extending the notion of divided differences to multi-loop integrands. In doing so, we will illustrate the main features and delicate issues that arise in the derivation of the \cLTD{} expression.

We  will consider three examples:
\begin{enumerate}
\item In sect.~\ref{sec:photon} we illustrate how physical numerators are treated in the framework we laid out in sections~\ref{sec:generic_partial_fractioning} and~\ref{sec:causal_ltd} with a photon one-loop self-energy diagram.
This also showcases how divided differences can be computed explicitly for polynomial numerators.

\item In sect.~\ref{sec:box} we investigate a box diagram that highlights the intricate way in which the number of propagators per loop line affects the structure of the linear propagators and divided differences.
Furthermore, using the same box diagram, we will consider the degenerate case of a tadpole diagram with a propagator raised to the fourth power, corresponding to the UV limit of the box itself, whose \cLTD{} representation is obtained directly as a limit of the non-degenerate case.

\item In sect.~\ref{sec:sunrise} and sect.~\ref{sec:n-loop-banana} we address the application of the partial fractioning procedure beyond one loop by giving explicit examples for the two-loop sunrise diagram as well as the generic $n$-loop banana topologies. These multi-loop diagrams yield especially compact representations, as they have only one single propagator per loop line.
\end{enumerate}
More examples can be obtained using our \textsc{python} code provided in the ancillary material, which generates the \cLTD{} expression for any Feynman diagram with an arbitrary external multiplicity, loop count, and numerator. We explain the usage of this code in Appendix~\ref{app:pyfractioning}.

\subsection{One-loop photon self-energy}
\label{sec:photon}
This first example investigates the one-loop photon self-energy:
\begin{equation*}
\begin{gathered}
\includegraphics[scale=.8,page=1]{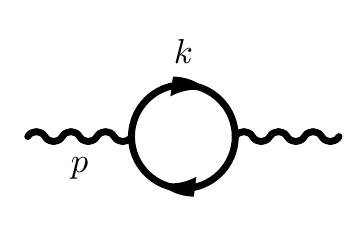}
\end{gathered}
\end{equation*}
Despite its simplicity, this case already highlights the subtleties arising when considering a non-trivial, physical numerator. It is also the simplest setup in which it is possible to see the basic pattern of dual cancellations.

The energy integration of the one-loop vacuum polarisation diagram is given by
\begin{align}
    \mathcal{I}^{(\Pi)}
    &=
     -\int \frac{dk^0}{2\pi \mathrm{i}}
    \frac{
        \epsilon_\mu(p,\lambda)
        \tr(
    (-\mathrm{i} e \gamma^\nu)
    \mathrm{i} (\slashed{k}+m)
    (-\mathrm{i} e \gamma^\mu)
    \mathrm{i} (\slashed{k}-\slashed{p}+m)
    )
        \epsilon^*_\nu(p,\lambda)
    }{
        (k^2-m^2+\mathrm{i}\delta)
        ((k-p)^2-m^2+\mathrm{i}\delta)
    },
\end{align}
whose numerator then reads:
\begin{align}
\begin{split}
    \mathcal{N}(k^0)
    &\equiv
    \epsilon_\mu(p,\lambda)
    \tr(
    (-\mathrm{i} e \gamma^\nu)
    \mathrm{i} (\slashed{k}+m)
    (-\mathrm{i} e \gamma^\mu)
    \mathrm{i} (\slashed{k}-\slashed{p}+m)
    )
    \epsilon^*_\nu(p,\lambda) \\
    &=4\pi\alpha\left(
    c_0 + c_1 k^0 + c_2 (k^0)^2
    \right),
\end{split}
\end{align}
where
\begin{align}
\begin{split}
    c_0 &= 4m^2 
            + 4\vec{k}\cdot\left(\vec{k}-\vec{p}\right) 
            + 8 \left(\vec{\epsilon}\cdot \vec{k}\right)\left(\vec{\epsilon^*}\cdot \vec{k}\right) \\
    c_1 &= 4 p^0
            -8\left[ 
            + \epsilon^0\left(\vec{\epsilon^*}\cdot \vec{k}\right)+(\epsilon^{0})^*\left(\vec{\epsilon}\cdot \vec{k}\right)
            \right]\\
    c_2 &= 8 \epsilon^{0}(\epsilon^{0})^*-4 \,.
\end{split}
\end{align}

We start by considering the usual LTD representation~\cite{Capatti:2019ypt} of the vacuum polarization bubble, which features two dual integrands, each corresponding to one of the two internal propagators of the fermion bubble being cut with positive on-shell energy:
\begin{equation}\label{starting-eq-bubble}
    \mathcal{I}^{(\Pi)}=4\pi\alpha \Bigg[ \frac{
    c_0 + c_1 E_1 + c_2 E_1^2}{2 E_1 ((E_1+p_0)^2-E_2^2)}+\frac{
    c_0 + c_1 (E_2-p_0) + c_2 (E_2 - p_0)^2}{2 E_1 ((E_2-p_0)^2-E_1^2)}\Bigg] \,.
\end{equation}
We can rewrite each propagator evaluated at the on-shell condition as the difference of two linear propagators multiplied by the inverse on-shell energy of the momentum flowing in that propagator:
\begin{equation}\label{p-frac-bubble}
    \frac{1}{(E_1+p_0)^2-E_2^2}=\frac{1}{2E_2}\Bigg(\frac{1}{E_1+p_0-E_2}-\frac{1}{E_1+p_0+E_2} \Bigg) \,,
\end{equation}
which makes it clear that the zeros of the inverse propagator is the union of the zeros of an E-surface $E_1+p_0+E_2=0$ and an H-surface $E_1+p_0-E_2=0$. The E-surface, characterised by its consistent sign across all involved on-shell energies, is bounded and corresponds to a physical threshold. 
Instead, the H-surface contains different on-shell energy signs and its imaginary prescription therefore depends on kinematics. 
We now show the explicit cancellation of this H-surface by substituting eq.~\eqref{p-frac-bubble} in eq.~\eqref{starting-eq-bubble} for both dual integrands:
\begin{equation}\label{p-frac-sunrise}
    \mathcal{I}^{(\Pi)}=\frac{\pi\alpha}{ E_1 E_2 }\Bigg[ \frac{
   \mathcal{N}(E_1)}{E_1+p_0+E_2}+\frac{
   \mathcal{N}(E_2)}{E_2-p_0+E_1}-\frac{c_1(E_1-E_2+p_0)+c_2(E_2^2-(E_2-p_0)^2)}{E_1-E_2+p_0}\Bigg] \,.
\end{equation}
Each dual integrand yields two new summands after this substitution: one corresponding to an H-surface singularity and one to an E-surface singularity.
We see that the H-surface summands have a common denominator and can be written as the last term of eq.~\eqref{p-frac-sunrise}. This combination is what manifestly realises the pairwise cancellation pattern that underlies dual cancellations.
Indeed, because the numerator is a polynomial in the energy, it is possible to explicitly perform the polynomial division. As a result, we obtain our final expression
\begin{equation}
    \mathcal{I}^{(\Pi)}=\frac{\pi\alpha}{ E_1 E_2 }\Bigg[ \frac{
   \mathcal{N}(E_1)}{E_1+p_0+E_2}+\frac{
   \mathcal{N}(E_2)}{E_2-p_0+E_1}-c_1-c_2(E_1 +E_2 -p_0)\Bigg] \,.
   \label{eq:final_expr_example_one}
\end{equation}
The expression of eq.~\eqref{eq:final_expr_example_one} is now manifestly free of spurious poles, and dual cancellations have been realized algebraically. The remaining singularities of $\Pi$ are located at the zeros of on-shell energies and E-surfaces.

\subsection{Box topology}
\label{sec:box}
As we increase the number of propagators per loop line (at one-loop this corresponds directly to an increase of the multiplicity of the external momenta), the cancellation of spurious singularities generates more terms in the \cLTD{} representation.

We demonstrate this by considering the box topology
\begin{equation*}
\begin{gathered}
\includegraphics[scale=.8,page=3]{img/pf_diags.pdf}
\end{gathered}
\end{equation*}
written as the loop energy integral with a generic numerator
\begin{align}
    \mathcal{I}^{(\mathrm{Box})}
    &=
    \frac{1}{2\pi\mathrm{i}}
    \int \mathrm{d} k^0
    \frac{\mathcal{N}(k^0)}{\prod_{i=1}^4\left((k+p_i)^2-m_i^2+\mathrm{i}\delta\right)}.
\end{align}
The corresponding \cLTD{} expression following from our recursive procedure of sect.~\ref{sec:loop-by-loop_iteration} reads:
\begin{align}
\begin{split}\label{eq:box_causal_LTD}
   \bigg(\prod_{i=1}^4 2E_i\bigg) \mathcal{I}^{(\mathrm{Box})}
    =
    &+\frac{\text{N}_{\mathcal{N}} [z_1 ] )}{d_{21} d_{31} d_{41}}
    + \frac{\text{N}_{\mathcal{N}} ( [z_1 ] )}{d_{21} d_{24} d_{34}}
    + \frac{\text{N}_{\mathcal{N}} ( [z_1 ] )}{d_{21} d_{31} d_{34}}
    + \frac{\text{N}_{\mathcal{N}} ( [z_1 ] )}{d_{21} d_{23} d_{43}}
    + \frac{\text{N}_{\mathcal{N}} ( [z_1 ] )}{d_{21} d_{41} d_{43}}
    \\[1em]
    &+ \frac{\text{N}_{\mathcal{N}} ( [z_1 ] )}{d_{21} d_{23} d_{24}}
    + \frac{\text{N}_{\mathcal{N}} ( [z_1 ] )}{d_{31} d_{32} d_{42}}
    + \frac{\text{N}_{\mathcal{N}} ( [z_1 ] )}{d_{31} d_{41} d_{42}}
    + \frac{\text{N}_{\mathcal{N}} ( [z_1 ] )}{d_{31} d_{32} d_{34}}
    + \frac{\text{N}_{\mathcal{N}} ( [z_1 ] )}{d_{41} d_{42} d_{43}}
    \\[1em]
    &+\frac{\text{N}_{\mathcal{N}} ( [z_2 ] )}{d_{12} d_{32} d_{42}}
    +\frac{\text{N}_{\mathcal{N}} ( [z_2 ] )}{d_{12} d_{14} d_{34}}
    +\frac{\text{N}_{\mathcal{N}} ( [z_2 ] )}{d_{12} d_{32} d_{34}}
    +\frac{\text{N}_{\mathcal{N}} ( [z_2 ] )}{d_{12} d_{13} d_{43}}
    +\frac{\text{N}_{\mathcal{N}} ( [z_2 ] )}{d_{12} d_{42} d_{43}}
    \\[1em]
    &+\frac{\text{N}_{\mathcal{N}} ( [z_2 ] )}{d_{12} d_{13} d_{14}}
    +\frac{\text{N}_{\mathcal{N}} ( [z_3 ] )}{d_{13} d_{23} d_{43}}
    +\frac{\text{N}_{\mathcal{N}} ( [z_3 ] )}{d_{13} d_{14} d_{24}}
    +\frac{\text{N}_{\mathcal{N}} ( [z_3 ] )}{d_{13} d_{23} d_{24}}
    + \frac{\text{N}_{\mathcal{N}} ( [z_4 ] )}{d_{14} d_{24} d_{34}}
    \\[1em]
    &-\frac{\text{N}_{\mathcal{N}} ( [z_1,z_2 ] )}{d_{32} d_{42}}
    -\frac{\text{N}_{\mathcal{N}} ( [z_1,z_2 ] )}{d_{32} d_{34}}
    -\frac{\text{N}_{\mathcal{N}} ( [z_1,z_2 ] )}{d_{42} d_{43}}
    -\frac{\text{N}_{\mathcal{N}} ( [z_1,z_3 ] )}{d_{23} d_{43}}
    -\frac{\text{N}_{\mathcal{N}} ( [z_1,z_3 ] )}{d_{23} d_{24}}
    \\[1em]
    &-\frac{\text{N}_{\mathcal{N}} ( [z_1,z_4 ] )}{d_{24} d_{34}}
    -\frac{\text{N}_{\mathcal{N}} ( [z_2,z_3 ] )}{d_{13} d_{43}}
    -\frac{\text{N}_{\mathcal{N}} ( [z_2,z_3 ] )}{d_{13} d_{14}}
    -\frac{\text{N}_{\mathcal{N}} ( [z_2,z_4 ] )}{d_{14} d_{34}}
    -\frac{\text{N}_{\mathcal{N}} ( [z_3,z_4 ] )}{d_{14} d_{24}}
    \\[1em]
    &+\frac{\text{N}_{\mathcal{N}} ( [z_1,z_2,z_3 ] )}{d_{43}}
    +\frac{\text{N}_{\mathcal{N}} ( [z_1,z_2,z_4 ] )}{d_{34}}
    +\frac{\text{N}_{\mathcal{N}} ( [z_1,z_3,z_4 ] )}{d_{24}}
    +\frac{\text{N}_{\mathcal{N}} ( [z_2,z_3,z_4 ] )}{d_{14}}
    \\[1em]
    &-\text{N}_{\mathcal{N}} ( [z_1,z_2,z_3,z_4 ] ) \,.
\end{split}
\end{align}
We observe that the original 4 dual integrands forming the LTD representation of ref.~\cite{Capatti:2019ypt} increase to up to $35$ summands in the \cLTD{} representation in eq.~\eqref{eq:box_causal_LTD}.
Although the sum contains $35$ terms, it features a simpler underlying structure, since there are only $12$ distinct linear propagators, denoted with
\begin{equation}
    d_{ij}= E_i + E_j + p_i^0 - p_j^0.
\end{equation}
At one loop, the four limits involved in the finite difference functional $\text{N}$ are simply defined by the respective poles of the propagators in the lower complex half-plane at
\begin{equation}
    z_i = E_i - p_i^0.
\end{equation}
It is interesting to note that some groups of terms in the expression~\eqref{eq:box_causal_LTD} can be written in alternative ways, making it clear that the representation is not unique, as discussed in both sect.~\ref{sec:generic_partial_fractioning} and sect.~\ref{sec:causal_ltd}. For example, one can change the denominator structure as in
\begin{align}
\label{eq:box_identity}
    \frac{\text{N}_\mathcal{N}([z_1])}{d_{21}d_{24}d_{34}}
    +
    \frac{\text{N}_\mathcal{N}([z_1])}{d_{21}d_{31}d_{34}}
    =
    \frac{\text{N}_\mathcal{N}([z_1])}{d_{21}d_{24}d_{31}}
    +
    \frac{\text{N}_\mathcal{N}([z_1])}{d_{24}d_{31}d_{34}} \,.
\end{align}
Moreover, the numerator can change its argument according to the identity
\begin{align}
    \frac{\text{N}_\mathcal{N}([z_1])}{d_{31}d_{32}d_{42}}
    -
    \frac{\text{N}_\mathcal{N}([z_1,z_2])}{d_{32}d_{42}}
    =
    \frac{\text{N}_\mathcal{N}([z_2])}{d_{31}d_{32}d_{42}}
    -
    \frac{\text{N}_\mathcal{N}([z_1,z_2])}{d_{31}d_{42}}\,.
\end{align}
The non-uniqueness can also be understood by noticing that the choice of edge labelling in Feynman diagrams is arbitrary. This means that although the final expression stays invariant under such a relabelling, the partial fractioning procedure results in a different but equivalent representation.

\par

When regulating the UV behaviour, it is convenient to introduce counterterm diagrams that have the same asymptotic behaviour as the original integrand and are simple enough to be integrated analytically.
These counterterms are subtracted from the integrand to make it suitable for numerical integration and their analytically integrated counterpart can be added back to the final result.
As counterterm diagrams often have degenerate edges, we show the degenerate limit of the box diagram which corresponds to a tadpole with a single propagator raised to the fourth power:
\begin{align}
    \mathcal{I}^{(\mathrm{Tadpole})}(4;m_1)
    &=
    \frac{1}{2\pi\mathrm{i}}
    \int \mathrm{d} k^0
    \frac{\mathcal{N}(k^0)}{\left((k+p_1)^2-m_1^2+\mathrm{i}\delta\right)^4} \,.
\end{align}
As pointed out in sect.~\ref{sec:generic_partial_fractioning}, the derivation for the manifestly causal LTD expression for diagrams with raised propagators does not require special treatment as it corresponds to the degenerate limit of pairwise distinct propagators.
We can therefore consider the limit $p_i=p_1$ and $m_i=m_1$, where $d_{ij}=2E_1\;\forall i,j$ and $z_i=E_1-p_1^0\;\forall i$, and apply the formula for degenerate divided differences given in eq. \eqref{eq:divided_difference_derivative}.
The general box expression of eq.~\eqref{eq:box_causal_LTD} then reduces to $20$ summands involving $\frac{\mathcal{N}(z_1)}{(2E_1)^3}$, $10$ summands involving $-\frac{\mathcal{N}'(z_1)}{(2E_1)^2}$, $4$ summands involving $\frac{\mathcal{N}''(z_1)}{2!(2E_1)}$ and one summand involving $-\frac{\mathcal{N}'''(z_1)}{3!}$, each with the overall prefactor $\frac{1}{(2E_1)^4}$, finally yielding
\begin{align}
      \mathcal{I}^{(\mathrm{Tadpole})}(4;m_1)=\frac{15 \mathcal{N}(z_1)- 15 E_1 \mathcal{N}'(z_1)+6 E_1^2 \mathcal{N}''(z_1) - E_1^3 \mathcal{N}'''(z_1)}{96 E_1^7}.
\end{align}
This integrand reproduces the asymptotic UV behaviour of the original $\mathcal{I}^{(\mathrm{Box})}$ representation.

\subsection{Sunrise topology}
\label{sec:sunrise}
The simplest example beyond one-loop is the sunrise diagram
\begin{equation*}
\begin{gathered}
\includegraphics[scale=.8,page=2]{img/pf_diags.pdf}
\end{gathered}
\end{equation*}
which is a two-loop two-point function with three propagators.
The integral over the loop energies reads
\begin{align}
    \mathcal{I}^{(\mathrm{Sunrise})}
    &=
    \frac{1}{(2\pi\mathrm{i})^2} \int \dd k_1^0 \dd k_2^0
    \frac{\mathcal{N}(k_1^0,k_2^0)}{
        \prod_{i=1}^3\left((\vec{\lambda}_i\cdot (k_1,k_2) + p_i)^2-m_i^2+\mathrm{i}\delta\right)
        } \,,
\end{align}
with loop line signatures $\vec{\lambda}_1=(1,0)$, $\vec{\lambda}_2=(1,-1)$ and $\vec{\lambda}_3=(0,1)$ with one propagator each. 
The corresponding \cLTD{} expression is given by the following four summands
\begin{multline}
\label{eq:sunrise_causal_LTD}
    \mathcal{I}^{(\mathrm{Sunrise})}
    =  -\frac{\text{N}_\mathcal{N}([z_2],[z_4])}{E_1+E_2+E_3+p_1^0-p_2^0-p_3^0}
    -\frac{\text{N}_\mathcal{N}([z_1],[z_3])}{E_1+E_2+E_3-p_1^0+p_2^0+p_3^0}\\
    +\text{N}_\mathcal{N}([z_1],[z_4,z_3]) 
    +\text{N}_\mathcal{N}([z_1,z_2],[z_4]) \,,
\end{multline}
where the numerator is evaluated with the following inputs
\begin{equation}
\begin{aligned}
    &z_1= E_1-p_1^0,
    &&z_2(k_2^0) = E_2+k_2^0-p_2^0,\\
    &z_3 = E_1+E_2-p_1^0+p_2^0,
    &&z_4 = E_3-p_3^0.
\end{aligned}
\end{equation}
Note that when evaluating the numerator $\text{N}_\mathcal{N}([z_2(k_2^0)],[z_4])$ (following app.~\ref{sec:multi_loop_notation}), the recursion in eq.~\eqref{eq: N_definition} is performed first in the first argument $[z_2(k_2^0)]$, which changes the numerator's dependence on $k_2^0$ since $\tilde{\mathcal{N}}(k_2^0)\equiv\mathcal{N}(z_2(k_2^0),k_2^0)$.
The evaluation in the second argument $[z_4]$ is then to be understood in the context of the functional recursion given in eq.~\eqref{eq: N_definition}, applied to the function $\tilde{\mathcal{N}}$.
This \cLTD{} expression for the sunrise diagram of eq.~\eqref{eq:sunrise_causal_LTD} agrees with the special case of a constant numerator presented recently in ref.~\cite{Verdugo:2020kzh}.
\par

In order to better understand how the iteration represented by the step of eq.~\eqref{eq:iterative-step-multi-loop} works, we unfold it here explicitly for a scalar sunrise vacuum bubble. Let us start by partial fractioning each propagator in its Minkowski representation, as done in eq.~\eqref{eq:Feynman_propagator}:
\begin{align}\label{p-fra-sunrise}
\begin{split}
    \mathcal{I}^{(\mathrm{Sunrise})}_{j=0}= &\int \frac{\dd k_1^0 \dd k_2^0/(2\pi \mathrm{i})^2}{8 E_1 E_2 E_3}\Bigg[\\
    &+\frac{1}{(k_1^0-E_1)(k_1^0-k_2^0-E_2)(k_2^0-E_3)}  -\frac{1}{(k_1^0-E_1)(k_1^0-k_2^0-E_2)(k_2^0+E_3)} \\ &-\frac{1}{(k_1^0-E_1)(k_1^0-k_2^0+E_2)(k_2^0-E_3)}  -\frac{1}{(k_1^0+E_1)(k_1^0-k_2^0-E_2)(k_2^0-E_3)} \\ &+\frac{1}{(k_1^0-E_1)(k_1^0-k_2^0+E_2)(k_2^0+E_3)}  +\frac{1}{(k_1^0+E_1)(k_1^0-k_2^0-E_2)(k_2^0+E_3)} \\
    &+\frac{1}{(k_1^0+E_1)(k_1^0-k_2^0+E_2)(k_2^0-E_3)} 
    -\frac{1}{(k_1^0+E_1)(k_1^0-k_2^0+E_2)(k_2^0+E_3)}\Bigg],
\end{split}
\end{align}
and let us contour integrate in the variable $k_1^0$ on the usual semi-circle spanning the lower half of the complex plane. We stress that the last two terms of~\eqref{p-fra-sunrise} vanish under this integration, whereas the integration of the first four terms contribute and yield eight new terms; of which many pairwise cancel, eventually yielding
\begin{align}
\begin{split}
    \mathcal{I}^{(\mathrm{Sunrise})}_{j=1}= -\int \frac{ \dd k_2^0/(2\pi \mathrm{i})}{8 E_1 E_2 E_3}\Bigg[-\frac{1}{(E_1-k_2^0+E_2)(k_2^0-E_3)}  -\frac{1}{(k_2^0+E_2+E_1)(k_2^0-E_3)} \\ +\frac{1}{(E_1-k_2^0+E_2)(k_2^0+E_3)}  +\frac{1}{(k_2^0+E_2+E_1)(k_2^0+E_3)}\Bigg] \,.
\end{split}
\end{align}
We now perform the energy integration in $k_2^0$ by closing the contour, again, in the lower-half of the complex plane. This last integral yields the final \cLTD{} expression for this example:
\begin{align}
    \mathcal{I}^{(\mathrm{Sunrise})}_{j=2}=\mathcal{I}^{(\mathrm{Sunrise})}=-\frac{1}{4 E_1 E_2 E_3}\frac{1}{E_1+E_2+E_3} \,.
\end{align}

\subsection{$n$-loop banana topologies}
\label{sec:n-loop-banana}
In this example we study a particular class of $n$-loop diagrams that feature only two vertices and one propagator per loopline. We refer to loop integrals of this class as $n$-loop banana integrals.
The very restrictive form of $n$-loop banana integrals allows us to give a closed form for its \cLTD{} expression.
To this end, we must set up some basic notation. First, we assign to each of the $n+1$ propagators of the $n$-loop banana an on-shell energy denoted by $E_i$. Let us furthermore write $p_i=0$ for $i\le n$ and $p_{n+1}=p_0$. The we define the following vector:
\begin{equation}
\vec{y}_i\in\mathbb{R}^{n}, \ \ (\vec{y}_i)_j=\begin{cases}
-E_j \text{ if } j<i-1 \\
E_{n+1}-p_0+\sum_{k=1}^{i-2}E_k-\sum_{k=i}^{n}E_k \text{ if } j=i-1 \\
E_j \text{ if } j>i-1 
\end{cases}.
\end{equation}
With this notation, we can write the following closed form for the unfolded \cLTD{} expression of the $n$-loop banana integral:
\begin{equation}\label{eq:n-loop-banana}
 \mathcal{I}^{(\mathrm{Banana})}_{j=n} =\frac{1}{\prod_{i=1}^{n+1}2 E_i}\Bigg[ \frac{\mathcal{N}(\vec{y}_1)}{\sum_{i=1}^{n+1} E_i -p_0}  +\frac{\mathcal{N}(\vec{y}_{n+1})}{\sum_{i=1}^{n+1} E_i +p_0}-\sum_{m=1}^{n} \mathrm{N}^m_{\mathcal{N}}([\vec{y}_m, \vec{y}_{m+1}])\Bigg] \,,
 \end{equation}
where $\mathrm{N}^m_{\mathcal{N}}$ contains the following divided differences 
 \begin{equation}\label{eq:banana-divided-difference}
 \mathrm{N}^m_{\mathcal{N}}([\vec{y}_m, \vec{y}_{m+1}])=\frac{\mathcal{N}(\vec{y}_m)-\mathcal{N}(\vec{y}_{m+1})}{(\vec{y}_m-\vec{y}_{m+1})_{m}}\,.
\end{equation}
We stress that eq.~\eqref{eq:banana-divided-difference} is now free of spurious singularities. 
Moreover, in that case $\mathcal{N}$ is polynomial, polynomial divisions can be performed explicitly and result in $\mathrm{N}^m_{\mathcal{N}}[\vec{y}_m, \vec{y}_{m+1}]$ becoming a simple polynomial in the on-shell energies whose rank is lowered by one w.r.t that of $\mathcal{N}$.

\section{Results}
\label{sec:results}
To assess the potential of the \cLTD{} representation for numerical applications, we compare it to the implementation of the LTD representation for several benchmark examples, both in terms of run-time speed (sect.~\ref{sec:res:computational_complexity}) and numerical stability (sect.~\ref{sec:res:numerical_stability}).

\subsection{Computational complexity}
\label{sec:res:computational_complexity}

As already discussed in sect.~\ref{sec:comparison_with_LTD}, the growth of the computational complexity of the LTD and \cLTD{} expressions differs significantly and is of concern for practical numerical applications.
Indeed, the naive implementation of the expression stemming from the iterative \cLTD{} procedure described in sect.~\ref{sec:loop-by-loop_iteration} would be slow, as the number of terms grows exponentially in the number of propagators.
An upper bound for the number of terms building the \cLTD{} expression of one-loop integrals with arbitrary numerators can be computed by considering eq.~\eqref{eq:total_number_of_terms_upper_bound} in relation with eq.~\eqref{eq:prop_pf_starting_expression}:
\begin{equation}
     \mqty(2\nu_{tot}-1\\\nu_{tot}),
\end{equation}
where $\nu_{tot}$ is the sum of all the powers of the denominators. 
However, since the denominators consist of only a limited number of E-surfaces and the numerators only depend on $P+E$ energies and dot products of $L+E$ spatial vectors, where $P$ is the number of propagators and $E$ the number of external momenta, a substantial amount of common sub-expressions are expected to be found across the terms in the \cLTD{} representation. 
In this section, we consider twelve examples of scalar integrals spanning different combinations of the computational complexity ordering parameters $P$, $E$ and polynomial numerator rank $r$. The particular polynomial chosen for each loop count $L$ and rank $r$ is given in table~\ref{tab:numerator_form}. We collect our results for these examples in table~\ref{tab:results}.
\begin{table}[t!]
    \centering
    \begin{tabular}{c|c}
         L=1 & $\left(k_1\cdot(p_1+p_2)\right)^{r}$ \\
         L=2 & $\left( \left(p_1\cdot(k_1+k_2)\right)^2+k_1\cdot k_2 \right)^{r/2}$ \\ 
         L=3 & $\left( \left(p_1\cdot(k_1+k_2+k_3)\right)^2+(k_1\cdot k_2+k_1\cdot k_3+k_2\cdot k_3) \right)^{r/2}$
    \end{tabular}
    \caption{Specific numerators selected for each loop count $L$ and rank $r$. The four-momenta denoted $k_i$ are loop momenta in some arbitrarily chosen loop momentum basis and those denoted $p_i$ are external momenta (with unspecified order).}
    \label{tab:numerator_form}
\end{table}

The optimisation procedure of the implementation (in the {\texttt{C}} language) of the \cLTD{} expression is carried out using the polynomial evaluation optimization feature of \textsc{form}~\cite{Kuipers:2013pba,Ruijl:2014spa}, using a 1000 iterations of the \texttt{Local Stochastic Search} optimisation procedure with additional \texttt{CSGreedy} optimisations (see the \texttt{opt} column of table~\ref{tab:results}).
We also report on the number of arithmetic operations obtained when bypassing this optimisation step (\texttt{nopt}=0) in order to monitor its relevance.
The evaluation time is typically linearly proportional to the number of multiplications ($N_x$ in table~\ref{tab:results}) although we note that for the simplest one-loop topologies the overhead from computing on-shell energies and individual denominators can bias this linear relationship. Moreover, the resulting \texttt{C}-code is compiled with the \texttt{GNU gcc 7.5.0} compiler and the \texttt{-O3 -fcx-fortran-rules -fcx-limited-range} optimisation flag, which brings in additional optimisations that can affect the LTD and \cLTD{} implementation differently, thus explaining small departures from this linear behaviour also for more complicated integrands.
Overall, we find that the \textsc{form} optimisation has a very significant impact on the computational complexity of the \cLTD{} expression and often brings it on par with the complexity of the LTD representation, especially for constant numerators (or even better than that for the simplest cases). Indeed, increasing the numerator rank appears to slightly disfavour the \cLTD{} representation w.r.t the LTD one, more so than the loop or propagator count.
This is to be expected given the recursive nature of divided differences that are at the core of the \cLTD{} procedure discussed in sect.~\ref{sec:loop-by-loop_iteration}, and for which the recursion depth is dictated by the rank of the polynomial numerator function.
We however note that master loop integrals typically appearing in the basis resulting from the reduction of scattering loop amplitudes with integration-by-part identities feature numerators with zero or low polynomial rank.
We only show results for $\textrm{n}_\textrm{opt}=0$ for the LTD expression in the case of constant numerators as our current implementation of it cannot benefit from optimisations in that case. In the presence of non-constant numerators however, we enable optimisation with  $\textrm{n}_\textrm{opt}=100$ as the LTD expression then also greatly benefits from it.

In sect.~\ref{sec:loop-by-loop_iteration} we discussed the many a priori arbitrary choices one must make when following our \cLTD{} procedure. It is clear that all choices lead to functionally different but valid expressions that locally evaluate to the same quantity, since they all correspond to the same defining energy integrals. However, we expect the resulting expressions to have a different number of terms, and therefore a different computational complexity. This is especially impactful for topologies with higher loop count since the differences in the treatment of the first energy integrals have a cumulative effect throughout the iterative process.

\begin{figure}
    \centering
    \includegraphics[scale=0.5]{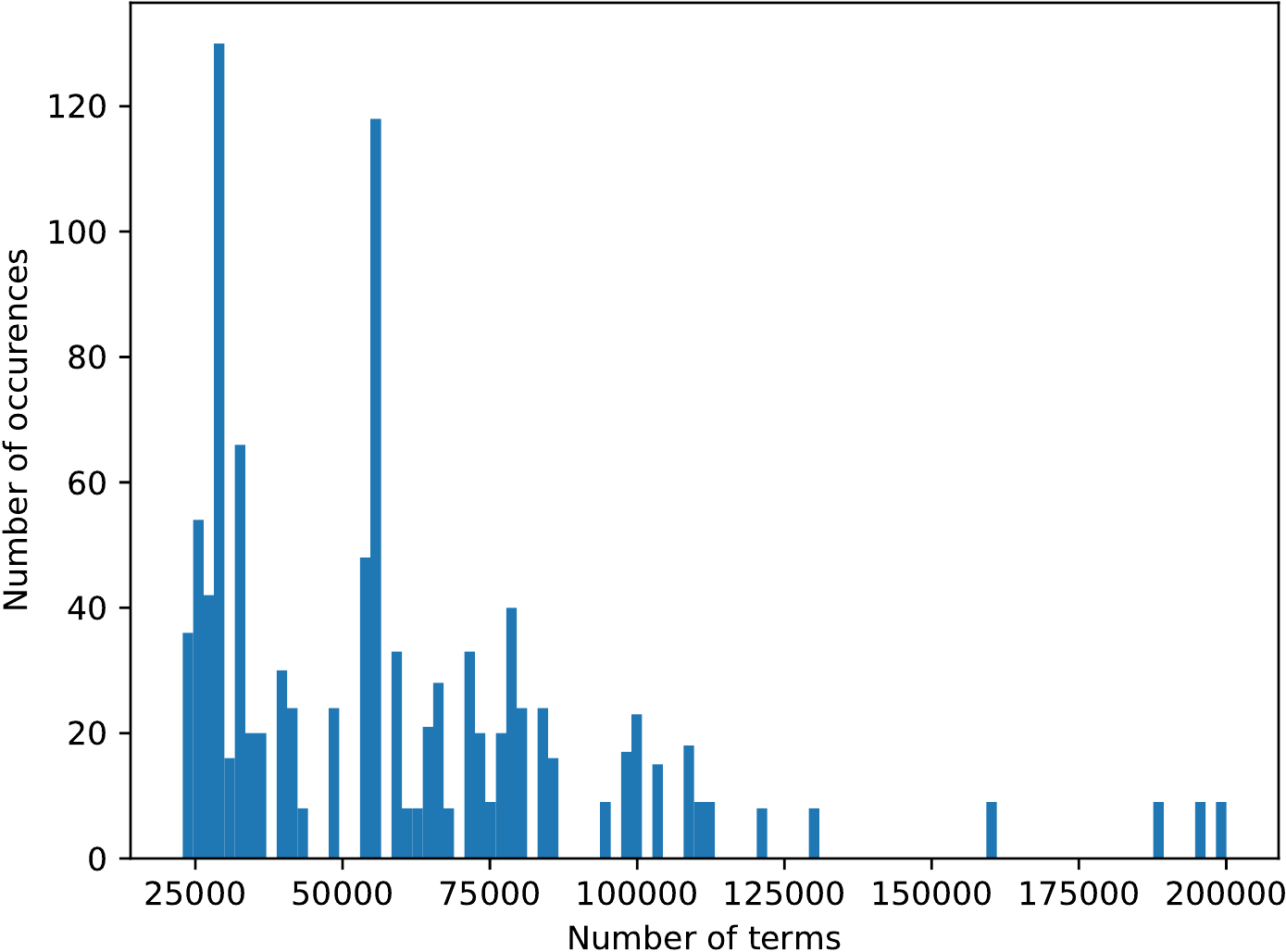}
    \caption{The distribution of the number of terms appearing in the \cLTD{} expression obtained for of the 1080 possible choices of loop line signatures and order applicable to the four-loop 2x2 fishnet topology with a constant numerator set to one.}
    \label{fig:fishnet_dist}
\end{figure}

In order to explore this aspect of the \cLTD{} representation, we adopt a compact notation for the choice of loop momentum basis and integration order. This notation corresponds to an \emph{ordered} list of loop momentum signatures together with the number of propagators that have this loop momentum signature. This matches the inputs given to the computer code \texttt{cLTD.py}, provided as ancillary material (see appendix \ref{app:pyfractioning}).
We then enumerate all 1080 possible choices of such inputs for the case of the four-loop 2x2 fishnet topology and report in fig.~\ref{fig:fishnet_dist} the number of terms appearing in the final \cLTD{} expression for each case.

We find that the worst and best set of inputs yield a number of terms in the \cLTD{} expression that differ by a factor 8.
It is interesting to note that the \textsc{form} optimisation of these different equivalent expressions cannot possibly restore optimal computational complexity, since the list of identified common sub-expressions is (and can only be) a local optimum. 
We highlight this fact by reporting in table~\ref{tab:worstbestLMB} results for the generation time and the number of arithmetic operations (additions $\textrm{N}_\textrm{+}$ and multiplications $\textrm{N}_\textrm{x}$) stemming from both a bad and good possible choice of inputs for the 2x2 fishnet topology.
We observe that the ratio of the number of multiplications necessary to evaluate the \cLTD{} expression of these two choices of inputs increases from 3.7 to about 6.0 after optimisation as expected since the local optimum found by the \texttt{Local Stochastic Search} becomes further away from the global optimum as the size of the input unoptimised expression increases.
Note that the count of terms reported in fig.~\ref{fig:fishnet_dist} is closely related but however not equivalent to neither the non-optimised nor the optimised number of arithmetic operations in the resulting expression. It can however be used as useful proxy for guiding the choice of an optimal loop momentum basis and integration order. 

\begin{table}[ht!]
    \centering
    \begin{tabular}{c|cc|cc|cccc}
        Loop momentum &\multicolumn{2}{c|}{Input} & \multicolumn{2}{c|}{$n_{\textrm{opt}}$=0} & \multicolumn{4}{c}{$n_{\textrm{opt}}$=1000} \\
        basis & signatures $\mathbf{\sigma}$ & $n_\textrm{prop}$ & $N_\textrm{+}$ & $N_\times$ & $t^{(\textrm{gen})}_{[\textrm{min}]}$ & $N_\textrm{+}$ & $N_\times$ & $t^{(\textrm{run})}_{[\mu\textrm{s}]}$\\
        \hline
        & & & & &\\
        \begin{tabular}{@{}c@{}}
\\[-0.4cm]
         \begin{tikzpicture}
            \begin{feynman}

            \tikzfeynmanset{every vertex={dot,minimum size=1mm}}

            \tikzfeynmanset{every vertex={empty dot,minimum size=0mm}}
             \vertex (a1);
            \vertex[right=0.5cm of a1] (a2);
            \vertex[right=0.5cm of a2] (a3);
            \vertex[below=1cm of a3] (a4);
            
            \vertex[below=0.5cm of a1] (t1);
            \vertex[below=0.5cm of a3] (t2);
            
            \vertex[left=0.5cm of a4] (a5);
            \vertex[left=0.5cm of a5] (a6);
                          
            \tikzfeynmanset{every vertex={dot,minimum size=0.8mm}}
            
            \vertex[right=0cm of a1] (e1);        
            \vertex[right=0cm of a3] (e2);        
            \vertex[right=0cm of a4] (e3);        
            \vertex[right=0cm of a6] (e4);    
            
            \tikzfeynmanset{every vertex={empty dot,minimum size=0mm}}                
  
            \vertex[left=0.15cm of e1] (d1);        
            \vertex[right=0.15cm of e2] (d2);        
            \vertex[right=0.15cm of e3] (d3);        
            \vertex[left=0.15cm of e4] (d4);    
            
            \tikzfeynmanset{every vertex={empty dot,minimum size=0.4mm}}         
                       \vertex[right=0.25cmof a1] ();        
            \vertex[right=0.75cm of a1] ();        
            \vertex[below right =1cm and 0.25cm of a1] ();        
            \vertex[below right =1cm and 0.75cm of a1] ();   
            
                        \node at (0.25,0.2) {\footnotesize $k_1$};
            \node at (0.75,0.2) {\footnotesize $k_3$};
            \node at (0.25,-1.2) {\footnotesize $k_4$};
            \node at (0.75,-1.2) {\footnotesize $k_2$};
            
                \diagram*[large]{	
                (a1)--(a2)--(a3)--(a4)--(a5)--(a6)--(a1), 
                (a2)--(a5),
             		(e1)--(d1),
		(e2)--(d2),
		(e3)--(d3),
		(e4)--(d4),
		(t1)--(t2),
                           }; 
            \end{feynman}
    \end{tikzpicture}\\[-0.1cm]
\end{tabular}
        &$\left[\begin{smallmatrix}
        \phantom{-}1 & \phantom{-}0 & \phantom{-}0 & \phantom{-}0 \\
        \phantom{-}0 & \phantom{-}1 & \phantom{-}0 & \phantom{-}0 \\
        \phantom{-}1 & -1 & \phantom{-}0 & \phantom{-}0 \\
        -1 & \phantom{-}0 & -1 & \phantom{-}0 \\
        \phantom{-}0 & -1 & \phantom{-}0 & -1 \\
        \phantom{-}0 & \phantom{-}0 & \phantom{-}1 & \phantom{-}0 \\
        \phantom{-}0 & \phantom{-}0 & -1 & \phantom{-}1 \\
        \phantom{-}0 & \phantom{-}0 & \phantom{-}0 & \phantom{-}1 \\
         \end{smallmatrix}\right]$ & $\left[\begin{smallmatrix}
        2 \\
        2 \\
        1 \\
        1 \\
        1 \\
        2 \\
        1 \\
        2 \\
        \end{smallmatrix}\right]$
        & 22'851 & 159'964 & 10.2 & 3'515 & 3'520 & 7.5 \\
        & & & & &\\
        \hline
        & & & & &\\
        \begin{tabular}{@{}c@{}}
\\[-0.4cm]
         \begin{tikzpicture}
            \begin{feynman}

            \tikzfeynmanset{every vertex={dot,minimum size=1mm}}

            \tikzfeynmanset{every vertex={empty dot,minimum size=0mm}}
             \vertex (a1);
            \vertex[right=0.5cm of a1] (a2);
            \vertex[right=0.5cm of a2] (a3);
            \vertex[below=1cm of a3] (a4);
            
            \vertex[below=0.5cm of a1] (t1);
            \vertex[below=0.5cm of a3] (t2);
            
            \vertex[left=0.5cm of a4] (a5);
            \vertex[left=0.5cm of a5] (a6);
                          
            \tikzfeynmanset{every vertex={dot,minimum size=0.8mm}}
            
            \vertex[right=0cm of a1] (e1);        
            \vertex[right=0cm of a3] (e2);        
            \vertex[right=0cm of a4] (e3);        
            \vertex[right=0cm of a6] (e4);    
            
            \tikzfeynmanset{every vertex={empty dot,minimum size=0mm}}                
  
            \vertex[left=0.15cm of e1] (d1);        
            \vertex[right=0.15cm of e2] (d2);        
            \vertex[right=0.15cm of e3] (d3);        
            \vertex[left=0.15cm of e4] (d4);    
            
            \tikzfeynmanset{every vertex={dot,minimum size=0.5mm}}
            
            \vertex[below right=0.25cm and 0.5cm of a1] ();        
            \vertex[below right=0.75cm and 0.5cm of a1] ();        
            \vertex[below =0.25cm of a1] ();        
            \vertex[below =0.75cm of a1] ();   
            
                        \node at (-0.2,-0.25) {\footnotesize $k_1$};
            \node at (0.30,-0.25) {\footnotesize $k_2$};
            \node at (-0.2,-0.75) {\footnotesize $k_3$};
            \node at (0.30,-0.75) {\footnotesize $k_4$};
            
                \diagram*[large]{	
                (a1)--(a2)--(a3)--(a4)--(a5)--(a6)--(a1), 
                (a2)--(a5),
             		(e1)--(d1),
		(e2)--(d2),
		(e3)--(d3),
		(e4)--(d4),
		(t1)--(t2),
                           }; 
            \end{feynman}
    \end{tikzpicture} \\[-0.1cm]
\end{tabular}
        &$\left[\begin{smallmatrix}
        \phantom{-}1 & \phantom{-}0 & \phantom{-}0 & \phantom{-}0 \\
        -1 & -1 & \phantom{-}0 & \phantom{-}0 \\
        \phantom{-}0 & \phantom{-}1 & \phantom{-}0 & \phantom{-}0 \\
        \phantom{-}1 & \phantom{-}0 & -1 & \phantom{-}0 \\
        \phantom{-}1 & \phantom{-}1 & -1 & -1 \\
        \phantom{-}0 & \phantom{-}0 & \phantom{-}1 & \phantom{-}0 \\
        \phantom{-}0 & \phantom{-}0 & \phantom{-}0 & \phantom{-}1 \\
        \phantom{-}0 & \phantom{-}0 & -1 & -1 \\
        \end{smallmatrix}\right]$ & $\left[\begin{smallmatrix}
        2 \\
        2 \\
        1 \\
        1 \\
        1 \\
        2 \\
        1 \\
        2 \\
        \end{smallmatrix}\right]$
        & 85'535 & 598'752 & 154.7 & 21'351 & 20'944 & 87
    \end{tabular}
    \caption{Single-core generation time $t^{(\textrm{gen})}_{[\textrm{min}]}$, evaluation time $t^{(\textrm{run})}_{[\mu\textrm{s}]}$ and number of arithmetic operations (additions $\textrm{N}_\textrm{+}$ and multiplications $\textrm{N}_\textrm{x}$) building the \cLTD{} expression of the four-loop 2x2 fishnet integral with a constant numerator set to one for a good (top row) and bad (bottom row) choice of inputs given to our iterative procedure. The inputs correspond to the loop momentum signatures of the \emph{ordered} list of loop lines featured by the topology, together with the number of propagator involved in each of them.}
    \label{tab:worstbestLMB}
\end{table}
\renewcommand{\arraystretch}{1.0}

The case of non-constant numerators further complicates the situation and the optimal choice of loop momentum basis and integration order then depends on the particular shape of the polynomial numerator.
In table~\ref{tab:results}, no particular attention was paid to the particular choice of inputs and we leave its optimal determination for future work.

Finally, while our results in this section highlight the competitive run-time evaluation speed of the \cLTD{} expression we must also comment on one potential drawback regarding code generation time.
Although code generation only needs to be processed once per particular integral, its processing time can still become a bottleneck if it is prohibitively slow.
In the results of table~\ref{tab:results}, we do not report quantitatively on that matter as we have not yet considered \emph{any} optimisation in this regard, thus leaving much room for improvement.
We will therefore limit ourselves to stating that the processing time of our current generation pipeline for the \cLTD{} expression of each the the scalar integrals shown in table~\ref{tab:results} goes from less than a minute for the simpler integrals to several hours for the more complicated ones.
This qualitative statement underlines that generation time will also need to be taken into consideration in future applications and developments of the \cLTD{} representation.

\subsection{Numerical stability}
\label{sec:res:numerical_stability}

One main benefit from the \cLTD{} expression is its improved numerical stability, as already discussed in sect.~\ref{sec:comparison_with_LTD}.
In this section we explore this statement more quantitatively.

\begin{figure}[t!]
\centering
\begin{subfigure}[b]{.49\linewidth}
\includegraphics[width=\linewidth]{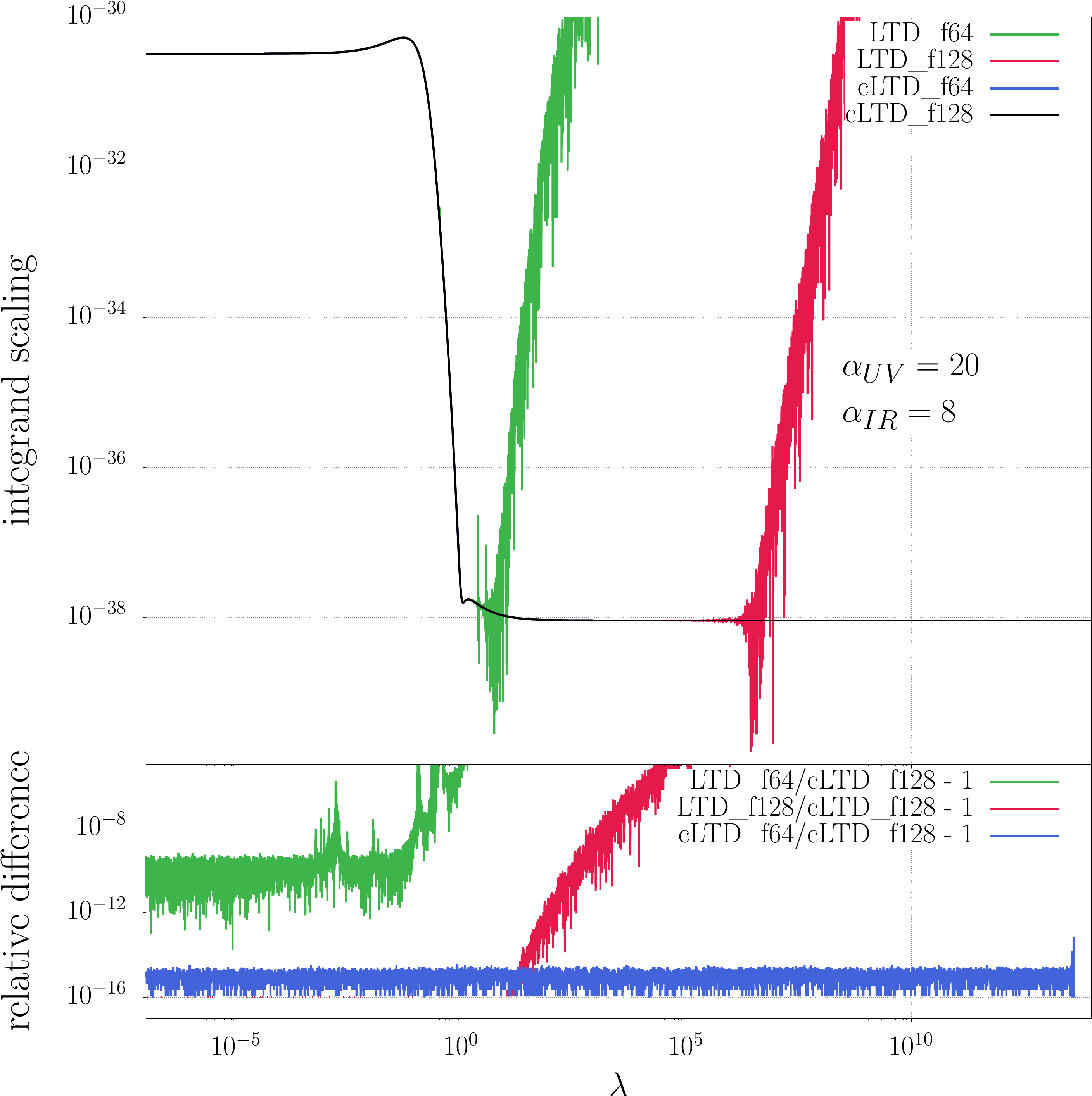}
\caption{
\begin{minipage}{.25\textwidth}
        \begin{tikzpicture}
            \begin{feynman}

            \tikzfeynmanset{every vertex={dot,minimum size=1mm}}

            \tikzfeynmanset{every vertex={empty dot,minimum size=0mm}}
             \vertex (a1);
            \vertex[right=0.5cm of a1] (a2);
            \vertex[right=0.5cm of a2] (a3);
            \vertex[below=1cm of a3] (a4);
            
            \vertex[below=0.5cm of a1] (t1);
            \vertex[below=0.5cm of a3] (t2);
            
            \vertex[left=0.5cm of a4] (a5);
            \vertex[left=0.5cm of a5] (a6);
                          
            \tikzfeynmanset{every vertex={dot,minimum size=0.8mm}}
            
            \vertex[right=0cm of a1] (e1);        
            \vertex[right=0cm of a3] (e2);        
            \vertex[right=0cm of a4] (e3);        
            \vertex[right=0cm of a6] (e4);    
            
            \tikzfeynmanset{every vertex={empty dot,minimum size=0mm}}                
  
            \vertex[left=0.15cm of e1] (d1);        
            \vertex[right=0.15cm of e2] (d2);        
            \vertex[right=0.15cm of e3] (d3);        
            \vertex[left=0.15cm of e4] (d4);    
            
                \diagram*[large]{	
                (a1)--(a2)--(a3)--(a4)--(a5)--(a6)--(a1), 
                (a2)--(a5),
             		(e1)--(d1),
		(e2)--(d2),
		(e3)--(d3),
		(e4)--(d4),
		(t1)--(t2),
                           }; 
            \end{feynman}
    \end{tikzpicture}
\end{minipage}
\begin{minipage}{.6\textwidth}
Stability of the 2x2 fishnet 4-loop integral with a constant numerator set to $1$.
\end{minipage}
}\label{fig:stability:fishnet}
\end{subfigure}
\begin{subfigure}[b]{.49\linewidth}
\includegraphics[width=\linewidth]{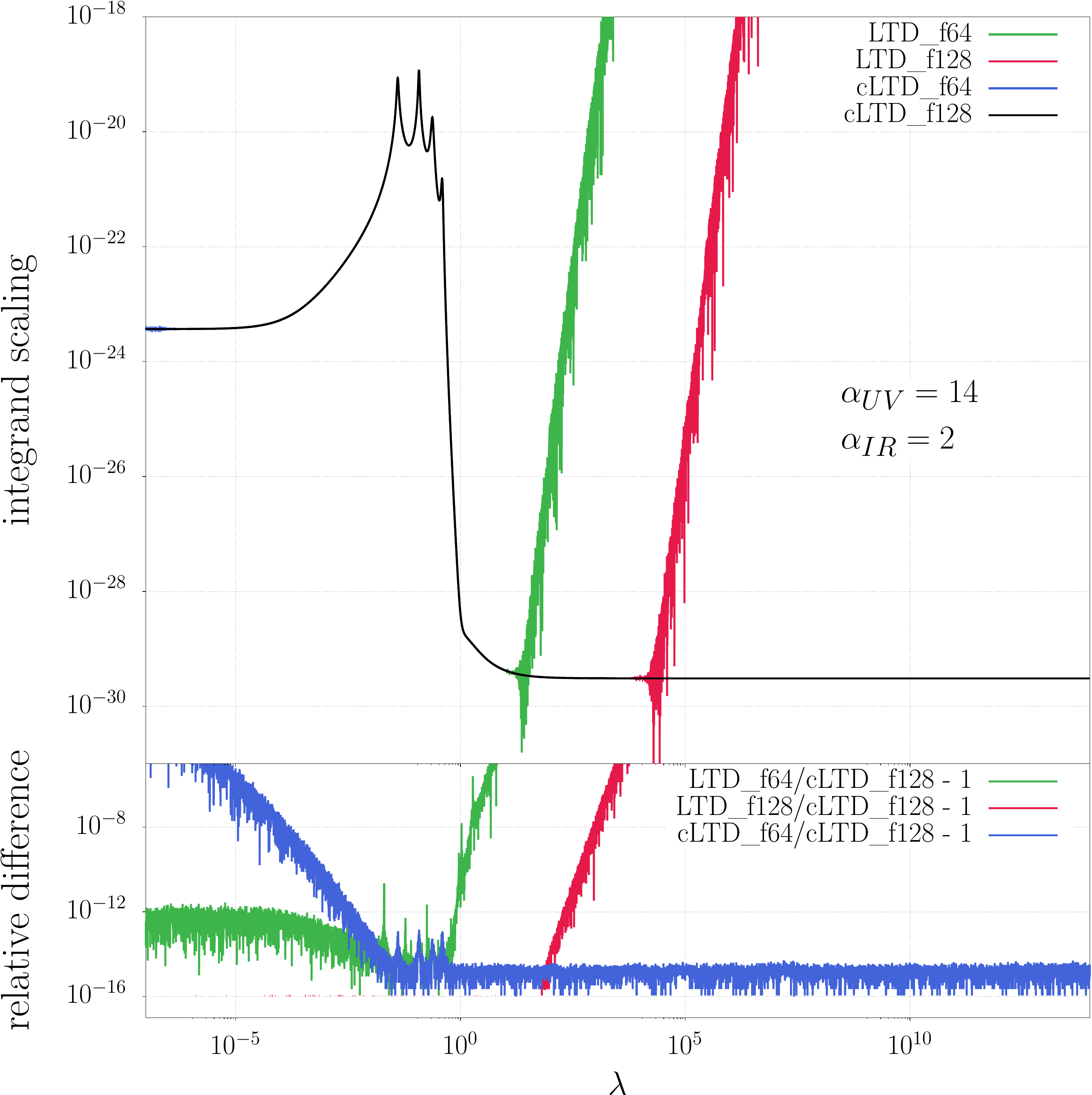}
\caption{ 
\begin{minipage}{.25\textwidth}
    \begin{tikzpicture}
    \begin{feynman}

    \tikzfeynmanset{every vertex={empty dot,minimum size=0mm}}
    \vertex (a1);
    
    \vertex[right=1cm of a1] (a3);
    \vertex[right=0.5cm of a1] (a2);
    \vertex[above=0.5cm of a2] (a4);
    \vertex[below=0.5cm of a2] (a5);
    
    \vertex[left=0.433cm of a2] (b1);
    \vertex[right=0.433cm of a2] (b2);
    
    \vertex[above=0.cm of a1] (c1);
    \vertex[above=0.25cm of b2] (c2);
    \vertex[below=0.25cm of b1] (d1);
    \vertex[below=0.cm of a3] (d2);
    \vertex[below=0.16666cm of a4] (l1);
    \vertex[below=0.33333cm of a4] (l2);
    \vertex[below=0.5cm of a4] (l3);
    \vertex[below=0.66666cm of a4] (l4);
    \vertex[below=0.83333cm of a4] (l5);
    
    \vertex[left=0.15cm of c1] (ec1);

    \vertex[right=0.15cm of d2] (ed2);
    \vertex[right=0.15cm of l1] (el1);
    \vertex[right=0.15cm of l2] (el2);
    \vertex[right=0.15cm of l3] (el3);
    \vertex[right=0.15cm of l4] (el4);
    \vertex[right=0.15cm of l5] (el5);

    \tikzfeynmanset{every vertex={dot,minimum size=0.8mm}}

    \vertex[below=0.cm of a3] (d2);
    
    \vertex[below=0.cm of l1] (l1);
    \vertex[below=0.cm of l2] (l2);
    \vertex[below=0.cm of l3] (l3);
    \vertex[below=0.cm of l4] (l4);
    \vertex[below=0.cm of l5] (l5);
    
        \diagram*[large]{	
        (a1)--[quarter left](a4) -- [quarter left](a3),
        (a5) --[quarter right](a3), 
        (a5)--[quarter left](a1),
        (a4) -- (a5),

        (d2) -- (ed2),
        (l1) -- (el1),
        (l1) -- (el1),
        (l2) -- (el2),
        (l3) -- (el3),
        (l4) -- (el4),
        (l5) -- (el5),
        }; 
    \end{feynman}
    \end{tikzpicture}
\end{minipage}
\begin{minipage}{.6\textwidth}
Stability of a 2-loop 6-point integral with the rank-2 numerator $(k_1+k_2)\cdot p_1 + k_1\cdot k_2$.
\end{minipage}
}\label{fig:stability:2L6P}
\end{subfigure} \\

\caption{\label{fig:stability} Study of the numerical stability of two different functional forms of the Loop-Tree Duality integrand: the original representation of ref.~\cite{Capatti:2019edf} (LTD) and the new manifestly causal one introduced in this work (\cLTD{}). Each representation is compared using double-precision arithmetic (\texttt{f64}) and quadruple-precision arithmetic (\texttt{f128}). The external momenta are fixed to some arbitrarily chosen real external momenta of order $\mathcal{O}(1)$. The integrands are then evaluated with spatial loop momenta $\vec{k}_i = \lambda \vec{c}_i$ where $\vec{c}_i$ are randomly chosen spatial constant directions with real-valued components of order $\mathcal{O}(1)$.
We report the resulting numerical evaluations of the absolute value of the integrands with a normalising factor $\left( 1+\lambda^{\alpha_\textrm{UV}} \right)\frac{\lambda^{\alpha_\textrm{IR}}}{1+\lambda^{\alpha_\textrm{IR}}}$, with $\alpha_\textrm{UV}$ and $\alpha_\textrm{IR}$ chosen so as to obtain a constant asymptotic behaviour in both the IR and UV limits.
}
\end{figure}

We consider the two scalar integrals depicted in the captions of fig.~\ref{fig:stability:fishnet} and fig.~\ref{fig:stability:2L6P}. 
Our selection is motivated by the fact that together these two integrals explore the numerical stability obtained in terms of the three relevant complexity parameters: the loop count, the multiplicity of external momenta and the rank of the polynomial numerator.

In order to probe the numerical stability in different relevant kinematic regions, we fix the external real four-momenta components to randomly chosen quantities of order $\mathcal{O}(1)$ and probe the LTD and \cLTD{} integrands along a particular 1-dimensional section of the spatial loop momenta phase-space defined by $\vec{k}_i = \lambda \vec{c}_i$ with a variable $\lambda$ spanning both the IR (small $\lambda$) and UV (large $\lambda$) regions. 
The defining direction $\vec{c}_i$ are chosen with random real-valued components of order $\mathcal{O}(1)$.
The numerical instabilities are correlated with the proximity of the sampling point to H-surfaces which are not isotropically distributed so that we expect quantitative differences when varying the choice of the directions $\vec{c}_i$.
In practice however, we tested several directions and found a similar qualitative behaviour of the stability for all of them. Moreover, results from direct integration (which explores the phase-space much more democratically) corroborate our findings presented in this section.
To facilitate the rendering of the results of both the IR and UV regions we choose to normalise the integrand (which does not include any jacobian) by multiplying it with the factor $\left( 1+\lambda^{\alpha_\textrm{UV}} \right)\frac{\lambda^{\alpha_\textrm{IR}}}{1+\lambda^{\alpha_\textrm{IR}}}$, with $\alpha_\textrm{UV}$ and $\alpha_\textrm{IR}$ chosen so as to obtain a constant asymptotic behaviour in both limits. 

We start by discussing results from fig.~\ref{fig:stability:fishnet} which explores the 2x2 4-loop fishnet topology with a constant numerator set to one.
We find that the double-precision result from the LTD representation (\texttt{LTD\_f64}) dramatically loses precision right after reaching the asymptotic flat behaviour (which loosely speaking coincides with a region of lesser interest as the integrals are UV finite).
This demonstrates that for such integrals, a numerical implementation in double-precision only is inapplicable, as the truncation of the UV region that would be necessary in this case would cause a non-negligible approximation error in the result of the Monte-Carlo integration. 
We notice small spikes in the numerical stability pattern of the \texttt{LTD\_f64} evaluation which we attribute to accrued proximity to dual canceling H-surfaces for such values of $\lambda$.
When considering an implementation of the LTD expression using quadruple precision arithmetic (\texttt{LTD\_f128}), the occurrence of numerical instabilities is delayed by about four order in magnitude in $\lambda$ but remain as severe as for \texttt{LTD\_f64}.
In ref.~\cite{Capatti:2019edf}, we considered numerical instability checks that included a fall-back onto the \texttt{LTD\_f128} implementation with subsequent dismissal of the sampling point if still deemed numerically unstable. In that case, the contribution of the UV region trimmed off can typically be safely neglected.
However, one is left with another severe problem: whenever a sampling point lies far in the UV region and the stability tests mischaracterise it as being stable\footnote{The shape of the evolution of the LTD integrands in fig.~\ref{fig:stability} shows that the numerically unstable evaluations are not uniformly distributed so that is not unlikely that stability tests mischaracterise an unstable point as stable.}, a very large incorrect weight can potentially be aggregated to the central value estimate of the Monte-Carlo procedure and spoil the integration.
Various strategies can be accommodated to cope with this problem, such as Taylor expansions around UV points as discussed in ref.~\cite{Capatti:2019edf} or some automated outlier detection. However, none of these approaches are entirely satisfactory as they entail some level of approximation or have problematic edge cases. Moreover, on modern computing hardware, performing arithmetic operations in quadruple precision is a hundred times slower than in double precision.
Instead, we see that the \cLTD{} representation completely removes the need for such regularisation procedure as it offers unconditional numerical stability over the entirety of the UV range. 
We remind the reader that the flooring at $10^{-15}$ of the relative accuracy of \texttt{cLTD\_f128} w.r.t \texttt{cLTD\_f64} is due to the fact the double-precision floating point representation only includes about 17 significant digits.
In the IR region and for the integral of fig.~\ref{fig:stability:fishnet} we find all representations and implementations to be numerically stable.

We now turn to the discussion of the results shown in fig.~\ref{fig:stability:2L6P}, where we consider the case of a 2-loop 6-point integrals with a rank-2 numerator which is arguably more relevant to collider phenomenology. Indeed, physics applications often involve a lower loop count than that of the 2x2 fishnet of fig.~\ref{fig:stability:fishnet} but include IR subtraction/cancellation patterns, complex-valued loop kinematics and complicated numerators.
The rank-2 numerator $(k_1+k_2)\cdot p_1 + k_1\cdot k_2$ chosen for the integral of fig.~\ref{fig:stability:2L6P} worsens its UV behaviour and is a first step towards exploring the benefits of the \cLTD{} expression in the presence of such complications.
In the UV region, we found the same qualitative behaviour as for the case of fig.~\ref{fig:stability:fishnet}, except that the numerical stability of the \texttt{LTD\_f128} implementation breaks down two orders of magnitude earlier.
In the IR region however, we observe a degradation of the numerical stability of the \texttt{cLTD\_f64} implementation compared to that of its LTD counterparts. 
It is harder to pin-point the origin of this feature, but it is important to note that this numerical stability breakdown is far less severe than that of the LTD representation in the UV region and that the \texttt{cLTD\_f128} implementation remains stable throughout the entire $\lambda$-range and can thus be considered to be a reliable ground truth.

We conclude this section by observing that our findings point towards a \emph{complementarity} of the cLTD{} and LTD representations. Whereas the numerical stability of the \cLTD{} is overall superior and completely solves numerical stability issues in the UV region, the implementation of the LTD representation is sometimes computationally cheaper and it \emph{can} also be more stable in the IR region.
We thus expect an evaluation stack consisting of both the LTD and \cLTD{} representations together with a fallback option to \texttt{f128} arithmetic to completely solve the stability issues that plagued numerical applications of LTD so far (see e.g.~\cite{Capatti:2019edf}).

\newpage
    
\clearpage
\begin{center}

\texttt{%
\begin{tabular}{@{}llrrrclccc@{}}
\hline
Topology & Rank & Method & opt & $N_+$ & $N_\times$ & $t\;[\mu s]$ & $\frac{N_{+}^{\cLTD{}}}{N_{+}^{LTD}}$  & $\frac{N_{\times}^{\cLTD{}}}{N_{\times}^{LTD}}$ & $\frac{t^{\cLTD{}}}{t^{LTD}}$ \\
\hline
\multirow{12}{*}{%
\begin{tabular}{@{}c@{}}
    \begin{tikzpicture}
    \begin{feynman}

        \tikzfeynmanset{every vertex={empty dot,minimum size=0mm}}
    \vertex (a1);

    \vertex[right=1cm of a1] (a3);
    
    \tikzfeynmanset{every vertex={empty dot,minimum size=0mm}}
    
    \vertex[right=0.5cm of a1] (a2);
    \vertex[above=0.5cm of a2] (a4);
    \vertex[below=0.5cm of a2] (a5);

    \vertex[left=0.433cm of a2] (b1);
    \vertex[left=0.433cm of a2] (b2);
    
    \vertex[right=0.433cm of a2] (b3);
    \vertex[right=0.433cm of a2] (b4);
    
    \vertex[left=0.433cm of a2] (d1);
    \vertex[left=0.433cm of a2] (d2);
    
    \vertex[right=0.433cm of a2] (d3);
    \vertex[right=0.433cm of a2] (d4);

    \vertex[above=0.255cm of b2] (c2);    
    \vertex[above=0.255cm of b3] (c3);
    \vertex[below=0.255cm of b2] (e2);   
    \vertex[below=0.255cm of b3] (e3);

    \tikzfeynmanset{every vertex={dot,minimum size=0.8mm}}
    
    \vertex[above=0.255cm of b1] (c1);
    \vertex[above=0.255cm of b4] (c4);    
    \vertex[below=0.255cm of b1] (e1);
    \vertex[below=0.255cm of b4] (e4);
    
    \tikzfeynmanset{every vertex={empty dot,minimum size=0mm}}
    
    \vertex[left=0.15cm of c1] (q1);
    \vertex[left=0.15cm of c2] (q2);
    \vertex[left=0.15cm of e1] (q3);
    \vertex[left=0.15cm of e2] (q4);
    \vertex[left=0.15cm of a1] (q5);

   \vertex[right=0.15cm of c3] (p1);
    \vertex[right=0.15cm of c4] (p2);
    \vertex[right=0.15cm of e3] (p3);
    \vertex[right=0.15cm of e4] (p4);
    \vertex[right=0.15cm of a3] (p5);
    
        \diagram*[large]{	
        (a1)--[quarter left](a4) -- [quarter left](a3),
        (a5) --[quarter right](a3), 
        (a5)--[quarter left](a1),        
               
        (c1) -- (q1),
        (e1) -- (q3),
        (c4) -- (p2),
        (e4) -- (p4),

        }; 
    \end{feynman}
    \end{tikzpicture} \\
Box
\end{tabular}}
& \multirow{3}{*}{0}& \multirow{1}{*}{LTD}& \multirow{1}{*}{\phantom{}0} & 3 & 24&  0.17  &  &  & \\
\cline{3-7}
& & \multirow{2}{*}{\cLTD{}}& 0 & 17 & 14 &   & 5.7 & 0.58 & \\
& & & 1000 &  15& 10 & 0.23 & 5.0 & 0.42 &  1.4\\
\cline{2-10}
& \multirow{3}{*}{2}& \multirow{1}{*}{LTD}& \multirow{1}{*}{\phantom{}1000} & 37& 75 &  3.7 & & &  \\
\cline{3-7}
& & \multirow{2}{*}{\cLTD{}}& 0 & 6.4$\cdot$10$^2$ & 3.4$\cdot$10$^3$&   & 17 & 45 & \\
& & & 1000 &  82 & 75 & 0.30  & 2.2 & 1.0 &  8.1$\cdot$10$^{-2}$\\
\cline{2-10}
& \multirow{3}{*}{4}& \multirow{1}{*}{LTD}& \multirow{1}{*}{\phantom{}1000} & 1.6$\cdot$10$^2$ & 2.8$\cdot$10$^2$ & 4.9   &  & & \\
\cline{3-7}
& & \multirow{2}{*}{\cLTD{}}& 0 & 5.4$\cdot$10$^3$ & 3.4$\cdot$10$^4$&   & 34 & 120 & \\
& & & 1000 &  3.9$\cdot$10$^2$ & 3.7$\cdot$10$^2$ & 4.4  & 2.4 & 1.3 & 0.90\\
\cline{2-10}
& \multirow{3}{*}{6}& \multirow{1}{*}{LTD}& \multirow{1}{*}{\phantom{}1000} & 5.0$\cdot$10$^2$ & 7.2$\cdot$10$^2$ & 7.2  &  & & \\
\cline{3-7}
& & \multirow{2}{*}{\cLTD{}}& 0 & 2.7$\cdot$10$^4$& 1.9$\cdot$10$^5$ &   & 54 & 2.6$\cdot$10$^2$ & \\
& & & 1000 &  1.2$\cdot$10$^3$ & 1.5$\cdot$10$^3$ & 14  & 2.4 & 2.1  & 1.9 \\
\hline
\multirow{9}{*}{%
\begin{tabular}{@{}c@{}}
    \begin{tikzpicture}
    \begin{feynman}
    \tikzfeynmanset{every vertex={dot,minimum size=0.8mm}}
    \vertex (a1);
    
    \vertex[right=1cm of a1] (a3);
    
    \tikzfeynmanset{every vertex={empty dot,minimum size=0mm}}
    
    \vertex[right=0.5cm of a1] (a2);
    \vertex[above=0.5cm of a2] (a4);
    \vertex[below=0.5cm of a2] (a5);

    \vertex[left=0.433cm of a2] (b1);
    \vertex[left=0.25cm of a2] (b2);
    
    \vertex[right=0.433cm of a2] (b3);
    \vertex[right=0.25cm of a2] (b4);
    
    \vertex[left=0.433cm of a2] (d1);
    \vertex[left=0.25cm of a2] (d2);
    
    \vertex[right=0.433cm of a2] (d3);
    \vertex[right=0.25cm of a2] (d4);

    \vertex[above=0.433cm of b2] (c2);    
    \vertex[above=0.255cm of b3] (c3);
    \vertex[below=0.433cm of b2] (e2);   
    \vertex[below=0.255cm of b3] (e3);
    
    \tikzfeynmanset{every vertex={dot,minimum size=0.8mm}}

    \vertex[above=0.433cm of b2] (c2);

    \vertex[above=0.433cm of b4] (c4);

    \vertex[below=0.433cm of b2] (e2);

    \vertex[below=0.433cm of b4] (e4);
    
    \tikzfeynmanset{every vertex={empty dot,minimum size=0mm}}
    
    \vertex[left=0.15cm of c1] (q1);
    \vertex[left=0.15cm of c2] (q2);
    \vertex[left=0.15cm of e1] (q3);
    \vertex[left=0.15cm of e2] (q4);
    \vertex[left=0.15cm of a1] (q5);

    \vertex[right=0.15cm of c3] (p1);
    \vertex[right=0.15cm of c4] (p2);
    \vertex[right=0.15cm of e3] (p3);
    \vertex[right=0.15cm of e4] (p4);
    \vertex[right=0.15cm of a3] (p5);
    
        \diagram*[large]{	
        (a1)--[quarter left](a4) -- [quarter left](a3),
        (a5) --[quarter right](a3), 
        (a5)--[quarter left](a1),

        (c2) -- (q2),
        (e2) -- (q4),
        (a1) -- (q5),
             
        (c4) -- (p2),
        (e4) -- (p4),
        (a3) -- (p5),
        }; 
    \end{feynman}
    \end{tikzpicture} \\
Hexagon
\end{tabular}}
& \multirow{3}{*}{0}& \multirow{1}{*}{LTD}& \multirow{1}{*}{\phantom{}0} & 5 & 60&  0.40  &  &  & \\
\cline{3-7}
& & \multirow{2}{*}{\cLTD{}}& 0 & 2.0$\cdot$10$^2$  & 2.2$\cdot$10$^2$ &   & 40 & 3.7 & \\
& & & 1000 &  1.3$\cdot$10$^2$ & 1.2$\cdot$10$^2$ & 0.35 & 26 & 2.0  & 0.88\\
\cline{2-10}
& \multirow{3}{*}{2}& \multirow{1}{*}{LTD}& \multirow{1}{*}{\phantom{}1000} & 89 & 1.8$\cdot$10$^2$&  2.9 &  &  & \\
\cline{3-7}
& & \multirow{2}{*}{\cLTD{}}& 0 & 3.8$\cdot$10$^3$& 3.9$\cdot$10$^3$&   & 43 & 22 & \\
& & & 1000 &  2.1$\cdot$10$^2$ & 2.0$\cdot$10$^2$ & 3.1  & 24 & 1.1 & 1.1 \\
\cline{2-10}
& \multirow{3}{*}{4}& \multirow{1}{*}{LTD}& \multirow{1}{*}{\phantom{}1000} & 4.0$\cdot$10$^2$ & 6.1$\cdot$10$^2$ &  3.3 &  & & \\
\cline{3-7}
& & \multirow{2}{*}{\cLTD{}}& 0 & 1.8$\cdot$10$^4$& 2.0$\cdot$10$^4$&   & 45 & 33 & \\
& & & 1000 &  1.2$\cdot$10$^3$ & 1.1$\cdot$10$^3$ & 8.1  & 3.0 & 1.8 & 2.5 \\
\hline
\multirow{9}{*}{%
\begin{tabular}{@{}c@{}}
     \begin{tikzpicture}
    \begin{feynman}

        \tikzfeynmanset{every vertex={empty dot,minimum size=0mm}}
    \vertex (a1);
    \vertex[right=0.5cm of a1] (a2);
    \vertex[right=1cm of a1] (a3);
    \vertex[above=0.5cm of a2] (a4);
    \vertex[below=0.5cm of a2] (a5);
    
    \vertex[below=0.2cm of a4] (b1);    
    \vertex[below=0.2cm of b1] (b2);    
    \vertex[below=0.2cm of b2] (b3);    
    \vertex[below=0.2cm of b3] (b4);    

    \tikzfeynmanset{every vertex={dot,minimum size=0.8mm}}
    
    \vertex[right=0.4cm of b1] (v1);
    \vertex[right=0.49cm of b2] (v2);
    \vertex[right=0.4cm of b4] (v3);
    \vertex[right=0.49cm of b3] (v4);
    
    \vertex[left=0.4cm of b1] (vv1);
    \vertex[left=0.49cm of b2] (vv2);    
    \vertex[left=0.4cm of b4] (vv3);
    \vertex[left=0.49cm of b3] (vv4);
        
    \tikzfeynmanset{every vertex={empty dot,minimum size=0mm}}

    \vertex[right=0.15cm of v1] (e1);
    \vertex[right=0.15cm of v2] (e2);
    \vertex[right=0.15cm of v3] (e3);
    \vertex[right=0.15cm of v4] (e4);
    
    \vertex[left=0.15cm of vv1] (ee1);
    \vertex[left=0.15cm of vv2] (ee2);    
    \vertex[left=0.15cm of vv3] (ee3);
    \vertex[left=0.15cm of vv4] (ee4);
     
        \diagram*[large]{	
        (a1)--[quarter left](a4) -- [quarter left](a3),
        (a5) --[quarter right](a3), 
        (a5)--[quarter left](a1),        
               
	(v1)--(e1),
	(v2)--(e2),
	(v3)--(e3),
	(v4)--(e4),
				
	(vv1)--(ee1),
	(vv2)--(ee2),
	(vv3)--(ee3),
	(vv4)--(ee4),

        }; 
    \end{feynman}
    \end{tikzpicture} \\
Octagon
\end{tabular}}
& \multirow{3}{*}{0}& \multirow{1}{*}{LTD}& \multirow{1}{*}{\phantom{}0} & 7  & 1.1$\cdot$10$^2$ &  0.72  &  & & \\
\cline{3-7}
& & \multirow{2}{*}{\cLTD{}}& 0 & 3.4$\cdot$10$^3$ & 2.2$\cdot$10$^4$ &   & 4.9$\cdot$10$^2$ & 2.0$\cdot$10$^2$ & \\
& & & 1000 &  6.0$\cdot$10$^2$ & 6.0$\cdot$10$^2$ & 1.1 & 86 & 5.5 & 1.5 \\
\cline{2-10}
& \multirow{3}{*}{2}& \multirow{1}{*}{LTD}& \multirow{1}{*}{\phantom{}1000} & 1.7$\cdot$10$^2$ & 3.4$\cdot$10$^2$ &  3.7 &  & & \\
\cline{3-7}
& & \multirow{2}{*}{\cLTD{}}& 0 & 1.3$\cdot$10$^5$ & 1.2$\cdot$10$^6$ &   & 7.6$\cdot$10$^2$ & 3.5$\cdot$10$^3$ & \\
& & & 1000 &  1.0$\cdot$10$^3$ & 9.4$\cdot$10$^2$ & 4.2 & 5.9 & 2.8 & 1.1 \\
\cline{2-10}
& \multirow{3}{*}{4}& \multirow{1}{*}{LTD}& \multirow{1}{*}{\phantom{}1000} & 6.3$\cdot$10$^2$ & 1.1$\cdot$10$^3$ &  5.1 &  & & \\
\cline{3-7}
& & \multirow{2}{*}{\cLTD{}}& 0 & 1.5$\cdot$10$^6$ & 1.5$\cdot$10$^7$ &   & 2.4$\cdot$10$^3$ & 1.4$\cdot$10$^4$ & \\
& & & 1000 &  2.1$\cdot$10$^3$ & 1.8$\cdot$10$^3$ & 6.4 & 3.3 & 1.6 & 1.2 \\
\hline
\multirow{9}{*}{%
\begin{tabular}{@{}c@{}}
     \begin{tikzpicture}
    \begin{feynman}

        \tikzfeynmanset{every vertex={empty dot,minimum size=0mm}}
    \vertex (a1);
       \tikzfeynmanset{every vertex={dot,minimum size=0.8mm}} 
    \vertex[right=1cm of a1] (a3);
    
    \tikzfeynmanset{every vertex={empty dot,minimum size=0mm}}
    
    \vertex[right=0.5cm of a1] (a2);
    \vertex[above=0.5cm of a2] (a4);
    \vertex[below=0.5cm of a2] (a5);

    \vertex[left=0.433cm of a2] (b1);
    \vertex[left=0.25cm of a2] (b2);
    
    \vertex[right=0.433cm of a2] (b3);
    \vertex[right=0.25cm of a2] (b4);
    
    \vertex[left=0.433cm of a2] (d1);
    \vertex[left=0.25cm of a2] (d2);
    
    \vertex[right=0.433cm of a2] (d3);
    \vertex[right=0.25cm of a2] (d4);
    
    \tikzfeynmanset{every vertex={dot,minimum size=0.8mm}}
    
    \vertex[above=0.cm of a1] (an);

    \tikzfeynmanset{every vertex={empty dot,minimum size=0mm}}
    
    \vertex[left=0.15cm of c1] (q1);
    \vertex[left=0.15cm of c2] (q2);
    \vertex[left=0.15cm of e1] (q3);
    \vertex[left=0.15cm of e2] (q4);
    \vertex[left=0.15cm of a1] (q5);
        \vertex[left=0.15cm of an] (q6);
    
     \vertex[above=0.15cm of a4] (ext);   
    \vertex[right=0.15cm of c3] (p1);

    \vertex[right=0.15cm of e3] (p3);

    \vertex[right=0.15cm of a3] (p5);
    
        \diagram*[large]{	
        (a1)--[quarter left](a4) -- [quarter left](a3),
        (a5) --[quarter right](a3), 
        (a5)--[quarter left](a1),

        (a3) -- (p5),
        (a4)--(a5),
        (an) -- (q6),

        }; 
        
    \tikzfeynmanset{every vertex={dot,minimum size=0.8mm}}

    \end{feynman}
    \end{tikzpicture} \\
DTriangle
\end{tabular}}
& \multirow{3}{*}{0}& \multirow{1}{*}{LTD}& \multirow{1}{*}{\phantom{}0} & 7 & 57 &  0.23  &  & & \\
\cline{3-7}
& & \multirow{2}{*}{\cLTD{}}& 0 & 19  & 60 &   & 2.7 & 1.1 & \\
& & & 1000 &  17 & 16 & 0.12 & 2.4 & 0.28 & 0.52 \\
\cline{2-10}
& \multirow{3}{*}{2}& \multirow{1}{*}{LTD}& \multirow{1}{*}{\phantom{}1000} & 85 & 1.7$\cdot$10$^2$ &  0.65 &  & & \\
\cline{3-7}
& & \multirow{2}{*}{\cLTD{}}& 0 & 5.2$\cdot$10$^2$ & 2.3$\cdot$10$^3$ &   & 6.1 & 14 & \\
& & & 1000 &  1.8$\cdot$10$^2$ & 1.6$\cdot$10$^2$ & 0.61 & 2.1 & 0.94 & 0.94 \\
\cline{2-10}
& \multirow{3}{*}{4}& \multirow{1}{*}{LTD}& \multirow{1}{*}{\phantom{}1000} & 3.2$\cdot$10$^2$ & 4.7$\cdot$10$^2$ &  3.6 &  & & \\
\cline{3-7}
& & \multirow{2}{*}{\cLTD{}}& 0 & 4.9$\cdot$10$^3$ & 3.4$\cdot$10$^4$ &  & 15 & 72 & \\
& & & 1000 &  9.0$\cdot$10$^2$ & 8.0$\cdot$10$^2$ & 4.0 & 2.8 & 1.7 & 1.1 \\
\hline
\end{tabular}%
}%
\end{center}

\begin{center}
\texttt{%
\begin{tabular}{@{}llrrrclccc@{}}
\hline
Topology & Rank & Method & opt & $N_+$ & $N_\times$ & $t\;[\mu s]$ & $\frac{N_{+}^{\cLTD{}}}{N_{+}^{LTD}}$  & $\frac{N_{\times}^{\cLTD{}}}{N_{\times}^{LTD}}$ & $\frac{t^{\cLTD{}}}{t^{LTD}}$  \\
\hline
\multirow{9}{*}{%
\begin{tabular}{@{}c@{}}
    \begin{tikzpicture}
    \begin{feynman}

        \tikzfeynmanset{every vertex={empty dot,minimum size=0mm}}
    \vertex (a1);

    \vertex[right=1cm of a1] (a3);
    
    \tikzfeynmanset{every vertex={empty dot,minimum size=0mm}}
    
    \vertex[right=0.5cm of a1] (a2);
    \vertex[above=0.5cm of a2] (a4);
    \vertex[below=0.5cm of a2] (a5);

    \vertex[left=0.433cm of a2] (b1);
    \vertex[left=0.433cm of a2] (b2);
    
    \vertex[right=0.433cm of a2] (b3);
    \vertex[right=0.433cm of a2] (b4);
    
    \vertex[left=0.433cm of a2] (d1);
    \vertex[left=0.433cm of a2] (d2);
    
    \vertex[right=0.433cm of a2] (d3);
    \vertex[right=0.433cm of a2] (d4);

    \vertex[above=0.255cm of b2] (c2);    
    \vertex[above=0.255cm of b3] (c3);
    \vertex[below=0.255cm of b2] (e2);   
    \vertex[below=0.255cm of b3] (e3);

    \tikzfeynmanset{every vertex={dot,minimum size=0.8mm}}
    
    \vertex[above=0.255cm of b1] (c1);
    \vertex[above=0.255cm of b4] (c4);    
    \vertex[below=0.255cm of b1] (e1);
    \vertex[below=0.255cm of b4] (e4);
    
    \tikzfeynmanset{every vertex={empty dot,minimum size=0mm}}
    
    \vertex[left=0.15cm of c1] (q1);
    \vertex[left=0.15cm of c2] (q2);
    \vertex[left=0.15cm of e1] (q3);
    \vertex[left=0.15cm of e2] (q4);
    \vertex[left=0.15cm of a1] (q5);

   \vertex[right=0.15cm of c3] (p1);
    \vertex[right=0.15cm of c4] (p2);
    \vertex[right=0.15cm of e3] (p3);
    \vertex[right=0.15cm of e4] (p4);
    \vertex[right=0.15cm of a3] (p5);
    
        \diagram*[large]{	
        (a1)--[quarter left](a4) -- [quarter left](a3),
        (a5) --[quarter right](a3), 
        (a5)--[quarter left](a1),        
               
        (c1) -- (q1),
        (e1) -- (q3),
        (c4) -- (p2),
        (e4) -- (p4),
        (a4) -- (a5),

        }; 
    \end{feynman}
    \end{tikzpicture}
     \\
DoubleBox\end{tabular}}
& \multirow{3}{*}{0}& \multirow{1}{*}{LTD}& \multirow{1}{*}{\phantom{}0} & 14 & 1.7$\cdot$10$^2$&  0.53  &  & & \\
\cline{3-7}
& & \multirow{2}{*}{\cLTD{}}& 0 & 2.5$\cdot$10$^2$ & 1.3$\cdot$10$^3$ &   & 18 & 7.6 & \\
& & & 1000 &  1.1$\cdot$10$^2$ & 1.2$\cdot$10$^2$ & 0.36 & 7.9 & 0.71 & 0.68 \\
\cline{2-10}
& \multirow{3}{*}{2}& \multirow{1}{*}{LTD}& \multirow{1}{*}{\phantom{}1000} & 2.6$\cdot$10$^2$& 5.1$\cdot$10$^2$&  1.0 &  & & \\
\cline{3-7}
& & \multirow{2}{*}{\cLTD{}}& 0 & 1.0$\cdot$10$^4$& 8.0$\cdot$10$^4$ &   & 38 & 1.6$\cdot$10$^2$ & \\
& & & 1000 &  6.0$\cdot$10$^2$ & 4.2$\cdot$10$^2$ & 2.7 & 2.3 & 0.82 & 2.7 \\
\cline{2-10}
& \multirow{3}{*}{4}& \multirow{1}{*}{LTD}& \multirow{1}{*}{\phantom{}1000} & 9.1$\cdot$10$^2$ & 1.5$\cdot$10$^3$ &  6.0 &  & & \\
\cline{3-7}
& & \multirow{2}{*}{\cLTD{}}& 0 & 1.3$\cdot$10$^5$ & 1.2$\cdot$10$^6$ &   & 1.4$\cdot$10$^2$ & 8.0$\cdot$10$^2$ & \\
& & & 1000 &  3.2$\cdot$10$^3$ & 2.5$\cdot$10$^3$ & 10 & 3.5 & 1.7 &  1.7 \\
\hline
\multirow{9}{*}{%
\begin{tabular}{@{}c@{}}
     \begin{tikzpicture}
    \begin{feynman}

    \tikzfeynmanset{every vertex={empty dot,minimum size=0mm}}
    \vertex (a1);
    
    \vertex[right=1cm of a1] (a3);
    \vertex[right=0.5cm of a1] (a2);
    \vertex[above=0.5cm of a2] (a4);
    \vertex[below=0.5cm of a2] (a5);
    
    \vertex[left=0.433cm of a2] (b1);
    \vertex[right=0.433cm of a2] (b2);
    
    \vertex[above=0.cm of a1] (c1);
    \vertex[above=0.25cm of b2] (c2);
    \vertex[below=0.25cm of b1] (d1);
    \vertex[below=0.cm of a3] (d2);
    \vertex[below=0.16666cm of a4] (l1);
    \vertex[below=0.33333cm of a4] (l2);
    \vertex[below=0.5cm of a4] (l3);
    \vertex[below=0.66666cm of a4] (l4);
    \vertex[below=0.83333cm of a4] (l5);
    
    \vertex[left=0.15cm of c1] (ec1);

    \vertex[right=0.15cm of d2] (ed2);
    \vertex[right=0.15cm of l1] (el1);
    \vertex[right=0.15cm of l2] (el2);
    \vertex[right=0.15cm of l3] (el3);
    \vertex[right=0.15cm of l4] (el4);
    \vertex[right=0.15cm of l5] (el5);

    \tikzfeynmanset{every vertex={dot,minimum size=0.8mm}}

    \vertex[below=0.cm of a3] (d2);
    
    \vertex[below=0.cm of l1] (l1);
    \vertex[below=0.cm of l2] (l2);
    \vertex[below=0.cm of l3] (l3);
    \vertex[below=0.cm of l4] (l4);
    \vertex[below=0.cm of l5] (l5);
    
        \diagram*[large]{	
        (a1)--[quarter left](a4) -- [quarter left](a3),
        (a5) --[quarter right](a3), 
        (a5)--[quarter left](a1),
        (a4) -- (a5),

        (d2) -- (ed2),
        (l1) -- (el1),
        (l1) -- (el1),
        (l2) -- (el2),
        (l3) -- (el3),
        (l4) -- (el4),
        (l5) -- (el5),
        }; 
    \end{feynman}
    \end{tikzpicture} \\
2L6P.a\end{tabular}}
& \multirow{3}{*}{0}& \multirow{1}{*}{LTD}& \multirow{1}{*}{\phantom{}0} & 19 & 3.0$\cdot$10$^2$&  0.94  &  & & \\
\cline{3-7}
& & \multirow{2}{*}{\cLTD{}}& 0 & 3.4$\cdot$10$^3$ & 2.1$\cdot$10$^4$ &   & 1.8$\cdot$10$^2$ & 70 & \\
& & & 1000 &  5.1$\cdot$10$^2$ & 5.5$\cdot$10$^2$ & 1.1 & 27 & 1.8 & 1.2 \\
\cline{2-10}
& \multirow{3}{*}{2}& \multirow{1}{*}{LTD}& \multirow{1}{*}{\phantom{}1000} & 4.2$\cdot$10$^2$ & 9.3$\cdot$10$^2$ &  4.1 &  & & \\
\cline{3-7}
& & \multirow{2}{*}{\cLTD{}}& 0 & 1.3$\cdot$10$^5$ & 1.1$\cdot$10$^6$ &  & 3.1$\cdot$10$^2$ & 1.2$\cdot$10$^3$ & \\
& & & 1000 &  2.0$\cdot$10$^3$ & 1.7$\cdot$10$^3$ & 5.9 & 4.8 & 1.8 & 1.4 \\
\cline{2-10}
& \multirow{3}{*}{4}& \multirow{1}{*}{LTD}& \multirow{1}{*}{\phantom{}1000} & 1.2$\cdot$10$^3$ & 2.0$\cdot$10$^3$ &  7.2 &  & & \\
\cline{3-7}
& & \multirow{2}{*}{\cLTD{}}& 0 & 1.7$\cdot$10$^6$& 1.7$\cdot$10$^7$ &   & 1.4$\cdot$10$^3$ & 8.5$\cdot$10$^3$ & \\
& & & 1000 & 8.2$\cdot$10$^3$ & 7.8$\cdot$10$^3$ & 22 & 6.8 & 3.9 & 3.1\\
\hline
\multirow{9}{*}{%
\begin{tabular}{@{}c@{}}
     \begin{tikzpicture}
    \begin{feynman}
    \tikzfeynmanset{every vertex={dot,minimum size=0.8mm}}
    \vertex (a1);
    
    \vertex[right=1cm of a1] (a3);
    
    \tikzfeynmanset{every vertex={empty dot,minimum size=0mm}}
    
    \vertex[right=0.5cm of a1] (a2);
    \vertex[above=0.5cm of a2] (a4);
    \vertex[below=0.5cm of a2] (a5);

    \vertex[left=0.433cm of a2] (b1);
    \vertex[left=0.25cm of a2] (b2);
    
    \vertex[right=0.433cm of a2] (b3);
    \vertex[right=0.25cm of a2] (b4);
    
    \vertex[left=0.433cm of a2] (d1);
    \vertex[left=0.25cm of a2] (d2);
    
    \vertex[right=0.433cm of a2] (d3);
    \vertex[right=0.25cm of a2] (d4);
    
    \tikzfeynmanset{every vertex={dot,minimum size=0.8mm}}

    \vertex[above=0.433cm of b2] (c2);

    \vertex[above=0.433cm of b4] (c4);

    \vertex[below=0.433cm of b2] (e2);

    \vertex[below=0.433cm of b4] (e4);
    
    \tikzfeynmanset{every vertex={empty dot,minimum size=0mm}}
    
    \vertex[left=0.15cm of c1] (q1);
    \vertex[left=0.15cm of c2] (q2);
    \vertex[left=0.15cm of e1] (q3);
    \vertex[left=0.15cm of e2] (q4);
    \vertex[left=0.15cm of a1] (q5);

    \vertex[right=0.15cm of c4] (p2);
    \vertex[right=0.15cm of e4] (p4);
    \vertex[right=0.15cm of a3] (p5);
    
        \diagram*[large]{	
        (a1)--[quarter left](a4) -- [quarter left](a3),
        (a5) --[quarter right](a3), 
        (a5)--[quarter left](a1),
        (a4)--(a5),

        (c2) -- (q2),
        (e2) -- (q4),
        (a1) -- (q5),
             
        (c4) -- (p2),
        (e4) -- (p4),
        (a3) -- (p5),
        }; 
    \end{feynman}
    \end{tikzpicture} \\
2L6P.c
\end{tabular}}
& \multirow{3}{*}{0}& \multirow{1}{*}{LTD}& \multirow{1}{*}{\phantom{}0} & 23 & 3.6$\cdot$10$^2$ &  1.1  &  & & \\
\cline{3-7}
& & \multirow{2}{*}{\cLTD{}}& 0 & 3.4$\cdot$10$^3$ & 2.4$\cdot$10$^4$ &  & 1.5$\cdot$10$^2$ & 67 & \\
& & & 1000 &  5.2$\cdot$10$^2$ & 5.6$\cdot$10$^2$ & 1.1 & 23 & 1.6 & 1 \\
\cline{2-10}
& \multirow{3}{*}{2}& \multirow{1}{*}{LTD}& \multirow{1}{*}{\phantom{}1000} & 3.0$\cdot$10$^2$ & 6.8$\cdot$10$^2$ &  3.6 &  & & \\
\cline{3-7}
& & \multirow{2}{*}{\cLTD{}}& 0 & 1.2$\cdot$10$^5$ & 1.2$\cdot$10$^6$ &   & 4.0$\cdot$10$^2$ & 1.8$\cdot$10$^3$ & \\
& & & 1000 &  4.0$\cdot$10$^3$ & 3.6$\cdot$10$^3$ & 14 & 13 & 5.3 & 3.9 \\
\cline{2-10}
& \multirow{3}{*}{4}& \multirow{1}{*}{LTD}& \multirow{1}{*}{\phantom{}1000} & 2.3$\cdot$10$^3$ & 3.7$\cdot$10$^3$ & 25 &  & & \\
\cline{3-7}
& & \multirow{2}{*}{\cLTD{}}& 0 & 1.7$\cdot$10$^6$ & 1.9$\cdot$10$^7$ &  & 7.4$\cdot$10$^2$ & 5.1$\cdot$10$^3$ & \\
& & & 1000 &  1.5$\cdot$10$^4$ & 1.0$\cdot$10$^4$ & 42 & 6.5 & 2.7 & 1.7 \\
\hline
\multirow{9}{*}{%
\begin{tabular}{@{}c@{}}
  \begin{tikzpicture}
    \begin{feynman}

    \tikzfeynmanset{every vertex={empty dot,minimum size=0mm}}
    \vertex (a1);
    
    \vertex[right=1cm of a1] (a3);
    \vertex[right=0.5cm of a1] (a2);
    \vertex[above=0.5cm of a2] (a4);
    \vertex[below=0.5cm of a2] (a5);
    
    \vertex[left=0.433cm of a2] (b1);
    \vertex[right=0.433cm of a2] (b2);
    
    \vertex[above=0.25cm of b1] (c1);
    \vertex[above=0.25cm of b2] (c2);
    \vertex[below=0.25cm of b1] (d1);
    \vertex[below=0.25cm of b2] (d2);
    \vertex[above=0.3333cm of a2] (l1);
    \vertex[below=0.3333cm of a2] (l2);
    
    \vertex[left=0.15cm of c1] (ec1);
    \vertex[right=0.15cm of c2] (ec2);
    \vertex[left=0.15cm of d1] (ed1);
    \vertex[right=0.15cm of d2] (ed2);
    \vertex[right=0.15cm of l1] (el1);
    \vertex[right=0.15cm of l2] (el2);     
    
    \tikzfeynmanset{every vertex={dot,minimum size=0.8mm}}
    
    \vertex[above=0.25cm of b1] (c1);
    \vertex[above=0.25cm of b2] (c2);
    
    \vertex[below=0.25cm of b1] (d1);
    \vertex[below=0.25cm of b2] (d2);
    
    \vertex[above=0.3333cm of a2] (l1);
    \vertex[below=0.3333cm of a2] (l2);
    
        \diagram*[large]{	
        (a1)--[quarter left](a4) -- [quarter left](a3),
        (a5) --[quarter right](a3), 
        (a5)--[quarter left](a1),
        (a4) -- (a5),
        
        (c1) -- (ec1),
        (c2) -- (ec2),
        (d1) -- (ed1),
        (d2) -- (ed2),
        (l1) -- (el1),
        (l2) -- (el2),

        }; 
    \end{feynman}
    \end{tikzpicture} \\
2L6P.f
\end{tabular}}
& \multirow{3}{*}{0}& \multirow{1}{*}{LTD}& \multirow{1}{*}{\phantom{}0} & 26 & 4.1$\cdot$10$^2$ &  1.4  &  & & \\
\cline{3-7}
& & \multirow{2}{*}{\cLTD{}}& 0 & 4.5$\cdot$10$^3$  & 3.2$\cdot$10$^4$ &  & 1.7$\cdot$10$^2$ & 78 & \\
& & & 1000 &  1.0$\cdot$10$^3$ & 9.5$\cdot$10$^2$ & 1.8 & 38 & 2.3 & 1.3 \\
\cline{2-10}
& \multirow{3}{*}{2}& \multirow{1}{*}{LTD}& \multirow{1}{*}{\phantom{}1000} & 6.2$\cdot$10$^2$ & 1.3$\cdot$10$^3$ &  6.6 &  & & \\
\cline{3-7}
& & \multirow{2}{*}{\cLTD{}}& 0 & 2.5$\cdot$10$^5$ & 2.5$\cdot$10$^6$ & & 4.0$\cdot$10$^2$ & 1.9$\cdot$10$^3$ & \\
& & & 1000 &  2.7$\cdot$10$^4$ & 2.0$\cdot$10$^3$ & 18 & 44 & 1.5 & 2.7 \\
\cline{2-10}
& \multirow{3}{*}{4}& \multirow{1}{*}{LTD}& \multirow{1}{*}{\phantom{}1000} & 2.2$\cdot$10$^3$ & 4.0$\cdot$10$^3$ &  8.5 &  & & \\
\cline{3-7}
& & \multirow{2}{*}{\cLTD{}}& 0 & 4.3$\cdot$10$^6$ & 4.9$\cdot$10$^7$ &  & 2.0$\cdot$10$^3$ & 1.2$\cdot$10$^4$ & \\
& & & 1000 &  3.1$\cdot$10$^4$ & 2.5$\cdot$10$^4$ &  1.5$\cdot$10$^2$ & 14 &  6.3 & 16\\
\hline
\end{tabular}%
}%
\end{center}

\begin{table}[tbp]
\begin{center}
\texttt{%
\begin{tabular}{@{}llrrrclccc@{}}
\hline
Topology & Rank & Method & opt & $N_+$ & $N_\times$ & $t\;[\mu s]$ & $\frac{N_{+}^{\cLTD{}}}{N_{+}^{LTD}}$  & $\frac{N_{\times}^{\cLTD{}}}{N_{\times}^{LTD}}$ & $\frac{t^{\cLTD{}}}{t^{LTD}}$  \\
\hline
\multirow{6}{*}{%
\begin{tabular}{@{}c@{}}
     \begin{tikzpicture}
    \begin{feynman}
    
    \tikzfeynmanset{every vertex={empty dot,minimum size=0mm}}
    
    \vertex (a1);
    
    \vertex[below=0.5cm of a1] (a3);
    \vertex[below=0.5cm of a3] (b1);
    \vertex[below=0.35355cm of a3] (a5);
    \vertex[left=0.35355cm of a5] (a6);
    
    \vertex[left=0.5cm of a3] (a7);
    \vertex[left=0.15cm of a7] (a8);
    
    \vertex[right=0.5cm of a3] (a9);
    \vertex[right=0.15cm of a9] (a10);

    \tikzfeynmanset{every vertex={dot,minimum size=0.8mm}}
    \vertex[left=0.5cm of a3] (a2);
    \vertex[right=0.5cm of a3] (a4);

        \diagram*[large]{	
        (a1)--[quarter left](a4) -- [quarter left](b1) -- [quarter left](a2)-- [quarter left](a1),
        (a3) -- (a1),
        (a3) -- (a6),
        (a3) -- (a4),
        (a7) -- (a8),
        (a9) -- (a10),
        }; 
    \end{feynman}
    \end{tikzpicture} \\
Mercedes
\end{tabular}}
& \multirow{3}{*}{0}& \multirow{1}{*}{LTD}& \multirow{1}{*}{\phantom{}0} & 23 & 2.4$\cdot$10$^2$ &  0.65  &  & & \\
\cline{3-7}
& & \multirow{2}{*}{\cLTD{}}& 0 & 87  & 3.5$\cdot$10$^2$ &   & 3.8 & 1.5 & \\
& & & 1000 &  62 & 53 & 0.23 & 2.7 & 0.22 & 0.35 \\
\cline{2-10}
& \multirow{3}{*}{4}& \multirow{1}{*}{LTD}& \multirow{1}{*}{\phantom{}0} & 2.2$\cdot$10$^3$ & 3.3$\cdot$10$^3$ &  11 &  & & \\
\cline{3-7}
& & \multirow{2}{*}{\cLTD{}}& 0 & 5.9$\cdot$10$^4$ & 4.9$\cdot$10$^5$ &   & 27 & 1.5$\cdot$10$^2$ & \\
& & & 1000 &  5.0$\cdot$10$^3$ & 4.1$\cdot$10$^3$ & 13 & 2.3 & 1.2 & 1.2 \\
\hline
\multirow{9}{*}{%
\begin{tabular}{@{}c@{}}
         \begin{tikzpicture}
            \begin{feynman}

            \tikzfeynmanset{every vertex={dot,minimum size=1mm}}

            \tikzfeynmanset{every vertex={empty dot,minimum size=0mm}}
             \vertex (a1);
            \vertex[right=0.33cm of a1] (a2);
            \vertex[right=0.33cm of a2] (a3);
             \vertex[right=0.33cm of a3] (a8);
            \vertex[below=1cm of a8] (a4);
            
            \vertex[left=0.33cm of a4] (a5);
            \vertex[left=0.33cm of a5] (a6);
            \vertex[left=0.33cm of a6] (a7);
                          
            \tikzfeynmanset{every vertex={dot,minimum size=0.8mm}}
            
            \vertex[right=0cm of a1] (e1);        
            \vertex[right=0cm of a8] (e2);        
            \vertex[right=0cm of a4] (e3);        
            \vertex[right=0cm of a7] (e4);    
            
            \tikzfeynmanset{every vertex={empty dot,minimum size=0mm}}                
  
            \vertex[left=0.15cm of e1] (d1);        
            \vertex[right=0.15cm of e2] (d2);        
            \vertex[right=0.15cm of e3] (d3);        
            \vertex[left=0.15cm of e4] (d4);    
            
                \diagram*[large]{	
                (a1)--(a2)--(a3)--(a8)--(a4)--(a5)--(a6)--(a7)--(a1), 
                (a2)--(a6),
                (a3)--(a5),
		(e1)--(d1),
		(e2)--(d2),
		(e3)--(d3),
		(e4)--(d4),
                           }; 
            \end{feynman}
    \end{tikzpicture} \\
TriBox
\end{tabular}}
& \multirow{3}{*}{0}& \multirow{1}{*}{LTD}& \multirow{1}{*}{\phantom{}0} & 55 & 9.0$\cdot$10$^2$ &  1.6  &  & & \\
\cline{3-7}
& & \multirow{2}{*}{\cLTD{}}& 0 & 3.4$\cdot$10$^3$  & 2.1$\cdot$10$^4$ & & 62 & 23 & \\
& & & 1000 &  5.9$\cdot$10$^2$ & 5.9$\cdot$10$^3$ & 1.1 & 11 & 6.6 & 0.69 \\
\cline{2-10}
& \multirow{3}{*}{2}& \multirow{1}{*}{LTD}& \multirow{1}{*}{\phantom{}1000} & 2.0$\cdot$10$^3$ & 4.0$\cdot$10$^3$ &  14 &  & & \\
\cline{3-7}
& & \multirow{2}{*}{\cLTD{}}& 0 & 2.5$\cdot$10$^5$ & 2.2$\cdot$10$^6$ &  & 1.3$\cdot$10$^2$ & 5.5$\cdot$10$^2$ & \\
& & & 1000 &  5.9$\cdot$10$^3$ & 4.5$\cdot$10$^3$ & 16 & 3.0 & 1.1 & 1.1 \\
\cline{2-10}
& \multirow{3}{*}{4}& \multirow{1}{*}{LTD}& \multirow{1}{*}{\phantom{}1000} & 1.1$\cdot$10$^4$ & 1.7$\cdot$10$^4$ &  59 &  & & \\
\cline{3-7}
& & \multirow{2}{*}{\cLTD{}}& 0 & 5.5$\cdot$10$^6$ & 5.6$\cdot$10$^7$& & 5.0$\cdot$10$^2$ & 3.3$\cdot$10$^3$ & \\
& & & 1000 &  5.5$\cdot$10$^4$ & 3.8$\cdot$10$^4$ & 2.1$\cdot$10$^2$ & 5.0 & 2.2 & 3.6 \\
\hline
\multirow{3}{*}{%
\begin{tabular}{@{}c@{}}
\\[-0.4cm]
        \begin{tikzpicture}
            \begin{feynman}
            \tikzfeynmanset{every vertex={dot,minimum size=1mm}}
            \tikzfeynmanset{every vertex={empty dot,minimum size=0mm}}
             \vertex (a1);
            \vertex[right=0.25cm of a1] (a2);
            \vertex[above=0.5cm of a2] (a4);
            \vertex[below=0.5cm of a2] (a5);

            \vertex[right=1cm of a1] (d1);

            \vertex[right=0.25cm of a4] (b1);
            \vertex[right=0.25cm of a5] (b2);
            \vertex[right=0.25cm of b1] (c1);
            \vertex[right=0.25cm of b2] (c2);
            
            \vertex[above=0.5cm of a1] (a7);
            \vertex[below=0.5cm of a1] (a8);
            \vertex[above=0.5cm of d1] (a9);
            \vertex[below=0.5cm of d1] (a10);
            
            \vertex[left=0.15cm of a7] (a11);
            \vertex[left=0.15cm of a8] (a12);
            \vertex[right=0.15cm of a9] (a13);
            \vertex[right=0.15cm of a10] (a14);
            
            \tikzfeynmanset{every vertex={dot,minimum size=0.8mm}}

            \vertex[above=0.5cm of a1] (e1);
            \vertex[below=0.5cm of a1] (e2);    
            
            \vertex[above=0.5cm of d1] (f1);
            \vertex[below=0.5cm of d1] (f2);

                \diagram*[large]{	
                (e1)--(a4), 
                (a5)--(e2),
		(e1)--(e2),

                (a4) -- (a5),
                (a4) -- (b1),
                (a5) -- (b2),
                (b1)--(b2),
                (b1)--(c1),
                (b2)--(c2),
                (c1)--(c2),

                (c1) --  (f1),
                (f2) -- (c2),
		        (f1)--(f2),
		        
		        (a7) -- (a11),
		        (a8) -- (a12),
		        (a9) -- (a13),
		        (a10) -- (a14),
                }; 
            \end{feynman}
    \end{tikzpicture} \\[-0.1cm]
QuadBox
\end{tabular}}
& \multirow{3}{*}{0}& \multirow{1}{*}{LTD}& \multirow{1}{*}{\phantom{}0} & 2.1$\cdot$10$^2$ & 4.4$\cdot$10$^3$ &  7.3  &  & & \\[0.2em]
\cline{3-7}
& & \multirow{2}{*}{\cLTD{}}& 0 & 4.9$\cdot$10$^4$  & 4.4$\cdot$10$^5$ & & 2.3$\cdot$10$^3$ & 1.0$\cdot$10$^2$ & \\[0.2em]
& & & 1000 &  2.8$\cdot$10$^3$ & 2.5$\cdot$10$^3$ & 4.4 & 13 & 0.57  & 0.60 \\[0.2em]
\hline
\multirow{3}{*}{%
\begin{tabular}{@{}c@{}}
\\[-0.4cm]
        \begin{tikzpicture}
            \begin{feynman}

            \tikzfeynmanset{every vertex={dot,minimum size=1mm}}

            \tikzfeynmanset{every vertex={empty dot,minimum size=0mm}}
             \vertex (a1);
            \vertex[right=0.5cm of a1] (a2);
            \vertex[right=0.5cm of a2] (a3);
            \vertex[below=1cm of a3] (a4);
            
            \vertex[below=0.5cm of a1] (t1);
            \vertex[below=0.5cm of a3] (t2);
            
            \vertex[left=0.5cm of a4] (a5);
            \vertex[left=0.5cm of a5] (a6);
                          
            \tikzfeynmanset{every vertex={dot,minimum size=0.8mm}}
            
            \vertex[right=0cm of a1] (e1);        
            \vertex[right=0cm of a3] (e2);        
            \vertex[right=0cm of a4] (e3);        
            \vertex[right=0cm of a6] (e4);    
            
            \tikzfeynmanset{every vertex={empty dot,minimum size=0mm}}                
  
            \vertex[left=0.15cm of e1] (d1);        
            \vertex[right=0.15cm of e2] (d2);        
            \vertex[right=0.15cm of e3] (d3);        
            \vertex[left=0.15cm of e4] (d4);    
            
                \diagram*[large]{	
                (a1)--(a2)--(a3)--(a4)--(a5)--(a6)--(a1), 
                (a2)--(a5),
             		(e1)--(d1),
		(e2)--(d2),
		(e3)--(d3),
		(e4)--(d4),
		(t1)--(t2),
                           }; 
            \end{feynman}
    \end{tikzpicture} \\[-0.1cm]
Fishnet
\end{tabular}}
& \multirow{3}{*}{0}& \multirow{1}{*}{LTD}& \multirow{1}{*}{\phantom{}0} & 1.9$\cdot$10$^2$ & 3.6$\cdot$10$^3$ &  6.6  &  & & \\[0.2em]
\cline{3-7}
& & \multirow{2}{*}{\cLTD{}}& 0 & 2.3$\cdot$10$^4$  & 1.6$\cdot$10$^5$ &   & 1.2$\cdot$10$^2$ & 44 & \\[0.2em]
& & & 1000 &  3.5$\cdot$10$^3$ & 3.5$\cdot$10$^3$ & 6.6 & 18 & 0.96 & 1.0 \\[0.2em]
\hline
\end{tabular}%
}%

\end{center}
\caption{
Number of operations (additions $N_+$ and multiplications $N_\times$) and evaluation time ($t$) in microseconds (on one Intel(R) Xeon(R) Gold 6136 CPU @ 3.00GHz core) of the implementation of the \cLTD{} and LTD representations for 12 topologies with varying propagator count and numerator rank. The \texttt{opt} column refers to the number of iteration of the \texttt{Local Stochastic Search} optimisation procedure used by the \textsc{form}~\cite{Ruijl:2017dtg, Kuipers:2013pba,Ruijl:2014spa} program to optimise the \texttt{C}-implementation of the \cLTD{} representation.
The case of \texttt{opt}=0 corresponds to the non-optimised output, matching what one obtains when directly reading off the expression stemming from applying the iterative \cLTD{} procedure presented in sect.~\ref{sec:loop-by-loop_iteration}.
}
\label{tab:results}
\end{table}
\clearpage

\newpage
\section{Conclusion}
\label{sec:conclusion}
In this work we have introduced a new iterative and systematic procedure that, given a multi-loop multi-scale Feynman diagram with arbitrary numerator, yields a representation of the integrand that is manifestly free of spurious poles. The resulting integrand has poles at physical thresholds only, which are regulated by a Feynman prescription with a definite sign. These physical thresholds, or E-surfaces, correspond to sets of particles whose four-momenta simultaneously become on-shell. In the original LTD representation there are spurious poles in the summands, which cancel in the sum. Our work introduces, for the first time, a general procedure to realise their cancellation \emph{within each individual term}.

The derivation of this representation involves two basic steps which are iterated over for the integration of each loop energy variable. At each iteration, the two steps are applied in succession: first, we analytically perform the integration in a loop energy variable using residue theorem, and then we apply a partial fractioning procedure on all the propagators involving the loop energy integrated over at this step. In this way, we show that it is possible to reabsorb spurious singularities in the definition of divided differences of the product of the numerator and physical poles of the loop integral. The analytic behaviour of divided differences of such quantities can be determined immediately, and in particular it allows us to conclude that the resulting expression is finite for all spurious poles. We refer to this new representation of the integral as the Manifestly Causal Loop-Tree Duality representation, or \cLTD{}. 

Our procedure readily applies to loop integrals with arbitrary numerators. In particular, when the numerator of the Feynman integral is a polynomial, as it in most cases of practical interest, we recognize that divided differences can be computed explicitly. This leads to an expression in which no spurious poles appear, neither explicitly nor implicitly. 
We studied the resulting numerical stability of the \cLTD{} expression and found the expected perfect numerical stability arbitrarily far in the UV region. In the IR region however, the \cLTD{} representation yields a marginally worse numerical stability when using double precision arithmetic, which we however show to be completely cured by the promotion to quadruple precision. This highlights the complementarity of the \cLTD{} and LTD representation for the numerical application of the Loop-Tree Duality.

One additional potential drawback of the \cLTD{} representation is that the number of terms generated when unfolding our iterative procedure grows exponentially in the number of propagators.
We however demonstrated with twelve benchmark integral examples of varying complexity that identifying common sub-expressions in these many terms can drastically reduce the numerical complexity of the implementation of the \cLTD{} representation and render it competitive, and sometime even advantageous, over that of its original LTD counterpart. 
This is especially true for integrals of lower numerator ranks, and together with the ease of accommodating raised propagators in the \cLTD{} expression, it is well suited to tackle the numerical computation of the more complicated master scalar integrals appearing in the reduction basis of the traditional analytic computation of scattering multi-loop amplitudes.
Furthermore, we provide as ancillary material a standalone program offering an automated generation of our novel \cLTD{} representation for an arbitrary loop integral.

The denominators involved in the \cLTD{} expression only feature E-surfaces, a structure also exhibited by the terms arising from Time Ordered Perturbation Theory.
We therefore investigated in more detail the connection between these two representations and observed that they locally evaluate to the same quantity but have a different functional form.
In particular, the factorial growth of TOPT  with the number of diagram vertices is not reflected in \cLTD{}. Instead, we find that it is possible to identify subsets of TOPT terms that combine into individual terms of \cLTD{}.
We conclude then that LTD, \cLTD{} and TOPT are all equivalent formulations of the same abstract object: the loop integral integrated over all loop momentum energies. Each of these formulations offer their particular benefits, and \cLTD{} features both causal denominator structures together with a computational complexity that can be made comparable to that of LTD for cases of practical interest.

The ability to systematically analyse and regularise the pole structure of Feynman diagrams, together with the engineering of integrands that are numerically stable and fast are two key steps towards the fully numerical computation of higher-order corrections to the prediction of collider observables. Our novel Manifestly Causal Loop-Tree Duality representation achieves both of these goals and offers a clear path to physics applications.

\section{Acknowledgements}

This project has received funding from the European Research Council (ERC) under grant agreement No 694712 (PertQCD) and SNSF grant No 179016.

\newpage
\appendix
\section{Recursive cancellation of spurious singularities}
\label{sec:recursion-derivation}

In what follows we derive the recursion relation presented in eq.~\eqref{eq:pf_general_step} that allows to remove in a systematic way all the spurious singularities from our starting expression in eq.\eqref{eq: F_integrated} with $n>0$.
Remember that if $n=1$ it is already free of spurious singularities.
We start by selecting an (arbitary) order in $\bx=(x_1,\dots,x_n)$, pick its first variable $x_1$ and work towards the intermediate goal of deriving an expression where all spurious singularities at $x_1=x_j$ for $1<j\leq \dim(\bx)$ are explicitly removed.
As we will see, this result can in turn be expressed as a recursive formula of terms similar to the starting expression, which are now manifestly free of singularities in $x_1$.
The application of this recursive formula is described in sect.~\ref{sec:generic_partial_fractioning}, where we apply this relation recursively to our starting expression in eq.~\eqref{eq: F_integrated} to render it manifestly free of all spurious singularities, resulting in eq.~\eqref{eq:pf_1loop_compact}. \par

More explicitly, we consider our starting expression in eq.~\eqref{eq: F_integrated} and split off the first summand, as
\begin{equation}
\label{eq:F_integrated_appendix}
    F\left(\mqty{\bx\\\bxbar}\,;\mathcal{N}\right)
    =
    \frac{\text{N}_\mathcal{N}([x_1]) E(x_1|\mathbf{\bar x})}{\prod_{j\neq 1}(x_1-x_j)}
    +
    \sum_{i=2}^n \frac{\text{N}_\mathcal{N}([x_i]) E(x_i|\mathbf{\bar x})}{\prod_{j\neq i}(x_i-x_j)}.
\end{equation}
We then use the identity
\begin{align}
\label{eq:partial_fractioning_identity_appendix}
    \sum_{i=1}^n \frac{1}{\prod_{j\neq i} (x_i-x_j)}
    = 0
\end{align}
and apply it to the first summand in eq.~\eqref{eq:F_integrated_appendix} and obtain
\begin{align}
\label{eq:pf_summand_appendix}
    \frac{
        \text{N}_\mathcal{N}([x_1]) E(x_1|\mathbf{\bar x})
        }{
        \prod_{j\neq 1}(x_1-x_j)}
    &=
    -
    \text{N}_\mathcal{N}([x_1])
    \sum_{i=2}^n
    \frac{
        E(x_1|\mathbf{\bar x})
        }{
        \prod_{j\neq i}(x_i-x_j)}.
\end{align}
Note that the identity in eq.~\eqref{eq:partial_fractioning_identity_appendix} allows us to move each of the poles $x_1=x_j$ for $1<j\leq n$ in the product on the left-hand side in eq.~\eqref{eq:pf_summand_appendix} into a separate summand on the right-hand side.
As a next step, we want to remove this pole from each of the summands.
This we will do through an iteration over $\bxbar$.
Therefore, we first pick an (arbitrary) order in $\bxbar=(\bar x_1,\dots,\bar x_{\bar n})$.
We observe that using the relation
\begin{align}
    E(x_1|\bar x_1) - E(x_i|\bar x_1)
    &=
    (x_i-x_1)E(x_1|\bar x_1)E(x_i|\bar x_1),
\end{align}
we can write
\begin{align}
\label{eq:removing_pole_step_appendix}
    \frac{E(x_1|\bar x_1)}{\prod_{j\neq i}(x_i-x_j)}
    &=
    \frac{E(x_1|\bar x_1)}{\prod_{j\neq i}(x_i-x_j)}
    -
    \frac{E(x_i|\bar x_1)}{\prod_{j\neq i}(x_i-x_j)}
    +
    \frac{E(x_i|\bar x_1)}{\prod_{j\neq i}(x_i-x_j)} \\
    &=
    \frac{
        E(x_1|\bar x_1)E(x_i|\bar x_1)
        }{\prod_{j\neq i,1}(x_i-x_j)}
    +
    \frac{
        E(x_i|\bar x_1)
        }{\prod_{j\neq i}(x_i-x_j)},
\end{align}
such that the pole at $x_i=x_1$ has been regulated in the first summand while remaining present in the second term with the numerator now evaluated at $x_i$ instead of $x_1$.

Using this result and combining factors as $E(x_1|\bxbar) = E(x_1|\bar x_1) E(x_1|\bxbar_{(2,)})$
we can express each summand on the right-hand side of eq.~\eqref{eq:pf_summand_appendix} as
\begin{align}
\label{eq:first_summand_pole_removed}
    \frac{
        E(x_1|\mathbf{\bar x})
        }{\prod_{j\neq i}(x_i-x_j)}
    &=
    \frac{
        E(x_1|\mathbf{\bar x}) E(x_i|\bar x_1)
        }{\prod_{j\neq i,1}(x_i-x_j)}
    +
    \frac{
        E(x_1|\bxbar_{(2,)})
        E(x_i|\bar x_1)
        }{\prod_{j\neq i}(x_i-x_j)}.
\end{align}
Analogously, we apply the relation in eq.~\eqref{eq:removing_pole_step_appendix} with $\bar x_1 \leftrightarrow \bar x_2$ to the second summand in eq.~\eqref{eq:first_summand_pole_removed}.
This step is performed for all $\bar x_r$ with $1\leq r \leq \bar n$ such that we arrive at
\begin{align}
\label{eq:pf_summand_no_spurious}
    \frac{
        E(x_1|\mathbf{\bar x})
        }{\prod_{j\neq i}(x_i-x_j)}
    &=
    \sum_{r=1}^{\bar n}
    \frac{
        E(x_1|\bxbar_{(r,)}) E(x_i|\bxbar_{(1,r)})
        }{\prod_{j\neq i,1}(x_i-x_j)}
    +
    \frac{
        E(x_i|\mathbf{\bar x})
        }{\prod_{j\neq i}(x_i-x_j)}.
\end{align}
Before we now use this result from eq.~\eqref{eq:pf_summand_no_spurious}, we first note that eq.~\eqref{eq:F_integrated_appendix} and
eq.~\eqref{eq:pf_summand_appendix} can be combined into a single sum as
\begin{align}
    F\left(\mqty{\bx\\\bxbar}\,;\mathcal{N}\right)
    =
    \sum_{i=2}^n
    \left(
    \frac{\text{N}_\mathcal{N}([x_i]) E(x_i|\mathbf{\bar x})}{\prod_{j \neq i}(x_i-x_j)}
    -
    \frac{
        \text{N}_\mathcal{N}([x_1]) E(x_1|\mathbf{\bar x})
        }{
        \prod_{j \neq i}(x_i-x_j)}
    \right)
    ,
\end{align}
such that plugging in the relation in eq.~\eqref{eq:pf_summand_no_spurious} yields
\begin{align}
\label{eq:implicit_recursion_appendix}
    F\left(\mqty{\bx\\\bxbar}\,;\mathcal{N}\right)
    =
    -
    \sum_{i=2}^n
    \sum_{r=1}^{\bar n}
    \frac{
        \text{N}_\mathcal{N}([x_1])
        E(x_1|\bxbar_{(r,)}) E(x_i|\bxbar_{(1,r)})
        }{\prod_{j\neq i,1}(x_i-x_j)}
    +
    \sum_{i=2}^n
    \frac{
        \text{N}_\mathcal{N}([x_1,x_i])
        E(x_i|\mathbf{\bar x})
        }{\prod_{j\neq i,1}(x_i-x_j)}
    ,
\end{align}
Note that all of the expressions on the right-hand side are manifestly free of singularities in $x_1$, as no explicit denominators $(x_1-x_j)$ with $1<j\leq n$ appear.
Of course, there are is a hidden dependence on the denominator $(x_i-x_1)$ in the divided difference $\text{N}_\mathcal{N}([x_1,x_i])$ for $1<i\leq n$, which however is regular as we discussed in sect.~\ref{sec:degenerate_case}.
This observation now reveals that the singularities in $x_1$ of the left-hand side, as they appear in each summand in eq.~\eqref{eq:F_integrated_appendix} are indeed spurious.
Futhermore, the arbitrary order in $\bx$, respectively the choice of $x_1$ at the beginning of this section makes it clear that the recursion can be used to explicitly remove spurious singularities in any variable $x_i$ for $1\leq i \leq n$.
We now observe that eq.~\eqref{eq:implicit_recursion_appendix} takes the implicit form of a recursion that we can make explicit by writing
\begin{equation}
    F\left(\mqty{\bx\\\bxbar}\,;\mathcal{N}(z)\right)
=   
-\text{N}_\mathcal{N}([x_1])
    \sum_{r=1}^{\bar n} E\left(x_1|\bxbar_{(r,)}\right) F\left(\mqty{\bx_{(2,)}\\\bxbar_{(1,r)}};1\right)
+F\left(\mqty{\bx_{(2,)}\\\bxbar};\text{N}_\mathcal{N}([x_1,z])\right).
\end{equation}
It is straightforward to make this recursion generic for an intermediate step with an arbitrary numerator $\mathcal{F}$ and vectors $\by\in(\mathbb{H}^*)^{\dim(\by)}$ and $\bybar\in(\mathbb{H}^*)^{\dim(\bybar)}$.
One then recovers precisely the recursion in eq.~\eqref{eq:pf_general_step}, which concludes its derivation.

\section{Multi-loop notation}\label{sec:multi_loop_notation}
It is convenient to have a way to represent the whole numerator expression also for the multi-loop case. 
This is particularly convenient if one wants to produce a symbolic expression defined by a choice of loop momenta routing and number of propagators.
As was already mentioned in sect.~\ref{sec:loop-by-loop_iteration}, one can derive an expression assuming all the propagators to be distinct and the corresponding expression will be valid also in the various degenerate cases.
\\
In order to organise the successive evaluations of the numerator function at each step of the iterative \cLTD{} procedure, one must keep track of all the limits in which the numerator has to be evaluated after taking each residue; this can easily be done by extending on the one-loop notation for the divided difference in eq.~\eqref{eq: N_definition}.
For the general case of a $L$-variate numerator function $\mathcal{N}(k^0_1,\dots, k^0_L)$ one can collect all the limits coming from integrating all the loop momenta as 
\begin{equation}
\text{N}_\mathcal{N}\left([\mathbf{z}_1],\dots,[\mathbf{z}_{L}]\right),
\end{equation}
where in general $\mathbf{z}_i=\mathbf{z}_i(k_{i+1},\dots,k_{L})$ have an implicit dependence on the integration variables $k_j$. In order to unfold this definition and take all the limits and differences in the correct order one must apply the following recursive steps,
\begin{align}
\begin{split}
&\text{N}_\mathcal{N}\left(y_1,\dots,y_m,[\mathbf{z}_1,x_1,x_2],\dots,[\mathbf{z}_n]\right)
    \\&\hspace{2.3cm}=\frac{
        \text{N}_\mathcal{N}\left(y_1,\dots,y_m,[\mathbf{z}_1,x_1],\dots,[\mathbf{z}_n]\right)
        -\text{N}_\mathcal{N}\left(y_1,\dots,y_m,[\mathbf{z}_1,x_2],\dots,[\mathbf{z}_n]\right)
        }{x_1-x_2},
\end{split} \\
&\text{N}_\mathcal{N}\left(y_1,\dots,y_m,[y_{m+1}],[\mathbf{z}_2],\dots,[\mathbf{z}_n]\right)
    =\lim_{k_{m+1}\rightarrow y_{m+1}}
    \text{N}_\mathcal{N}\left(y_1,\dots,y_m,y_{m+1},[\mathbf{z}_2],\dots,[\mathbf{z}_n]\right).
\end{align}
The limit ensures that all the implicit dependencies are evaluated on the correct point.
Once all the limits have been taken one can substitute the definition of the numerator function as
\begin{equation}
\text{N}_\mathcal{N}\left(y_1,\dots,y_{n+m}\right)  = \mathcal{N}(y_1,\dots,y_{n+m}).
\end{equation}
For explicit polynomials, one can optimize this evaluation by applying sequentially the set of limits $\mathbf{z}_i$ by means of eq.~\eqref{eq:pf_poly_numerator}, thus obtaining at each step a polynomial that depends on one less loop momentum energy, hence minimising the depth of the recursion.

\section{Usage of {\texttt{cLTD.py}}}
\label{app:pyfractioning}

The Python script \texttt{cLTD.py} supplied in ancillary material of this work contains the function \texttt{integrate\_energies} that generates the \cLTD{} expression fully automatically. Its input is a topology, described in terms of a list of loop lines. A loop line is the collection of all propagators with the same loop momenta flowing through it.

All two-loop Feynman diagrams can be expressed in only three loop lines. For a particular choice of loop momentum routing, these loop lines have the loop momentum dependence $k$, $l$, and $k-l$, or written in terms of signatures for the basis $(k, l)$: \texttt{[[1, 0], [0, 1], [1, -1]]}.

For example, the two-loop pentabox topology $\begin{gathered}  \scalebox{0.9}{
    \begin{tikzpicture}[baseline,every node/.style={inner sep=0,outer sep=0}]
    \begin{feynman}
    
     \tikzfeynmanset{every vertex={empty dot,minimum size=0mm}}
    \vertex (a1);
    
    \tikzfeynmanset{every vertex={dot,minimum size=0.8mm}}
    \vertex[right=1cm of a1] (a3);
    
    \tikzfeynmanset{every vertex={empty dot,minimum size=0mm}}
    
    \vertex[right=0.5cm of a1] (a2);
    \vertex[above=0.5cm of a2] (a4);
    \vertex[below=0.5cm of a2] (a5);
    
        \vertex[left=0.433cm of a2] (b1);
    \vertex[left=0.25cm of a2] (b2);
    
    \vertex[right=0.433cm of a2] (b3);
    \vertex[right=0.25cm of a2] (b4);
    
    \vertex[left=0.433cm of a2] (d1);
    \vertex[left=0.25cm of a2] (d2);
    
    \vertex[right=0.433cm of a2] (d3);
    \vertex[right=0.25cm of a2] (d4);
    
    \tikzfeynmanset{every vertex={dot,minimum size=0.8mm}}
    
    \vertex[above=0.433cm of b2] (c2);

    \vertex[above=0.433cm of b4] (c4);

    \vertex[below=0.433cm of b2] (e2);

    \vertex[below=0.433cm of b4] (e4);
    
    \tikzfeynmanset{every vertex={empty dot,minimum size=0mm}}

    \vertex[left=0.15cm of c2] (q2);
    \vertex[left=0.15cm of e2] (q4);

    \vertex[right=0.15cm of c4] (p2);
    \vertex[right=0.15cm of e4] (p4);
    \vertex[right=0.15cm of a3] (p5);
    
        \diagram*[]{	
        (a1)--[quarter left](a4) -- [quarter left](a3),
        (a5) --[quarter right](a3), 
        (a5)--[quarter left](a1),
        (a4)--(a5),
        
       (c2) -- (q2),
        (e2) -- (q4),
             
       (c4) -- (p2),
        (e4) -- (p4),
        (a3) -- (p5),
        }; 
    \end{feynman}
    \end{tikzpicture}
    }\end{gathered}$ can be denoted in terms of the above signature together with the specification of the number of propagators per loop line: \texttt{[3, 1, 4]}.

The input to \texttt{integrate\_energies} is then:
\begin{lstlisting}
signatures = [[1, 0], [1, -1], [0, 1]]
n_props = [1, 3, 4]

res = integrate_energies(n_props, signatures,
                        verbose=False,
                        name="PentaBox",
                        output_type="mathematica")
\end{lstlisting}
which yields the \cLTD{} expression output in the above example via the Mathematica file\\\noindent \texttt{cLTD\_2l\_pentabox.m} and \texttt{cLTD\_2l\_pentabox.den}. Other output formats are \texttt{yaml}, \texttt{pickle} (binary), and \texttt{FORM}~\cite{Ruijl:2017dtg}.

\newpage
For the simpler topology of the 2-loop sunrise diagram we have as input:
\begin{lstlisting}
signatures = [[1, 0], [1, -1], [0, 1]]
n_props = [1, 1, 1]

res = integrate_energies(n_props, signatures,
                         verbose=False,
                         name='Sunrise',
                         output_type='FORM')
\end{lstlisting}
and we obtain as \textsc{form} output:
\begin{lstlisting}
L F = 
  -1*invd0*num(ncmd(+1*k1+1*E1-1*p1), ncmd(+1*E2-1*p2))
  -1*invd1*num(ncmd(+1*E0-1*p0), ncmd(+1*E0-1*p0+1*E1+1*p1))
  -1*num(ncmd(+1*E0-1*p0), ncmd(+1*E2-1*p2, +1*E0-1*p0+1*E1+1*p1))
  -1*num(ncmd(+1*E0-1*p0, +1*k1+1*E1-1*p1), ncmd(+1*E2-1*p2))
;
\end{lstlisting}
and
\begin{lstlisting}
d0 = +1*E0+1*p0+1*E1-1*p1+1*E2-1*p2;
d1 = +1*E0-1*p0+1*E1+1*p1+1*E2+1*p2;
\end{lstlisting}
where \texttt{invd`i'= 1/d`i'}. The \textsc{form} expression \texttt{F} gives the most generic form of the numerator that has to be matched to the powers of $k^0$ and $l^0$ in the input, and has to be unfolded using eq.~\eqref{eq:pf_poly_numerator} and eq~\eqref{eq:pf_poly_numerator2}, where \texttt{num(ncmd(x1), ncmd(x2))} represents $\text{N}_\mathcal{N}([x_1],[x_2])$, as described in app.~\ref{sec:multi_loop_notation}.

An example \textsc{form} code on how to unfold the numerator equations is provided in the ancillary material.

\newpage

\bibliographystyle{JHEP}

\bibliography{biblio}

\end{document}